\documentclass[%
 aip,
 jmp,%
 amsmath,amssymb,
 reprint
]{revtex4-2}

\usepackage{graphicx}
\usepackage{dcolumn}
\usepackage{bm}
\usepackage[usenames,dvipsnames]{color}
\usepackage{stmaryrd} 

\usepackage{cancel}
\usepackage[dvipsnames]{xcolor}

\usepackage[normalem]{ulem}

\begin{document}

\setlength\parindent{0pt}

\preprint{AIP/123-QED}

\title[Girsanov reweighting for underdamped Langevin dynamics]{Girsanov reweighting for simulations of underdamped Langevin dynamics. Theory.}%
\author{Stefanie Kieninger}
\altaffiliation{equal contribution}
\affiliation{ 
Freie Universität Berlin, Department of Biology, Chemistry and Pharmacy, Arnimallee 22, 14195 Berlin
}

\author{Simon Ghysbrecht}
\altaffiliation{equal contribution}
\affiliation{ 
Freie Universität Berlin, Department of Biology, Chemistry and Pharmacy, Arnimallee 22, 14195 Berlin
}
%
\author{Bettina G.~Keller}%
\email{bettina.keller@fu-berlin.de}
\affiliation{ 
Freie Universität Berlin, Department of Biology, Chemistry and Pharmacy, Arnimallee 22, 14195 Berlin
}%

\date{\today}

%
%
\begin{abstract}
The critical step in a molecular process is often a rare-event and has to be simulated by an enhanced sampling protocol.
Recovering accurate dynamical estimates from such biased simulation is challenging. 
Girsanov reweighting is a method to reweight dynamic properties formulated as path expected values. 
The path probability is calculated at the time-step resolution of the molecular-dynamics integrator.
But the theory is largely limited to overdamped Langevin dynamics.
For underdamped Langevin dynamics, the absolute continuity of the path probability ratio for the biased and unbiased potential is not guaranteed, but it depends on the Langevin integrator. 
We develop a general approach to derive the path probability ratio for Langevin integrators and to analyze whether absolute continuity is fulfilled.
We demonstrate our approach on symmetric splitting methods for underdamped Langevin dynamics. 
For methods that obey absolute continuity, and thus can be used for Girsanov reweighting, we provide an expression for the relative path probability.
\end{abstract}

\maketitle

\section{Introduction}

Understanding rare events in molecular systems on an atomistic resolution would have great impact in many areas, such as the binding of drug molecules to receptors, protein-protein interactions in molecular machines, aggregation processes in biomolecular systems or artificial materials, phase transitions and chemical reactions.
In principle, these processes can be studied by molecular dynamics (MD) simulations \cite{Karplus:2002,Dror:2012,Lane:2013}.
However, the timescale of molecular rare events are often well beyond the timescales that can be reached by direct MD simulations. 
Even if occasional rare-event transitions can be observed in the course of a direct MD simulation, the estimates of thermodynamic or kinetic properties are often not statistically meaningful.
This is because passage times across a barrier into a target state are long-tailed distributed, and the tail contributes to the rare-event estimate. 
Furthermore, the dynamics in the fast degrees of freedom influence the free-energy surface and the diffusion constant of the rare-event transition in ways that are hard to predict, and any rare event can consist of multiple separate transition paths with different intermediate states and transition states. %
In short: it is important to sample the full path ensemble that contributes to a rare-event.

One approach to speed up the sampling of rare events is to add a bias to the molecular interaction potential.
Enhanced sampling methods like metadynamics\cite{Huber:1994, Grubmuller:1995, Darve:2001, Laio:2002, Barducci:2008} and umbrella sampling\cite{Torrie:1977, Kaestner:2005} add energy to the system in order to steer the simulation away from states which have already sufficiently been explored.
Since enhanced sampling changes the dynamics of the system, estimates of thermodynamics and dynamic properties are distorted and need to be unbiased.
For thermodynamic properties, estimators that accurately reweight the enhanced sampling simulations, such as weighted histogram analysis method \cite{Swendsen:1992, Kumar:1992}, are available.
With these reweighting techniques, the statistical certainty of thermodynamic properties is drastically increased compared to direct simulations. 
As a result, the field is moving from direct simulations to combining enhanced sampling with thermodynamic reweighting techniques.
For dynamic properties, on the other hand, one cannot yet routinely use enhanced sampling simulations.
In fact, dynamic reweighting techniques are currently a very active field of research \cite{Kieninger:2020}. 
Dynamic properties are path expected values weighted by the path probability density which depends on the potential energy function.
Suppose, the system has been simulated at a potential $V$, and one would like to know a dynamic property at a target potential $\widetilde{V} = V + U$.
To reweight the corresponding path expected value, one needs the relative path probability, i.e the ratio of the path probabilities at $\widetilde{V}$ and at $V$.
Furthermore, the relative path probability needs to obey absolute continuity, i.e. any path that is possible the target potential $\widetilde{V}$ also needs to be possible at the simulation potential $V$. 

Several methods to reweight simulations with bias potentials have been proposed in recent years. 
Usually an effective model of the dynamics is assumed, either a two-state dynamics in which transition state theory or Kramers' rate theory holds \cite{deOliveira:2007, Doshi:2011, Tiwary:2013, Frank:2016, Palacio:2022b} or a Markov state model on a discretized state space\cite{Bicout:1998, Wu:2014, Mey:2014, Rosta:2015, Wu:2016, Stelzl:2017}. 
Since the validity of these effective models changes with the potential energy function, they cannot be equally valid at $V$ and at $\widetilde{V}$. 
It is difficult to judge how this affects the accuracy of the reweighted estimate.
We argue that a more accurate approach is to consider the calculation of the path probability ratio as part of the enhanced sampling simulation. 
In the subsequent analysis, one would use the simulated paths along with the time-series of the relative path probability to reweight the desired dynamic property. 
For this the relative path probability ratio needs to be calculated at the time-step resolution of the MD simulation, and the corresponding equations need to match the MD integrator. 
Additionally, at this high time-resolution, the question of absolute continuity needs to be addressed.
Based on works by L.~Onsager and S.~Machlup \cite{Onsager:1953} and, independently, by I.V.~Girsanov \cite{Girsanov1960}, one can derive an exact reweighting technique, in which the path probability ratio is calculated at the time-step resolution of the MD simulation. 
For overdamped Langevin dynamics, the Girsanov theorem guarantees that the relative path probability does not violate absolute continuity, as long as the biasing forces do not approach infinity.
This guarantee holds even for continuous solutions of the stochastic differential equation \cite{Girsanov1960, Oeksendal:2003}.
Additionally, the expression for the relative path probability for time-discretized paths generated by the Euler-Maruyama algorithm is well established \cite{Kloeden:1992, Oeksendal:2003}.
Since the late 1990s, it has been shown several times that Girsanov reweighting or, equivalently, dynamic importance sampling can be used to unbias overdamped Langevin dynamics, both for model potentials \cite{Mazonka:1998, Woolf:1998, Zuckerman:1999, Zuckerman:2000, Adib:2008} and for small molecular systems \cite{Jang:2006, Schuette:2015, Bolhuis:2022}.
But the use of overdamped Langevin dynamics to model molecular rare events is limited.  
It can be used in the mesoscopic molecular regime, in which molecules are (partly) treated as rigid bodies, to study molecular crowding effects, association processes between large molecules, and even the dynamics of coarse-grained polymers \cite{Huber:2019, Cholko:2022}
But because overdamped Langevin dynamics suppresses the fast intramolecular fluctuations, it cannot be used to model conformational transitions at atomistic resolutions.
By contrast, underdamped Langevin dynamics, often under the name ,,Langevin thermostats'', is an accurate and frequently used equation of motion for atomistic MD \cite{Huenenberger:2005}.
For continuous solutions of underdamped Langevin dynamics, the path probability can be formulated\cite{Kwon:2022}, but one cannot guarantee that the relative path probability obeys absolute continuity.
This, at first, seems like an impasse in the attempt to unbias dynamic estimates: one needs to sample the molecular rare events by underdamped Langevin dynamics, but for underdamped Langevin dynamics the relative path probability might not exist. 
Fortunately, when reweighting a MD simulation, the path expected value is not calculated for time-continuous paths but for time-discretized paths. 
As explained above, an accurate reweighting method needs to calculate the relative path probability at the time resolution at which the path is produced. 
Thus, depending on the integrator used to propagate the underdamped Langevin dynamics, the relative path probability density may exist after all, and Girsanov reweighting may become possible. 
Our approach is therefore not to discretize the continuous path integral. \cite{Adib:2008}. 
Instead we start from already existing algorithms to propagate the equation of motion for underdamped Langevin dynamics and derive the relative path probability for the resulting time-discretized paths. 
Girsanov reweighting for underdamped Langevin dynamics has first been reported in Refs.~\onlinecite{Athenes:2004} and  \onlinecite{Xing:2006}.
In Ref.~\onlinecite{Donati:2017} we introduced a formulation of the path probability ratio as a function of the random numbers generated during the simulation potential at $V$ and a random number difference to $\widetilde V$, which we called reweighting on-the-fly.
This allowed us on the one hand to efficiently calculate part of the relative path probability already during the simulation.
On the under hand, it allowed us to approximate the relative path probability for underdamped Langevin dynamics. 
With this approximation, we could reweight metadynamics simulations of the folding equilibrium of $\beta$-hairpin peptide \cite{Donati:2018}.
In Ref.~\onlinecite{Kieninger:2021} we analyzed this approximation and derived the path probability ratio for a simple Langevin integrator.
The goal of this contribution is to develop a general approach to derive the path probability for Langevin integrators and to analyze whether the relative path probability obeys absolute continuity.
We focus on  symmetric splitting methods for underdamped Langevin dynamics \cite{Bussi:2007, Leimkuhler:2012, Leimkuhler:2013, Sivak:2013}, and additionally include a closely related\cite{Kieninger:2022} variant\cite{BouRabee:2010, Goga:2012} which is used as default Langevin thermostat in several MD simulation programs \cite{Eastman:2017, Case:2005, VanDerSpoel:2005}. 
For methods that obey absolute continuity, and thus can be used for Girsanov reweighting, we provide an expression for the relative path probability.
We chose symmetric splitting methods, because their derivation is documented in great detail \cite{Bussi:2007, Leimkuhler:2012, Leimkuhler:2013, Sivak:2013, Leimkuhler:2015}, which provides an easy entry point for our analysis. 
Additionally, the convergence properties of this class of integrators are well-understood
\cite{Leimkuhler:2013, Sivak:2014, Leimkuhler:2016b, Fass:2018}. 
However, the development of integrators for underdamped Langevin dynamics has been a very active field for decades and many other algorithms have been proposed 
\cite{Brunger:1984, vanGunsteren:1988, Pastor:1988, Hershkovitz:1998, Paterlini:1998, Skeel:2002, Ricci:2003, VandenEijnden:2006, Melchionna:2007, Izaguirre:2010, Hongen:2011, GronbechJensen:2013, Peters:2014, Zhang:2019, Gronbech:2020, Finkelstein:2021}. 
We believe our approach to derive the path probability ratio can be applied to these Langevin integrators, too.  
The article is structured as follows: In sections \ref{sec:GirsanovReweighting} and \ref{sec:LangevinIntegrators} we review the theory of Girsanov reweighting and of symmetric splitting algorithms.
Sections \ref{sec:image} and \ref{sec:AbsoluteContinuity} contain our analysis of these integrators and the derivation of the relative path probabilities.  
In section \ref{sec:graphicalRepresentation} we introduce a graphical representation of Langevin integrators that helped us to visualize and classify the effects of changeing the potential energy on the behaviour of the Langevin integrator.
Section \ref{sec:conclusion} contains a short discussion and outlook.

\section{Girsanov reweighting}
\label{sec:GirsanovReweighting}

\subsection{Equation of motion, path probability density, and  path integral}
Consider a particle with mass $m$ that moves in a one-dimensional position space $q \in \mathbb{R}$ according to underdamped Langevin dynamics
\begin{align}
    \dot{q} &= \frac{p}{m}   \cr
    \dot{p} &= - \nabla_q V(q) - \xi p + \sqrt{2\xi k_BTm} \, \eta(t) \, ,
 \label{eq:underdamped_Langevin_01}    
\end{align}
In eq.~\ref{eq:underdamped_Langevin_01},
 $V(q)$ is the potential energy function,
$\nabla_q = \partial/\partial q$ denotes the gradient with respect to the position coordinate,
$\xi$  is a collision or friction rate (in units of s$^{-1}$),
$T$ is the temperature and $k_B$ is the  Boltzmann constant. 
$\eta \in \mathbb{R}$ is an uncorrelated Gaussian white noise with unit variance centered at zero
$ \langle \eta(t)\eta(t') \rangle = \delta(t-t')$,
where $\delta(t-t')$ is the Dirac delta-function.
We use the dot-notation for derivatives with respect to time: $\dot q = \partial q / \partial t$.
$x(t) = (q(t), p(t)) \in \Gamma \subset \mathbb{R}^2$ denotes the state of the system at time $t$, which consists of positions $q(t)$ and conjugated momenta $p(t) = m\dot{q}(t)$.
$\Gamma$ is called state space or phase space of the system. 

A Langevin integrator yields a time-discretised solution of eq.~\ref{eq:underdamped_Langevin_01}: $\mathbf{x} = (x_0, x_1, x_2, \dots x_n)$. 
The path  $\mathbf{x}$ is a sequence of $n+1$ states $x_k = (q_k, p_k) \in \Gamma$, 
where consecutive states $x_k$ and $x_{k+1}$ are separated by a (small) time step $\Delta t$, 
$q_k$ is the position at time $k\Delta t$, and
$p_k$ is the momentum at time $k\Delta t$.
$x_0 =(q_0, p_0)$ is the initial state of the path, and $\tau = n \Delta t$ is the path length.
A time-discretized path $\mathbf{x}$ is an element of the space $\mathcal{S} = \Gamma^{n+1}$. 
Its path probability density is
$
\mathcal{P}[\mathbf{x}]	= p(x_0) \cdot \mathcal{P}[x_1 \dots x_n | x_0] \, ,
$
where we wrote the path probability as a product of the probability density of the intial state $p(x_0)$ and the conditional probability of the path $\mathcal{P}[x_1 \dots x_n | x_0]$, given that the path starts in $x_0$.
The Langevin integrators considered in this contribution implement a Markov process.
That is, the probability to observe the state $x_{k+1}$ at time $t= (k+1) \Delta t$ depends only on the previous state $x_k$ at time $t = k\Delta t$ and not on any of the states before that, i.e.
$
	p(x_{k+1} | x_k, x_{k-1} \dots x_0) =p(x_{k+1} | x_k) 
$.
The path probability density to observe a particular path $\mathbf{x}$ can then be written as a product of the single-step transition probabilities $p(x_{k+1} | x_k)$
\begin{eqnarray}
	P[\mathbf{x}] = p(x_0)  \cdot P[x_1 \dots x_n | x_0] &=& p(x_0)  \cdot \prod_{k=0}^{n-1} p(x_{k+1} | x_k)\, .
\label{eq:pathProbability_timeDiscretized}	
\end{eqnarray}
The probability density of the initial state depends on the setup of the computational experiment. 
Here, we assume that the path $\mathbf{x}$ is a short snippet of a long equilibrium trajectory.
Then we can assume that $p(x_0)$ is distributed according to the Boltzmann distribution
\begin{eqnarray}
    p(x_0) &=& p(q_0, p_0) 
            =   \frac{1}{Z}\exp\left(-\frac{V(q_0)}{k_BT}\right) \cdot 
                \frac{1}{\sqrt{2 \pi k_BT m}}\exp\left(-\frac{1}{k_BT} \frac{p_0^2}{2m} \right) \, ,
\label{eq:BoltzmannDist}
\end{eqnarray}
where $Z= \int_{-\infty}^{\infty} \mathrm{d}q_0\, \exp\left(-\frac{V(q_0)}{k_BT}\right)$ is the classical partition function.
The functional form of the single-step transition probability depends on the Langevin integrator, and it is the aim of this contribution to analyze $ p(x_{k+1} | x_k)$ for various integrators.
The path probability density $P[\mathbf{x}] $ is normalised as
\begin{eqnarray}
	\int_{\mathcal{S}} \mathcal{D}\mathbf{x} \,\mathcal{P}[\mathbf{x}] 
	=	\int_{x_0 \in \Gamma} \int_{x_1 \in \Gamma} \dots \int_{x_n \in \Gamma} \mathrm{d}x_0 \mathrm{d}x_1 \dots \mathrm{d}x_n \;\mathcal{P}[\mathbf{x}] 
	&=& 1 \, ,
\label{eq:pathIntegral_timeDiscretised}	
\end{eqnarray}
where the first equality defines the path integral $\int_{\mathcal{S}} \mathcal{D}\mathbf{x} \dots$.
Let $s: \mathcal{S} \rightarrow \mathbb{R}$ be a path observable, i.e.~a function that maps a time-discretised path to a number. 
The path expected value of this observable is
\begin{eqnarray}
	\langle s \rangle = \int \mathcal{D}\mathbf{x} \,\mathcal{P}[\mathbf{x}] s[\mathbf{x}]
	=	\int_{x_0 \in \Gamma} \int_{x_1 \in\Gamma} \dots \int_{x_n \in \Gamma} \mathrm{d}x_0 \mathrm{d}x_1 \dots \mathrm{d}x_n \;\mathcal{P}[\mathbf{x}] s[\mathbf{x}] \, ,
\label{eq:pathExpectedValue_timeDiscretised}		
\end{eqnarray}
Assuming ergodicity, this path expected value can be estimated from a set of $N$ paths $(\mathbf{x}_1, \mathbf{x}_2 \dots \mathbf{x}_N)$ that has been sampled according to $\mathcal{P}[\mathbf{x}]$
\begin{eqnarray}
	\langle s \rangle  &=& 	 \lim_{N \rightarrow \infty}  \frac{1}{N}\, \sum_{i=1}^N s[\mathbf{x}_i]\, .
\label{eq:pathExpectedValue_estimator_timeDiscretised}	
\end{eqnarray}

A short comment on notation: we denote functions of paths $\mathbf{x}$ with square brackets around the argument, and functions of states $x =(q,p)$  with round brackets.

%
%
\subsection{Girsanov reweighting}
\label{sec:PathReweighting}
In Girsanov reweighting, one samples paths at a simulation potential $V(q)$, and from this sample one estimates path expected values at a target potential 
\begin{eqnarray}
    \widetilde{V}(q) &=& V(q) + U(q)\, ,
\label{eq:targetPotential}    
\end{eqnarray}
where $U(q)$ is called bias or perturbation potential. 
Let us point out the sign convention. 
In eq.~\ref{eq:targetPotential}, we add the perturbation $U$ to the simulation potential $V$, which is the convention used in the literature on path reweighting. 
The literature on enhanced sampling simulations uses the opposite sign convention, i.e.~the bias is subtracted from the simulation potential to obtain to the true (target) potential: $\widetilde{V}  = V - U_{\mathrm{enh.}\, \mathrm{samp.}}$. 
Thus, if Girsanov reweighting is used to unbias enhanced sampling simulations, the bias in eq.~\ref{eq:targetPotential} is $U = - U_{\mathrm{enh.}\, \mathrm{samp.}}$.
We assume that $\widetilde{V}$ represents a typical molecular potential, i.e. a function that has infinite discontinuities whenever two atoms occupy the same position, but is continuous and differentiable everywhere else.  
We further assume that $U(q)$ respects these discontinuities and does not introduce any new discontinuities.
In practice this means that one cannot use soft-core potentials \cite{Beutler:1994, Zacharias:1994} to construct $U(q)$.
Formally, one can reweight the path expected value at the target potential as
\begin{eqnarray}
	\widetilde{\langle s \rangle} 
	= \int \mathcal{D}\mathbf{x} \,\widetilde{\mathcal{P}}[\mathbf{x}] s[\mathbf{x}]
	= \int \mathcal{D}\mathbf{x} \,g(x_0)M[\mathbf{x}|x_0]\, \mathcal{P}[\mathbf{x}] s[\mathbf{x}]
	\, ,
\label{eq:pathExpectedValue_reweightedPathIntegral}		
\end{eqnarray}
where $\widetilde{\mathcal{P}}[\mathbf{x}]$ is the path probability density at $\widetilde{V}$, and 
\begin{eqnarray}
    g(x_0) \cdot M[\mathbf{x} | x_0] &=& 
    \frac{\widetilde{p}(x_0)}{p(x_0)} \cdot
    \frac{\widetilde{\mathcal{P}}[x_1 \dots x_n | x_0]}{\mathcal{P}[x_1 \dots x_n | x_0]}
\label{eq:gM}    
\end{eqnarray}
is the path reweighting factor. 
If $\widetilde{p}(x_0)$ and $p(x_0)$  are given by eq.~\ref{eq:BoltzmannDist} (for $\widetilde{V}$ and $V$, respectively), then the relative weight of the initial state is
\begin{eqnarray}
    g(x_0) &=& \frac{\widetilde p(x_0)}{p(x_0)} = \frac{Z}{\widetilde{Z}}\exp\left(-\frac{1}{k_BT} U(q_0)\right)\, .
\label{eq:relativeStationaryProbability}    
\end{eqnarray}
If $M[\mathbf{x} | x_0]$ exists and one can find a computable expression for it, then one can estimate the path expected value at the target potential, $\widetilde{\langle s \rangle}$ from paths sampled at the simulation potential $(\mathbf{x}_1, \mathbf{x}_2 \dots \mathbf{x}_N)$ by reweighting their contribution to the estimator
\begin{eqnarray}
   \widetilde{\langle s \rangle}
    &=& \lim_{N \rightarrow \infty} \frac{1}{N}\, \sum_{i=1}^N g(x_{i,0}) M[\mathbf{x}_i| x_{i,0}] \cdot s[\mathbf{x}_i] \,.
\label{eq:pathExpectedValue_estimator_reweighted}    
\end{eqnarray}
where $g(x_{i,0}) M[\mathbf{x}_i| x_{i,0}]$ is the relative weight of path $\mathbf{x}_i$ a the target potential $\widetilde V$.
Whether $M[\mathbf{x} | x_0]$ exists, depends on the underlying dynamics, specifically on the condition of absolute continuity.

%
%
\subsection{Absolute continuity}
The path probability density $\widetilde{\mathcal{P}}[\mathbf{x}]$ at the target potential and the path probability density $\mathcal{P}[\mathbf{x}]$ at the simulation potential
are absolutely continuous with respect to each other \cite{Girsanov1960, Oeksendal:2003, Donati:2017, Donati:2022} if 
\begin{eqnarray}
    \widetilde{\mathcal{P}}[S] = \int_{S \subset \mathcal{S}} \mathcal{D}\mathbf{\mathbf{x}}\,  \widetilde{\mathcal{P}}[\mathbf{x}] = 0 
    &\Leftrightarrow&
    \mathcal{P}[S] = \int_{S \subset \mathcal{S}} \mathcal{D}\mathbf{\mathbf{x}}\,  \mathcal{P}[\mathbf{x}] = 0 \, ,
\label{eq:absoluteContinuity}    
\end{eqnarray}
where $S = S_0 \times S_1 \times \dots \times S_n$ is a small subset of the path space $\mathcal{S}$, a
and $S_i$ is a small region in phase space $\Gamma$ in which $x_i$ may be found. 
That is, any region of the path space $\mathcal{S}$ that is sampled by the dynamics at $\widetilde{V}(x)$ also needs to be sampled by the dynamics at $V(x)$, and vice versa. 
Otherwise the relative path probability density in eq.~\ref{eq:gM} is not defined.
(Strictly speaking only $\widetilde{\mathcal{P}}$ needs to be absolutely continuous with respect to $\mathcal{P}$. But since this almost surely implies\cite{Oeksendal:2003} that also $\mathcal{P}$ is absolutely continuous with respect to $\widetilde{\mathcal{P}}$, we here omit this distinction.)

If $P[\mathbf{x}]$ and $\widetilde{P}[\mathbf{x}]$ can be decomposed as in eq.~\ref{eq:pathProbability_timeDiscretized}, absolute continuity is fulfilled if 
\begin{eqnarray}
    \widetilde{p}(S_0) = \int_{S_0} \mathrm{d}x_0\, \widetilde{p} (x_0)\, = 0
    &\Leftrightarrow&
    p(S_0) = \int_{S_0} \mathrm{d}x_0\, p (x_0)\, = 0
\label{eq:absoluteContinuity_initialState}    
\end{eqnarray}
and if
\begin{eqnarray}
    \widetilde{p}(S_{k+1} | x_k) = \int_{S_{k+1}} \mathrm{d}x_{k+1}\, \widetilde{p} (x_{k+1} | x_k)\, = 0
    &\Leftrightarrow&
    p(S_{k+1} | x_k) = \int_{S_{k+1}} \mathrm{d}x_{k+1}\, p(x_{k+1} | x_k)\, = 0 \, .
\label{eq:absoluteContinuity_transition}        
\end{eqnarray}
for all $k$ with $k=0,1\dots,n-1$.
Eq.~\ref{eq:absoluteContinuity_initialState} is fulfilled if $\widetilde{p}(x_0)$ and $p(x_0)$  are given by eq.~\ref{eq:BoltzmannDist} (for $\widetilde{V}$ and $V$, respectively), but other choices are also possible. 
For overdamped Langevin dynamics, the Girsanov theorem guarantees that - for sensible choices of $U(q)$
(see section \ref{sec:PathReweighting}) - eq.~\ref{eq:absoluteContinuity_transition} is fulfilled.
Thus for overdamped Langevin dynamics, $M[\mathbf{x} | x_0]$ exists, and for overdamped time-discretized paths generated by the Euler-Maruyama scheme \cite{Kloeden:1992, Oeksendal:2003}, the computable expression for $M[\mathbf{x} | x_0]$ is well-established
\cite{Oeksendal:2003, Zuckerman:1999, Xing:2006, Adib:2008, Schuette:2015, Donati:2017, Donati:2018}.
By contrast, the existence of $M[\mathbf{x} | x_0]$ cannot be guaranteed for underdamped Langevin dynamics (eq.~\ref{eq:underdamped_Langevin_01}).
It may however exist for time-discretized paths of underdamped Langevin dynamics. 
In sections~\ref{sec:image} and \ref{sec:AbsoluteContinuity}, we will discuss the existence of $M[\mathbf{x} | x_0]$ for symmetric splitting schemes 
\cite{Bussi:2007, Leimkuhler:2012, Leimkuhler:2013, Sivak:2013}
for underdamped Langevin dynamics.

\subsection{Reweighting on-the-fly}
To derive $M[\mathbf{x}| x_0]$, we use reweighting on-the-fly (proposed in  Ref.~\onlinecite{Donati:2017} and discussed in more detail in Ref.~\onlinecite{Kieninger:2021}).
In this approach, the reweighting factor is formulated in terms of the random numbers $\eta_k$, which are generated during the simulation at $V$, and a random number difference $\Delta \eta_k$, which depends on the gradient of the bias $\nabla U$. 
Both properties are easily accessible during the simulation,  and part of the reweighting factor can be pre-calculated on-the-fly during the simulation.
This makes the actual reweighting (eq.~\ref{eq:pathExpectedValue_estimator_reweighted}) computationally simple and efficient.
A stochastic integrator generates a sequence of random numbers $\boldsymbol{\eta} = (\eta_0, \eta_1, \dots, \eta_{n-1})$ to represent the random force in a stochastic equation of motion (e.g.~eq.~\ref{eq:underdamped_Langevin_01}).
The random numbers $\eta_{k}$ are usually drawn from a Gaussian normal distribution with zero mean and unit variance
\begin{eqnarray}
    P(\eta_k)  &=& \sqrt{\frac{1}{2 \pi}}  \exp \left( - \frac{1}{2}\eta_{k}^2 \right) \, .
\label{eq:GaussianDist}
\end{eqnarray}
If the initial state $x_0$ has been set and $\boldsymbol{\eta}$ has been generated, the path $\mathbf{x} |x_0 = (x_1 \dots x_n | x_0)$ is fully determined. 
Thus, the stochastic integrators can be viewed as a map from a random number sequence $\boldsymbol{\eta}$ to a path $\mathbf{x}|x_0$.
If the map is bijective (one-to-one mapping between random number sequence and path), the conditional path probability density is equal to the probability of drawing the corresponding sequence of random numbers:
\begin{eqnarray}
    \mathcal{P}[\mathbf{x}|x_0] 
    &=& \mathcal{P} [\boldsymbol{\eta}]
    = \frac{1}{\sqrt[n]{2 \pi}}  \exp \left( - \frac{1}{2}\sum_{k=0}^{n-1}\eta_{k}^2 \right) \, .
\label{eq:P_oneEta}    
\end{eqnarray}
The stochastic integrator evaluates the gradient of the potential $V$, and therefore the map depends on $V$. 
If the potential is modified from $V$ to the target potential $\widetilde V$, the map changes accordingly, and a different random number sequence $\boldsymbol{\widetilde \eta}$ is needed to generate the same path $\mathbf{x} |x_0$.
The conditional path probability density at $\widetilde V$ is then equal to the probability of drawing $\boldsymbol{\widetilde \eta}$:
$\widetilde{\mathcal{P}}[\mathbf{x} |x_0] = \mathcal{P} [\widetilde{\boldsymbol{\eta}}]$. 
At each integration step $k$, the random numbers $\eta_k$ and $\tilde \eta_k$ differ by $\Delta \eta_k$: 
\begin{eqnarray}
    \tilde \eta_k &=& \eta_k + \Delta \eta_k \, .
\label{eq:DeltaEta01}      
\end{eqnarray}
This allows us to write the relative conditional path probability density as \cite{Kieninger:2021}
\begin{eqnarray}
    M[\mathbf{x}|x_0]  = 
    \frac{\widetilde{\mathcal{P}}[\mathbf{x}|x_0] }{\mathcal{P}[\mathbf{x}|x_0] }
    =
	\frac{\mathcal{P}[\widetilde{\boldsymbol{\eta}}]}{\mathcal{P}[\boldsymbol{\eta}]} 
	&=&	 \exp\left(- \sum_{k=0}^{n-1} \eta_k \cdot \Delta \eta_k \right) \cdot \exp\left(-\frac{1}{2} \sum_{k=0}^{n-1} (\Delta \eta_k)^2 \right) \, .
\label{eq:M_oneEta}   
\end{eqnarray}
The random numbers $\eta_k$ are recorded during the simulation at $V$; the expression for $\Delta \eta_k$ depends on the stochastic integrator.
The discussion so far applies to stochastic integrators that generate one random number $\eta_k$ per integration step and degree of freedom. 
Some symmetric splitting methods for underdamped Langevin dynamics however generate two random numbers $(\eta_k^{(1)}, \eta_k^{(2)})$ per integration step and degree of freedom.
The two random numbers are drawn independently from a Gaussian normal distribution with zero mean and unit variance, i.e.~$P(\eta_k^{(1)}, \eta_k^{(2)}) = P(\eta_k^{(1)})P(\eta_k^{(2)})$, and $P(\eta_k^{(1)})$ and $P(\eta_k^{(2)})$ are given by 
eq.~\ref{eq:GaussianDist}.
Thus, an integrator with two random numbers per integration step can be viewed as a map from two sequences of random numbers $(\boldsymbol{\eta}^{(1)}, \boldsymbol{\eta}^{(2)})$ to a path $\mathbf{x}|x_0$, 
where 
$\boldsymbol{\eta}^{(1)} = (\eta_0^{(1)}, \eta_1^{(1)}, \dots, \eta_{n-1}^{(1)})$ and
$\boldsymbol{\eta}^{(2)} = (\eta_0^{(2)}, \eta_1^{(2)}, \dots, \eta_{n-1}^{(2)})$.
If the map is bijective, the conditional path probability density is equal to the probability of drawing the two random number sequences:
\begin{eqnarray}
    \mathcal{P}[\mathbf{x}|x_0] 
    &=& \mathcal{P} [\boldsymbol{\eta}^{(1)}, \boldsymbol{\eta}^{(2)}]
    = \frac{1}{\sqrt[n]{2 \pi}}  \exp \left( - \frac{1}{2}\sum_{k=0}^{n-1}\left(\eta_{k}^{(1)}\right)^2 \right) \cdot
    \frac{1}{\sqrt[n]{2 \pi}}  \exp \left( - \frac{1}{2}\sum_{k=0}^{n-1}\left(\eta_{k}^{(2)}\right)^2 \right)\, .
\label{eq:P_twoEta}    
\end{eqnarray}
At the target potential $\widetilde V$, two different random number sequences
$\widetilde{\boldsymbol{\eta}}^{(1)} = (\widetilde\eta^{(1)}_0, \widetilde\eta^{(1)}_1, \dots, \widetilde\eta^{(1)}_{n-1})$ 
and
$\widetilde{\boldsymbol{\eta}}^{(2)} = (\widetilde\eta^{(2)}_0, \widetilde\eta^{(2)}_1, \dots, \widetilde\eta^{(2)}_{n-1})$
are needed to generate the same path $\mathbf{x}$.
At each integration step $k$, the random numbers differ by $\Delta \eta^{(1)}_k$ and $\Delta \eta^{(2)}_k$:
\begin{eqnarray}
    \widetilde \eta^{(1)}_k &= & \eta^{(1)}_k +\Delta \eta^{(1)}_k  \cr
    \widetilde \eta^{(2)}_k &= & \eta^{(2)}_k +\Delta \eta^{(2)}_k  \, .
\end{eqnarray}
The relative conditional path probability density can then be written as 
\begin{eqnarray}
    M[\mathbf{x}|x_0] &=& 
    \frac{\widetilde{\mathcal{P}}[\mathbf{x}\, |\, x_0]}{\mathcal{P}[\mathbf{x}\, |\, x_0]}
    =
    \frac{P[\widetilde{\boldsymbol{\eta}}^{(1)}, \widetilde{\boldsymbol{\eta}}^{(2)}] }{P[\boldsymbol{\eta}^{(1)}, \boldsymbol{\eta}^{(2)}]} \cr
    &=& \exp \left( - \sum\limits_{k=0}^{n-1} \eta^{(1)}_{k} \cdot \Delta \eta^{(1)}_{k} \right) \cdot
        \exp \left( - \frac{1}{2} \sum\limits_{k=0}^{n-1} (\Delta \eta^{(1)}_{k})^2 \right) \cdot \cr
    && \exp \left( - \sum\limits_{k=0}^{n-1} \eta^{(2)}_{k} \cdot \Delta \eta^{(2)}_{k} \right) \cdot
        \exp \left( - \frac{1}{2} \sum\limits_{k=0}^{n-1} (\Delta \eta^{(2)}_{k})^2 \right) \, .
\label{eq:M_twoEta}
\end{eqnarray}
As in eq.~\ref{eq:M_oneEta}, the random numbers $\eta_k^{(1)}$ and $\eta_k^{(2)}$ are recorded during the simulation at $V$, and the expressions for  $\Delta \eta_k^{(1)}$ and $\Delta \eta_k^{(2)}$ depend on the integrator.
At this point the obstacle in constructing path reweighting factors for underdamped Langevin dynamics becomes noticeable.
In underdamped Langevin dynamics,  the state $x_k = (q_k, p_k)$ is a two-dimensional vector.
Thus, integrators with two random numbers per integration step map two real numbers to a two-dimensional state space, whereas integrators with one random number map a single real number to a two-dimensional state space. 
In both cases, $P[\mathbf{x}|x_0]$ can be derived from the integrator equations (see section \ref{sec:LangevinIntegrators} and supplementary material), but  $M[\mathbf{x}|x_0]$ does not exist for all integrators (see sections \ref{sec:image} and \ref{sec:AbsoluteContinuity}).

%
%

\section{Langevin integrators}
\label{sec:LangevinIntegrators}
%

%
%
\subsection{Equation of motion and splitting methods}
Eq.~\ref{eq:underdamped_Langevin_01} can be reformulated as a vector field
\begin{eqnarray}
     \left(
     \begin{array}{c}
        \dot q\\ \dot p
     \end{array}
     \right)
     &= 
     \underbrace{
     \left(\begin{array}{c}
        p/m \\ 0 
     \end{array} \right)}_{A} 
     +
     \underbrace{
     \left(\begin{array}{c}
        0 \\ - \nabla_q V(q) 
     \end{array} \right)}_{B} 
     +
     \underbrace{     
     \left(\begin{array}{c}
        0 \\ -\xi p + \sqrt{2\xi k_B T m} \, \eta(t) 
     \end{array} \right)}_{O} \, .
 \label{eq:underdamped_Langevin_02}         
\end{eqnarray}
Each of the three terms, $A$, $B$ and $O$, can be integrated separately
to yield the following time-discretized update operators
\begin{subequations}
\begin{align}
     \mathcal{A}
     \left(\begin{array}{c} q_{k}\\p_{k}\end{array}\right) 
     &= \left(\begin{array}{c} q_k+\Delta t \frac{1}{m}p_k \\p_k\end{array}\right) 
     = \left(\begin{array}{c} q_k+a p_k \\p_k\end{array}\right)
     \label{eq:A}\\
     \mathcal{B}\left(\begin{array}{c} q_{k}\\p_{k}\end{array}\right) 
     &= \left(\begin{array}{c}q_k \\ p_k-\Delta t\nabla V(q_k)\end{array}\right) 
     = \left(\begin{array}{c}q_k \\ p_k+b(q_k)\end{array}\right) 
     \label{eq:B}\\
     \mathcal{O}\left(\begin{array}{c} q_{k}\\ p_{k}\end{array}\right) 
     &= \left(\begin{array}{c} q_k \\ e^{-\xi\Delta t}p_k + \sqrt{k_BTm(1-e^{-2\xi\Delta t})}\, \eta_k\end{array}\right) 
     = \left(\begin{array}{c} q_k \\ d \, p_k + f\, \eta_k\end{array}\right)
     \label{eq:O}\, ,
\end{align}
\end{subequations}
with time step $\Delta t$ and random number $\eta_k \sim \mathcal{N}(0,1)$, 
where $\mathcal{N}(\mu,\sigma^2)$ denotes the Gaussian normal distribution with mean $\mu$ and variance $\sigma^2$.
Eqs.~\ref{eq:A} -- \ref{eq:O} are the time-discretized solutions of their respective parts in the differential equation (eq.~\ref{eq:underdamped_Langevin_02}), where \ref{eq:O} is the result known from the Ornstein-Uhlenbeck process (see Chapter 7.3.1. in Ref.~\onlinecite{Leimkuhler:2015}).
The second equality introduces the following abbreviations to keep the notation manageable
\begin{subequations}
\begin{eqnarray}
    a       &=& \Delta t \frac{1}{m} \label{eq:a}\\
    b(q_k)  &=& -\Delta t\nabla V(q_k) \label{eq:b}\\
    d       &=& e^{-\xi\Delta t}  \label{eq:d}\\
    f       &=& \sqrt{k_BTm(1-e^{-2\xi\Delta t})} \label{eq:f}
\end{eqnarray}
\end{subequations}
where $d$ stands for dissipation and $f$ for the thermal fluctuation.
Particularly accurate Langevin integration schemes can be derived using the (symmetric) operator splitting method, or Strang splitting \cite{Trotter:1959, Hongen:2011}.
In these algorithms, some of the update operators (eqs.~\ref{eq:A}, \ref{eq:B}, and \ref{eq:O}) 
are carried out twice during an integration step, but each time with only half a time step $\frac{\Delta t}{2}$.
If a step is carried out for only half a time step $\frac{\Delta t}{2}$, we denote the corresponding operator with a prime, e.g.
\begin{align}
     \mathcal{A'}\left(\begin{array}{c}q_{k}\\ p_{k}\end{array}\right) &= \left(\begin{array}{c}q_k+\frac{\Delta t}{2}\frac{1}{m}p_k \\ p_k\end{array}\right) 
     = \left(\begin{array}{c} q_k+a' p_k \\p_k\end{array}\right)\, .
\end{align}
Correspondingly, $a'$, $b'$, $d'$ and $f'$ are obtained by replacing $\Delta t$ with $\Delta t/2$ in eqs.~\ref{eq:a}-\ref{eq:f}.
For an in-depth discussion on the theory of splitting operators we refer to Refs.~\onlinecite{Trotter:1959, Tuckerman:1992, Hongen:2011, Leimkuhler:2012, Leimkuhler:2013} and Chapter 7 in Ref.~\onlinecite{Leimkuhler:2016}.
In this contribution, we will use the ABO method to illustrate our approach, and we will apply the approach to the following integrators based on eq.~\ref{eq:underdamped_Langevin_02}:
ABOBA,
BAOAB,
OABAO,
AOBOA,
OBABO (Bussi-Parrinello thermostat), and
BOAOB \cite{Bussi:2007, Leimkuhler:2012, Leimkuhler:2013, Sivak:2013, Leimkuhler:2016b}.
These algorithsm are implemented in OpenMM \cite{Eastman:2017} via the package OpenMMTools \cite{Fass:2018}.
We additionally include BAOA \cite{BouRabee:2010}, because it is implemented in several MD simulations packages \cite{Eastman:2017, Case:2005, VanDerSpoel:2005}, where it is sometimes called Verlet-Middle integrator \cite{Zhang:2019}. GROMACS implements
the GROMACS stochastic dynamics (GSD) method \cite{Goga:2012}, which is equivalent to BAOA \cite{Kieninger:2022}.

%
\subsection{Example: ABO method}
The Langevin integrator that is constructed by carrying out eqs.~\ref{eq:A}, \ref{eq:B} and \ref{eq:O} consecutively is called ABO method
(sequential splitting \cite{Trotter:1959, Hongen:2011}). 
The ABO algorithm yields a less accurate approximation to the actual dynamics than symmetric splitting methods  (or equivalently: a small time step $\Delta t$ is needed to obtain a given accuracy). Here, we use it to illustrate concepts.
The integrator equations of the ABO algorithm are
\begin{subequations}
\begin{align}
   q_{k+1}    &= q_k+ap_k \label{eq:ABO01}\\
   p_{k+1/2}  &= p_k+b(q_{k+1}) \label{eq:ABO02}\\
   p_{k+1}    &= dp_{k+1/2} + f\eta_k  \, .\label{eq:ABO03}
\end{align}
\end{subequations}
$p_{k+1/2}$ should not be interpreted as an integration by half a time step, but rather as 50\% progress in the update of the momenta. 
Namely, in eq.~\ref{eq:ABO02} the momentum is updated according to the forces due to the potential energy function, and in  eq.~\ref{eq:ABO03} the momentum update due to friction and random forces is carried out.
The contributions of these two steps to the total momentum update are by no means equal.

The joint update of the position and momentum, i.e.~the update operator of the ABO method $\mathcal{U}_{\mathrm{ABO}}$, is obtained by sequentially applying operators $\mathcal{A}$, $\mathcal{B}$, and $\mathcal{O}$ to the current state $(q_k, p_k)^{\top}$.
\begin{eqnarray}
    \left(\begin{array}{c} q_{k+1}\\p_{k+1}\end{array}\right) 
    &=& \mathcal{U}_{\mathrm{ABO}}\left(\begin{array}{l} q_{k}\\p_{k}\end{array}\right) \cr
    &=& \mathcal{O}\mathcal{B}\mathcal{A}\left(\begin{array}{l} q_{k}\\p_{k}\end{array}\right) \cr
    &=& \mathcal{O}\mathcal{B}\left(\begin{array}{l} q_{k} + ap_k\\p_{k}\end{array}\right) \cr
    &=& \mathcal{O}\left(\begin{array}{l} q_{k} + ap_k\\p_{k} + b(q_{k} + ap_k)\end{array}\right) \cr
    &=& \left(\begin{array}{l} q_{k} + ap_k\\dp_{k} + db(q_{k} + ap_k) +f\eta_k\end{array}\right) \, .
\label{eq:ABO_update_operator01}         
\end{eqnarray}
As usual the operators are applied from right to left, and hence the order of the operators in eq.~\ref{eq:ABO_update_operator01} is reversed compared to the name.
The algorithms for each of the integrators in Tab.~\ref{tab:integrators} as well as their update operators are reported in the supplementary material.

%
%
\section{The images of Langevin update functions}
\label{sec:image}
The update operator $\mathcal{U}$ of a Langevin integrator depends parametrically on the random number $\eta_k$: when $\eta_k$ is varied, the updated state $(q_{k+1}, p_{k+1})$ changes. 
One can thus interpret the update operator in eq.~\ref{eq:ABO_update_operator01} as a function $\mathcal{U}$ that maps a random number $\eta_k \in \mathbb{R}$ to a point in state space $x_{k+1} = (q_{k+1}, p_{k+1})^{\top} \in \Gamma$. 
The current state $(q_k, p_k)$  is treated as a parameter of $\mathcal{U}$. 
The image of the update function $\mathcal{U}$, i.e.~``the set of all output values it may produce.'',
is the set of all points in state space that can theoretically be reached from $(q_k, p_k)$ within a single integration time step.
Note that the image only reflects whether or not a certain point can be reached. 
The probability with which this point would be generated during an integration time step does not play a role when discussing the image of $\mathcal{U}$.
Besides $\eta_k$ and $(q_k, p_k)$, the update function $\mathcal{U}$ also depends on the potential energy function $V$. 
%
Thus the image of $\mathcal{U}$ might change if $V$ is varied.

\subsection{One random number per integration time step}
\begin{figure}
    \centering
    \includegraphics[width=6cm]{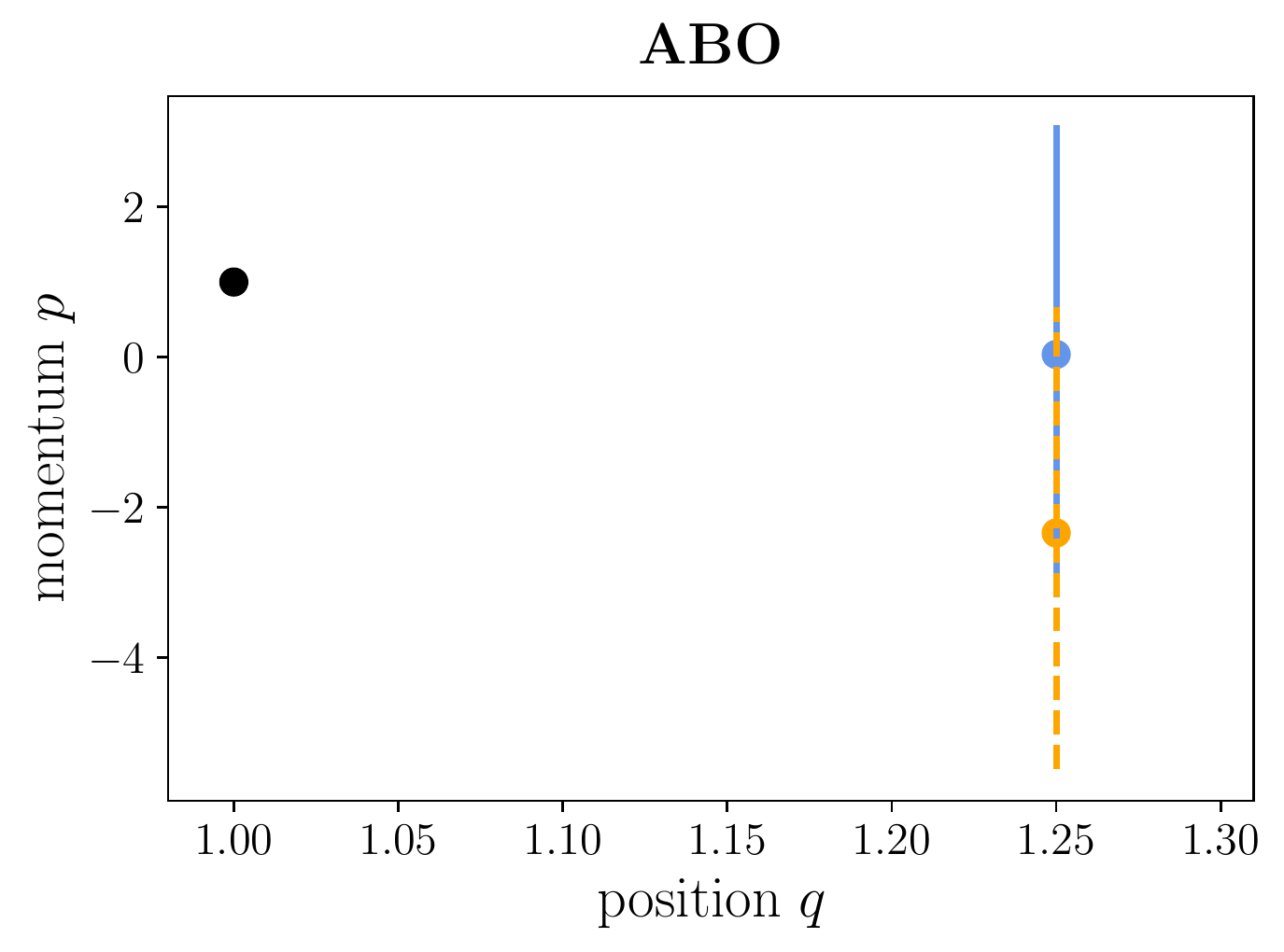}
    \caption{Initial state $(q_k, p_k)^\top$ (black dot) and image of the update function $\mathcal{U}_{\mathrm{ABO}}(\eta_k; x_k, V)$ for an unscaled potential (blue line) and a scaled potential (orange line) for the ABO method. The image of the update function contains all possible states $(q_{k+1}, p_{k+1})^\top$ that can be reached from $(q_k, p_k)^\top$ within one integration time step $\Delta t$.\\
    Parameters: $k_B, m, T, \xi=1$, $\Delta t=0.25$, $(q_k, p_k)^\top = (1,1)^\top$, $V(q) = (q^2 - 1)^2 + q$, $\widetilde{V}(q) = 4.2 \cdot (q^2 - 1)^2 + q$ and $\eta \in [-5,5]$.}
    \label{fig:update_image_ABO}
\end{figure}
First, we consider Langevin integrators that use a single random number per integration step, i.e.~one $\mathcal{O}$-step (eq.~\ref{eq:O}) per integration step $k$.
In Tab.~\ref{tab:integrators}, these are ABO, ABOBA, BAOAB and BAOA/GSD.
%

%
%

%
The following realization is crucial: 
given an initial state $x_k$, the image of $\mathcal{U}$ is not the entire state space $\Gamma$, but a one-dimensional curve $C_{1d}$ within $\Gamma$.
This one-dimensional curve is parametrized by $\eta_k$.
Formally, we can characterize the function $\mathcal{U}$ as
\begin{eqnarray}
    \mathcal{U}:\;\; \mathbb{R} &\rightarrow& C_{1d} \subset \Gamma \cr
    \mathcal{U}: \; \eta_k &\mapsto& x_{k+1} \, .
    \label{eq:U_function}
\end{eqnarray}
See Ref.~\onlinecite{Arens2018} 
for an overview of parametrized curves and parametric equations.

Mathematically, it is not surprising that the image is one-dimensional, since it is impossible to map the one-dimensional number line $\mathbb{R}$ to a two-dimensional space. 
Algorithmically this means that, given an initial state $x_k$, not all states in state space $\Gamma$ can be reached by a single iteration of the integrator. 
Rather, the accessible states lie on $C_{1d}$, and which precise point on this line is obtained depends on $\eta_k$.
For the ABO method, this parametrization of $C_{1d}$ by $\eta_k$ can be made explicit by reformulating eq.~\ref{eq:ABO_update_operator01} as
\begin{eqnarray}
    \left(\begin{array}{c} q_{k+1}\\p_{k+1}\end{array}\right) 
    &=& \mathcal{U}_{\mathrm{ABO}}(\eta_k; x_k, V)
    =   \left(\begin{array}{l} q_{k+1}\\ \bar{p}_{k+1}\end{array}\right) + 
        \left(\begin{array}{l} 0\\f\end{array}\right) \eta_k
\label{eq:ABO_update_operator02}           
\end{eqnarray}
with
$q_{k+1} = q_{k} + ap_k$
and
$\bar{p}_{k+1} = dp_{k} + db(q_{k} + ap_k)$.
That is, the algorithm first moves the system to $(q_{k+1}, \bar{p}_{k+1})^{\top}$ and then adjusts the momenta by $(0,f)^{\top}\eta_k$. 
Consequently, all accessible states of $\mathcal{U}_{\mathrm{ABO}}(\eta_k; x_k, V)$ lie on a vertical through $(q_{k+1}, \bar{p}_{k+1})^{\top}$.

Fig.~\ref{fig:update_image_ABO} illustrates the image of $\mathcal{U}_{\mathrm{ABO}}(\eta_k; x_k, V)$. 
The initial point $(q_k, p_k) =(1,1)$ is marked as a black dot. 
The point $(q_{k+1}, \bar{p}_{k+1})^{\top}$ for the potential $V(q) = (q^2 - 1)^2 + q$ is shown as a blue dot and acts as the support point for the image of $\mathcal{U}_{\mathrm{ABO}}(\eta_k; x_k, V).$
Varying $\eta_k$ from -5 to +5 then yields the blue line, i.e.~the image of $\mathcal{U}_{\mathrm{ABO}}(\eta_k; x_k, V)$ for $\eta_k \in [-5,5]$.
When the potential is changed to $\widetilde{V}(q) = 4.2 \cdot (q^2 - 1)^2 + q$, the point $(q_{k+1}, \bar{p}_{k+1})^{\top}$ shifts to the orange point.
Varying $\eta_k$ from -5 to +5 yields the image of $\mathcal{U}_{\mathrm{ABO}}(\eta_k; x_k, \widetilde{V})$ for $\eta_k \in [-5,5]$, shown as the dashed orange line in Fig.~\ref{fig:update_image_ABO}.
For arbitrary values of $\eta_k$, the images of the two functions coincide. 
Thus, the image of the update function of the ABO method, i.e.~the set of possible updated states $(q_{k+1}, p_{k+1})$, does not depend on the potential $V$.
\begin{figure}
    \centering
    \includegraphics[width=\textwidth]{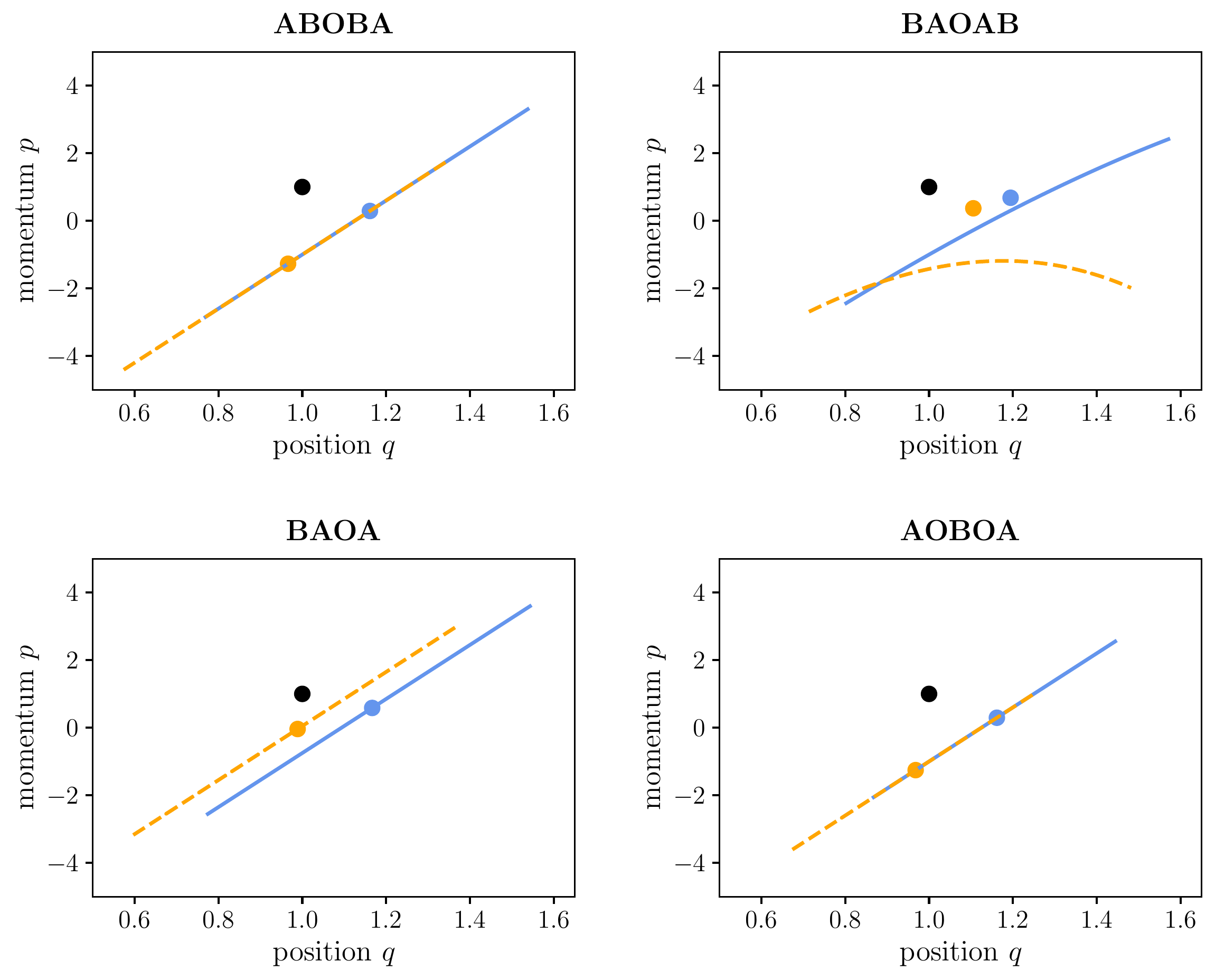}
    \caption{Initial state $(q_k, p_k)^\top$ (black dot) and image of the update function $\mathcal{U}(\eta_k; x_k, V)$ for an unscaled potential (blue line) and a scaled potential (orange line) for the ABOBA, BAOAB, BAOA/GSD and AOBOA method. The image of the update function contains all possible states $(q_{k+1}, p_{k+1})^\top$ that can be reached from $(q_k, p_k)^\top$ within one integration time step $\Delta t$.\\
    Parameters: $k_B, m, T, \xi=1$, $\Delta t=0.25$, $(q_k, p_k)^\top = (1,1)^\top$, $V(q) = (x^2 - 1)^2 + x$, $\widetilde{V}(q) = 4.2 \cdot (x^2 - 1)^2 + x$, $\eta \in [-5,5]$, or $\eta^{\mathrm{comb}} \in  [-5,5]$ .}
    \label{fig:update_image_all_integrators}
\end{figure}
The update functions for ABOBA, BAOAB and BAOA/GSD are derived in the supplementary material.
Their images are summarized in the second column of Tab.~\ref{tab:integrators} and illustrated in Fig.~\ref{fig:update_image_all_integrators}.
As in Fig.~\ref{fig:update_image_ABO}, the current state $(q_k, p_k)$ is shown as a black dot, the support points for the image for two different potentials are shown as blue and orange dots, and the corresponding images for $\eta_k \in [-5, +5]$ are shown as blue and orange lines.
Fig.~\ref{fig:update_image_all_integrators} also contains the diagram for a Langevin integrator with two random numbers, AOBOA, which will be discussed in the following section. 
Two aspects of the images of the update functions in Fig.~\ref{fig:update_image_all_integrators} are important to point out. 
First, for ABOBA, BAOA/GSD and AOBOA, the possible updated states depend linearly on $\eta_k$, and thus the image is a line $L_{1d}$ in the state space $\Gamma$.
By contrast, the possible updated states of BAOAB depend non-linearly on $\eta_k$, and the image is a curve $C_{1d}$.
Second, in ABOBA and AOBOA the image does not depend on the potential, whereas in BAOAB and BAOA/GSD the image changes if $V$ is varied.

\subsection{Two random numbers per integration time step}
Next, we consider Langevin integrators that generate two random numbers per integration step, i.e.~two $\mathcal{O}$-half-steps (eq.~\ref{eq:O} with $\Delta t/2$ instead of $\Delta t$) per integration step $k$.
In Tab.~\ref{tab:integrators}, these are AOBOA, BOAOB, OBABO/BP, and OABAO.
Their update functions are derived in the supplementary material.
The image of update functions of Langevin integrators with two random numbers  can be the entire phase space $\Gamma$.
That is, given an initial state $(q_k, p_k)$, any point in phase space can be reached within a single integration time step, albeit mostly with very low probability. 
For BOAOB and OBABO/BP this is indeed the case. 
Both of their update functions have the following form 
\begin{eqnarray}
     \left(\begin{array}{c} q_{k+1}\\p_{k+1}\end{array}\right) 
    &=& \left(\begin{array}{l} \bar{q}_{k+1}\\\bar{p}_{k+1} \end{array}\right) +
        \left(\begin{array}{l} 0\\c \cdot b'(\bar{q}_{k+1}+ af'\eta_k^{(1)})\end{array}\right) + 
        \left(\begin{array}{l} af'\\d'f'\end{array}\right)\eta_k^{(1)} + 
        \left(\begin{array}{l} 0\\ f' \end{array}\right)\eta_k^{(2)} \, ,
\label{eq:BOAOB_update_function}        
\end{eqnarray}
where $\eta_k^{(1)}$ and $\eta_k^{(2)}$ are the two random numbers, 
and $c = 1$ for BOAOB, and $c=d'$ for OBABO/BP.
The functions first move the system deterministically to a support point $(\bar{q}_{k+1}, \bar{p}_{k+1})$.
BOAOB and OBABO/BP differ in the way this deterministic update is calculated (see supplementary material), but have analogous subsequent terms in the update function.
(BOAOB: $\bar{q}_{k+1} = q_{k} + ad'p_{k} + ad'b'(q_k)$
and
$\bar{p}_{k+1} = d'd'p_{k} + d'd'b'(q_k)$.
OBABO/BP:
$\bar{q}_{k+1} = q_{k} + ad'p_{k} + ab'(q_k)$
and
$\bar{p}_{k+1} = d'd'p_{k} + d'b'(q_k)$.)
From the support point, the momentum is slightly adjusted in the second term 
eq.~\ref{eq:BOAOB_update_function}, which however depends on the random number $\eta_k^{(1)}$ and thus represents a partly randomized update of the momentum.
The last two terms in eq.~\ref{eq:BOAOB_update_function} randomize the position and the momentum.
Note that by scaling $\eta_k^{(1)}$ between $-\infty$ and $+\infty$, one can access any position, and by scaling $\eta_k^{(2)}$ one can access any momentum. 
Thus, the image of the update functions of BOAOB and OBABO/BP is the entire phase space $\Gamma$ (Tab.~\ref{tab:integrators}).
The update function of OABAO differs in a crucial point from eq.~\ref{eq:BOAOB_update_function}: in the second term both position and the momentum get a partly randomized update.
This partly randomized update depends on the potential energy function via $b(q)$.
One can construct cases, in which the update of the positions in the second term and the update in the third term compensate each other, and the images collapses from $\Gamma$ to a line 
(see supplementary material).
Thus, the image of the update function of OABAO depends on the potential, and the existence of $M[\mathbf{x} | x_0]$ and analytical expressions for $\Delta \eta_k^{(1)}$ and $\Delta \eta_k^{(2)}$ need to be discussed in the context of specific potentials $V$ and $\widetilde{V}$.
We will therefore exclude OABAO from the subsequent analysis.
AOBOA generates two random numbers $\eta_k^{(1)}$ and $\eta_k^{(2)}$ per integration time step. 
But $\eta_k^{(1)}$ and $\eta_k^{(2)}$ do not affect the position and the momentum independently as in eq.~\ref{eq:BOAOB_update_function}, and several combinations of $\eta_k^{(1)}$ and $\eta_k^{(2)}$ lead to the same updated state $(q_{k+1}, p_{k+1})$. 
In fact, one can combine $\eta_k^{(1)}$ and $\eta_k^{(2)}$ into an effective random number 
\begin{eqnarray}
    \eta^{\mathrm{comb}}_k &=& d'\eta_k^{(1)} + \eta_k^{(2)} \sim \mathcal{N}(0,d'^2+1) \, .
\label{eq:AOBOA_combined_eta}     
\end{eqnarray}
and formulate the update function in terms of $\eta^{\mathrm{comb}}_k$
\begin{eqnarray}
     \left(\begin{array}{c} q_{k+1}\\p_{k+1}\end{array}\right) 
    &=& \left(\begin{array}{l} \bar{q}_{k+1}\\\bar{p}_{k+1} \end{array}\right) +
        \left(\begin{array}{l} a'\\ 1\end{array}\right) f'\eta^{\mathrm{comb}}
\label{eq:AOBOA_update_function}        
\end{eqnarray}
with
$\bar{q}_{k+1} = q_{k} + (a' + a'd'd')p_k  + a'd'b(q_{k} + a' p_k)$ and
$\bar{p}_{k+1} = d'd'p_{k} + d'b(q_{k} + a' p_k)$ 
(see supplementary material).
The AOBOA method first generates a deterministic update and moves the system to the support point $(\bar{q}_{k+1}, \bar{p}_{k+1})$, and then randomizes the position and momentum along a line with slope $(a',1)$.
Thus, the image of this update function is a line $L_{1d}$ through the state space $\Gamma$ (see Fig.~\ref{fig:update_image_all_integrators}).
Note that any combination of $\eta_k^{(1)}$ and $\eta_k^{(2)}$ that yields the same value $\eta^{\mathrm{comb}}_k$, will yield the same updated state $(q_{k+1}, p_{k+1})$.

\renewcommand{\arraystretch}{2.0}
\begin{table}[]
    \centering
    \begin{tabular}{l|l |l | l | l}
         &&\textbf{image at}    &\textbf{absolute} &\textbf{reweighting in}\\
         \textbf{integrator}    &\textbf{update} $u$ &$V$ \textbf{and} $\widetilde V$ 
                                &\textbf{continuity} &\textbf{phase space}\\    
         \hline
         ABO&$\mathbb{R} \rightarrow L_{1d}$  
                    &$L_{1d} = \widetilde{L}_{1d}$
                    &yes
                    &$\Delta \eta_k = \frac{d}{f} \Delta t\nabla U(q_{k+1})$\\
         \hline 
         ABOBA      &$\mathbb{R} \rightarrow L_{1d}$  
                    &$L_{1d} = \widetilde{L}_{1d}$         
                    &yes
                    &$\Delta \eta_k  = \frac{(d+1)}{f} \frac{\Delta t}{2} \nabla U(q_{k+1/2})$\\
        \hline 
         BAOAB      &$\mathbb{R} \rightarrow C_{1d}$  
                    &$C_{1d} \ne \widetilde{C}_{1d}$
                    &no
                    &n/a\\
        \hline            
         BAOA/GSD   &$\mathbb{R} \rightarrow L_{1d}$  
                    &$L_{1d} \ne \widetilde{L}_{1d}$
                    &no
                    &n/a\\
         \hline
         AOBOA      &$\mathbb{R}^2 \rightarrow L_{1d}$ 
                    &$L_{1d} = \widetilde{L}_{1d}$
                    &yes
                    &$\Delta \eta^{\mathrm{comb}}_k =  \frac{d'}{f'}\Delta t \nabla U(q_{k+1/2})$\\
                    &&&& $\eta^{\mathrm{comb}}_k = d'\eta_k^{(1)} +\eta_k^{(2)}$ \\    
         \hline 
         BOAOB      &$\mathbb{R}^2 \rightarrow \Gamma$ &$\Gamma$ in both cases
                    &yes
                    &$\Delta \eta_k^{(1)} =  \frac{d'}{f'} \, \frac{\Delta t}{2}\nabla U(q_k)$ \\
                    &&&& $\Delta \eta_k^{(2)} = \frac{1}{f'}\frac{\Delta t}{2} \nabla U(q_{k+1})$\\
        \hline            
         OBABO/BP   &$\mathbb{R}^2 \rightarrow \Gamma$ &$\Gamma$ in both cases
                    &yes
                    &$\Delta \eta_k^{(1)} = \frac{1}{f'}\frac{\Delta t}{2} \nabla U(q_{k})$ \\
                    &&&& $\Delta \eta_k^{(2)} =  \frac{d'}{f'} \, \frac{\Delta t}{2}\nabla U(q_{k+1})$\\ 
         \hline
         OABAO      &\multicolumn{4}{c}{depends on $V$ and $\widetilde{V}$.}\\   
         \hline
    \end{tabular}
    \caption{Image of integrator update function and corresponding expressions for the random number difference}
    \label{tab:integrators}
\end{table}
\renewcommand{\arraystretch}{1.0}

\section{Absolute continuity and path reweighting factor}
\label{sec:AbsoluteContinuity}

The notion of absolute continuity is closely related to the image of update function. 
Recall that absolute continuity implies that the same regions of path space are sampled by the dynamics at $\widetilde{V}(x)$ and by the dynamics at $V(x)$.
If the image of the update function depends on the potential energy, transitions $(q_k, p_k) \rightarrow (q_{k+1}, p_{k+1})$ that are possible at $V$ are impossible at $\widetilde{V}$.
Thus, the corresponding path probabilities are not absolutely continuous with respect to each other, and $M[\mathbf{x}|x_0]$ does not exist for the corresponding integrators. 
Among the integrators we considered here, this is the case for BAOAB and BAOA/GSD (Tab.~\ref{tab:integrators}).
Phase-space trajectories generated by these two integrators cannot be reweighted.
By contrast, the images of the update functions of ABO, ABOBA, AOBOA, BOAOB and OBABO/BP do no depend on the potential energy function.
Consequently, any transition $(q_k, p_k) \rightarrow (q_{k+1}, p_{k+1})$ that is possible at $V$ is also possible at $\widetilde{V}$.
The corresponding path probabilities are absolutely continuous with respect to each other, and $M[\mathbf{x}|x_0]$ exists. 
Phase-space trajectories generated by these integrators can be reweighted (Tab.~\ref{tab:integrators}).
Having identified integrators of underdamped Langevin dynamics for which Girsanov reweighting is possible, we next derive computable expressions for $M[\mathbf{x}|x_0]$ using the reweighting-on-the-fly approach (eqs.~\ref{eq:M_oneEta} and \ref{eq:M_twoEta}).
The mathematically formal way to derive an expression for $\Delta \eta_k = \widetilde{\eta}_k - \eta_k$ (eq.~\ref{eq:DeltaEta01}) goes as follows. 
We denote the update function of integrator $I$ at the simulation potential by 
$\mathcal{U}_I(\eta_k; x_k, V)$, 
and the update function of the same integrator at the target potential by
$\mathcal{U}_I(\widetilde{\eta}_k;x_k, \widetilde{V})$.
We require that both update operators yield the same update $(q_{k+1}, p_{k+1})^{\top}$ given the initial state $x_k=(q_{k}, p_{k})^{\top}$, i.e.~the path remains unchanged,
\begin{align}
    \left(\begin{array}{c} q_{k+1}\\ p_{k+1} \end{array}\right)
    &= \mathcal{U}_I(\eta_k; x_k, V) = \mathcal{U}_I(\widetilde{\eta}_k;x_k, \widetilde{V}) \, .
\end{align}
Thus, we need to solve
\begin{align}
    \left(\begin{array}{l} 0\\  0 \end{array}\right) &= \mathcal{U}_I(\eta_k; x_k, V) - \mathcal{U}_I(\widetilde{\eta}_k;x_k, \widetilde{V})
\label{eq:Delta_update_operator}    
\end{align}
for $\Delta \eta_k$, i.e.~we determine the change in the random number that yields the same path even though the potential has changed.
\subsection{One random number per integration time step}
We again use the ABO method to illustrate how to derive an expression for $\Delta \eta_k$. 
Its update function at the simulation potential $V$ is eq.~\ref{eq:ABO_update_operator01}, and at the target potential $\widetilde{V}(q)$  it is
\begin{eqnarray}
    \left(\begin{array}{c} q_{k+1}\\p_{k+1}\end{array}\right) 
    = \mathcal{U}_{\mathrm{ABO}}(\widetilde{\eta}_k; x_k, \widetilde{V})
    &=& \left(\begin{array}{l} q_{k} + ap_k\\dp_{k} + d\widetilde{b}(q_{k} + ap_k) +f\widetilde{\eta}_k\end{array}\right)  \, ,
\label{eq:ABO_update_operator03}         
\end{eqnarray}
where
\begin{eqnarray}
    \widetilde{b}(q) &=& b(q) -\Delta t\nabla U(q) \, .
    \label{eq:b_tilde}
\end{eqnarray}
Inserting eq.~\ref{eq:ABO_update_operator01} and \ref{eq:ABO_update_operator03} into eq.~\ref{eq:Delta_update_operator} yields
\begin{eqnarray}
    %
    \left(\begin{array}{c} 0\\0\end{array}\right) 
    &=& \mathcal{U}_{\mathrm{ABO}}(\eta_k; x_k, V) - \mathcal{U}_{\mathrm{ABO}}(\widetilde{\eta}_k; x_k, \widetilde{V}) \cr
    &=& \left(\begin{array}{l} 0\\ d\cdot (b(q_{k} + ap_k)- \widetilde{b}(q_{k} + ap_k)) + f\cdot(\eta_k -\widetilde{\eta}_k)\end{array}\right)  \cr
    &=& \left(\begin{array}{l} 0\\ d \Delta t\nabla U(q_{k+1}) - f\cdot\Delta \eta_k \end{array}\right)  \, ,
\end{eqnarray}
where, in the last line, we  replaced $q_{k} + ap_k$ by $q_{k+1}$ (eq.~\ref{eq:ABO01}).
Thus, for 
\begin{eqnarray}
    \Delta \eta_k &=& \frac{d}{f} \Delta t\nabla U(q_{k+1})\, ,
\label{eq:ABO_Delta_eta}
\end{eqnarray}
the two update functions yield the same state $(q_{k+1}, p_{k+1})^{\top}$.
The same calculation for ABOBA yields
\begin{eqnarray}
\label{eq:ABOBA_Delta_eta}
     \Delta \eta_k  &=& \frac{(d+1)}{f} \frac{\Delta t}{2} \nabla U(q_{k+1/2})\, 
\end{eqnarray}
(see supplementary material).
The relative conditional path probability densities $M[\mathbf{x} |x_0]$ for these two integrators can now be calculated according to eq.~\ref{eq:M_oneEta}.

%
%
\subsection{Two random numbers per integration time step}
The condition 
\begin{eqnarray}
    \left(\begin{array}{l} 0\\  0 \end{array}\right) &=& \mathcal{U}_{\mathrm{BOAOB}}(\eta_k^{(1)}, \eta_k^{(2)}; x_k ,V) -
    \mathcal{U}_{\mathrm{BOAOB}}(\widetilde{\eta}_k^{(1)}, \widetilde{\eta}_k^{(2)}; x_k ,\widetilde{V})
\end{eqnarray}
yields the random number differences for the BOAOB method
\begin{subequations}
\begin{eqnarray}
    \Delta \eta_k^{(1)} &=&  \frac{d'}{f'} \, \frac{\Delta t}{2}\nabla U(q_k)
\label{eq:BOAOB_Delta_eta1} \\     
    \Delta \eta_k^{(2)} &=& \frac{1}{f'}\frac{\Delta t}{2} \nabla U(q_{k+1}) \, .
\label{eq:BOAOB_Delta_eta2}        
\end{eqnarray}
\end{subequations}
Similarly, the condition 
\begin{eqnarray}
    \left(\begin{array}{l} 0\\  0 \end{array}\right) &=& \mathcal{U}_{\mathrm{OBABO}}(\eta_k^{(1)}, \eta_k^{(2)}; x_k ,V) -
    \mathcal{U}_{\mathrm{OBABO}}(\widetilde{\eta}_k^{(1)}, \widetilde{\eta}_k^{(2)}; x_k ,\widetilde{V})
\end{eqnarray}
yields the random number differences for the OBABO/BP method
\begin{subequations}
\begin{align}
    \Delta \eta_k^{(1)} &= \frac{1}{f'}\frac{\Delta t}{2} \nabla U(q_{k}) \label{eq:OBABO_Delta_eta1} \\
    \Delta \eta_k^{(2)} &=  \frac{d'}{f'} \, \frac{\Delta t}{2}\nabla U(q_{k+1}) \, . \label{eq:OBABO_Delta_eta2}      
\end{align}
\end{subequations}
For both methods, the intermediate steps of this calculation are reported in the supplementary material. 
The relative conditional path probability densities $M[\mathbf{x} |x_0]$ for these two integrators can now be calculated according to eq.~\ref{eq:M_twoEta}.

%
%
\subsection{AOBOA: two random numbers, but one-dimensional image}
The condition 
\begin{eqnarray}
    \left(\begin{array}{l} 0\\  0 \end{array}\right) &=& \mathcal{U}_{\mathrm{AOBOA}}(\eta_k^{\mathrm{comb}}; x_k ,V) -
    \mathcal{U}_{\mathrm{AOBOA}}(\widetilde{\eta}_k^{\mathrm{comb}}; x_k ,\widetilde{V})\, , 
\end{eqnarray}
where $\mathcal{U}_{\mathrm{AOBOA}}(\eta_k^{\mathrm{comb}}; x_k ,V)$ is given by eq.~\ref{eq:AOBOA_update_function}, yields the difference of the combined random number
\begin{eqnarray}
\label{eq:AOBOA_Delta_eta}
    \Delta \eta^{\mathrm{comb}}_k &=& \frac{d'}{f'}\Delta t \nabla U(q_{k+1/2}) \, .
\end{eqnarray}
The intermediate steps of this calculation are reported in the supplementary material. 

To reweight trajectories generated by the AOBOA integrator, we need to formulate $M[\mathbf{x}|x_0]$ as a function of the combined random numbers $\eta^{\mathrm{comb}}_k$.
From the update function of AOBOA (eq.~\ref{eq:AOBOA_update_function}), it follows that the conditional path probability is 
\begin{eqnarray}
    \mathcal{P}[\mathbf{x}|x_0] 
    &=& \mathcal{P} [\boldsymbol{\eta}^{\mathrm{comb}}]
    = \frac{1}{\sqrt[n]{2 (d'^2+1) \pi}}  \exp \left( - \sum_{k=0}^{n-1}\frac{\left(\eta_{k}^{\mathrm{comb}}\right)^2}{2(d'^2+1)} \right) \, .
\end{eqnarray}
Note that the probability density of the weighted sum of two normally distributed random numbers is again a normal distribution
with adjusted mean $\mu$ and variance $\sigma^2$. 
For the combined random number (eq.~\ref{eq:AOBOA_combined_eta}): 
$\mu^{\mathrm{comb}} = d' \mu^{(1)} + \mu^{(2)} = 0$ and
$(\sigma^{\mathrm{comb}})^2 = d'^2 (\sigma^{(1)})^2 + (\sigma^{(2)})^2 = d'^2+1$.
The relative conditional path probability density for AOBOA is
\begin{eqnarray}
    M[\mathbf{x}|x_0]  = 
	\frac{\mathcal{P}[\widetilde{\boldsymbol{\eta}}^{\mathrm{comb}}]}{\mathcal{P}[\boldsymbol{\eta}^{\mathrm{comb}}]} 
	&=&	 \exp\left(- \sum_{k=0}^{n-1} \frac{\eta_k^{\mathrm{comb}} \cdot \Delta \eta_k^{\mathrm{comb}}}{d'^2+1} \right) \cdot \exp\left(- \sum_{k=0}^{n-1} \frac{\left(\Delta \eta_k^{\mathrm{comb}}\right)^2}{2(d'^2+1)} \right) \, .
\label{eq:M_combEta}   
\end{eqnarray}
The random numbers $\eta_k^{(1)}$ and $\eta_k^{(2)}$ are recorded during the simulation at $V$. 
The combined random number $\eta_k^{\mathrm{comb}}$ is calculated according to eq.~\ref{eq:AOBOA_combined_eta}, 
and the expression for $\Delta \eta_k^{\mathrm{comb}}$ is given by eq.~\ref{eq:AOBOA_Delta_eta}.
%

\section{Graphical representation of Langevin integrators}
\label{sec:graphicalRepresentation}
In this section, we introduce a graphical representation of splitting methods for Langevin integrators. 
This representation helps in reasoning why some methods have potential-independent images, and why others do not. 
It also shows which parts of the integration algorithm are affected by a change in the potential and how this influences $\Delta \eta_k$.
\begin{figure}[h]
    \centering
    \includegraphics[width=14cm]{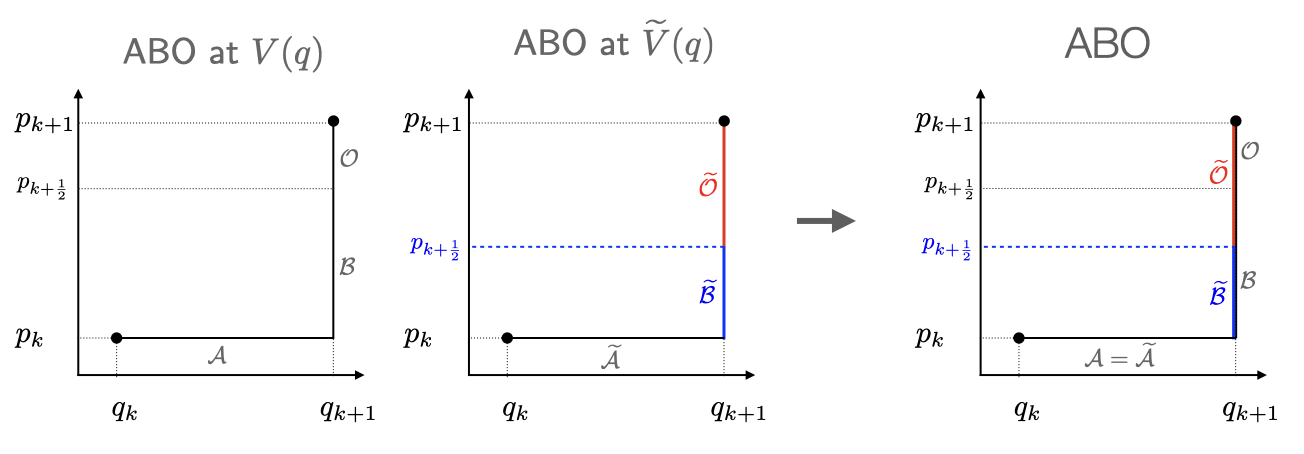}
    \caption{Update from $(q_k, p_k)$ to $(q_{k+1}, p_{k+1})$ for the ABO method at the simulations potential $V(x)$ and at the target potential $\widetilde{V}(x)= V(x)+U(x)$. Update operators and intermediate results that are affected by the change in potential are shown in colour.}
    \label{fig:stepsInPhaseSpace_ABO}
\end{figure}

Fig.~\ref{fig:stepsInPhaseSpace_ABO} illustrates this graphical representation for the ABO method.
The graphs show the phase space $(q, p)$, and the black dots represent the initial state $(q_k, p_k)^{\top}$ and the state $(q_{k+1}, p_{k+1})^{\top}$after one iteration of the ABO method.
The left-hand graph shows the update from $(q_k, p_k)$ to $(q_{k+1}, p_{k+1})$ for the simulation potential $V$ decomposed into the three update operators $\mathcal{A}$, $\mathcal{B}$, and $\mathcal{O}$. 
The $\mathcal{A}$-step depends on the momentum $p_k$ and updates the position from $q_k$ to $q_{k+1}$. 
This is followed by two momentum updates: the $\mathcal{B}$-step which depends on the updated position $q_{k+1}$ and potential $V$ and yields the intermediate momentum $p_{k+1/2}$, followed by an $\mathcal{O}$-step which depends on this intermediate momentum and a random number.
The graph in the middle shows how the same update from $(q_k, p_k)$ to $(q_{k+1}, p_{k+1})$ is achieved at the target potential.
Update operators and intermediate results that are affected by the change in potential are shown in colour.
The $\mathcal{A}$-step only depends on the initial momentum $p_k$ and therefore does not change. 
The $\mathcal{B}$-step evaluates the gradient of the potential and thus yields a different intermediate momentum $p_{k+1/2}$, shown in blue.
In order to reach the state $(q_{k+1}, p_{k+1})$, the random number in the $\mathcal{O}$-step needs to be adjusted such that $\mathcal{O}$ covers the remaining distance to $p_{k+1}$.
Colloquially: we have to adjust the random number in the $\mathcal{O}$-step such that $\sqrt{k_BTm(1-e^{-2\xi\Delta t})}\Delta \eta_{k}$  compensates ``for the mess that $-\nabla U$ created''.
The right-hand graph combines the two previous graphs, so that the update at $V(x)$ and at $\widetilde{V}(x)$ can be compared directly. 
Because the $\mathcal{A}$-step in the ABO method is unaffected by a change in the potential, the equation to determine $\Delta \eta_k$ simplifies.
Instead of the condition 
\begin{eqnarray}
    \left[
    \mathcal{O}\mathcal{B}\mathcal{A}
    -
    \widetilde{\mathcal{O}} \widetilde{\mathcal{B}}\widetilde{\mathcal{A}}
    \right]
    \left(\begin{array}{l} q_k\\p_k\end{array}\right)
    &=&
    \left(\begin{array}{c} 0 \\ 0 \end{array}\right)   
\end{eqnarray}
we only need to solve the condition
\begin{eqnarray}
    \left[
    \mathcal{O}\mathcal{B}
    -
    \widetilde{\mathcal{O}} \widetilde{\mathcal{B}}
    \right]
    \left(\begin{array}{c} q_{k+1} \\ p_k \end{array}\right)
    &=&
    \left(\begin{array}{c} 0 \\ 0 \end{array}\right)   \, .
\label{eq:ABO_cond}    
\end{eqnarray}
which yields the same expression for $\Delta \eta_k$ as eq.~\ref{eq:ABO_Delta_eta}.
\begin{figure}[h]
    \centering
    \includegraphics[width=14cm]{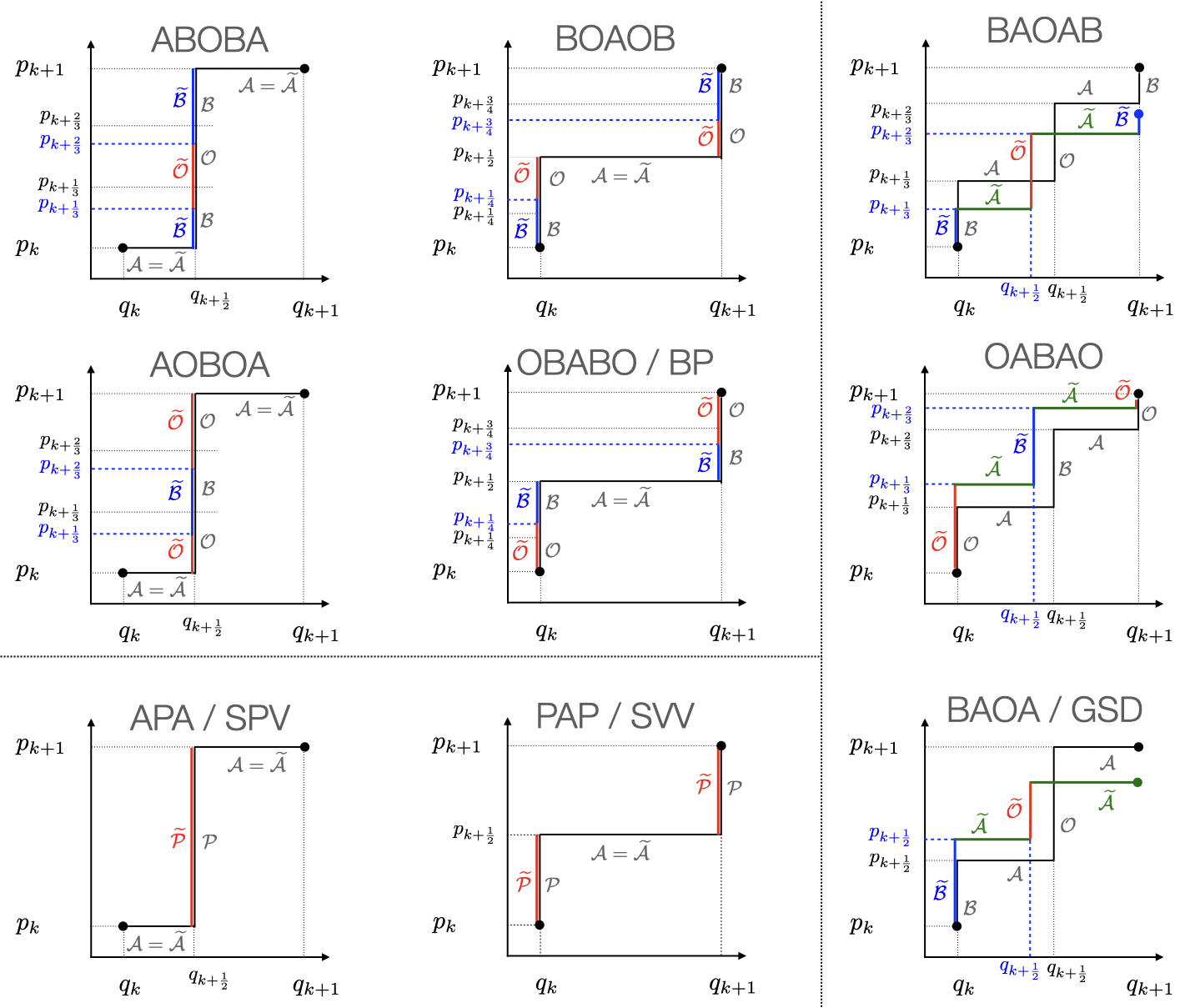}
    \caption{Decomposition of the transition from $(q_k, p_k)$ to $(q_{k+1}, p_{k+1})$ for Langevin integrators derived via the operator splitting method.
    Methods that are not suited to reweight position trajectories are highlighted in gray.\\
    }
    \label{fig:stepsInPhaseSpace}
\end{figure}

Fig.~\ref{fig:stepsInPhaseSpace} shows the graphical representation of the six symmetric splitting methods \cite{Bussi:2007, Leimkuhler:2012, Leimkuhler:2013, Sivak:2013}. as well as for BAOA/GSD.
These Langevin integrators can be classified according to the route they take through phase space during an integration time step from $(q_k, p_k)$ to $(q_{k+1}, p_{k+1})$.
In the integrators in the first columns of Fig.~\ref{fig:stepsInPhaseSpace}, the position update occurs in two steps, and in between the steps the momentum is updated. 
Stochastic position Verlet (SPV) \cite{Melchionna:2007}, a splitting method that has been proposed prior to the symmetric splitting methods, takes the same route through phase space.
In the algorithm in the second column, the momentum update occurs in two steps, and in between the steps the position is updated.  
Also in this case, there is an older algorithm that takes the same route through phase space: stochastic velocity Verlet (SVV) \cite{Melchionna:2007}.
Algorithms in which neither position nor momentum update is calculated in one go, but position and momentum update are interspersed, are shown in the third column.
Using Fig.~\ref{fig:stepsInPhaseSpace}, one can derive simplified conditions for $\Delta \eta_k$ by considering which sub-steps in the integrator are affected by a change in $V$.
In the ABOBA method, the first $\mathcal{A}$-step depends only on $p_k$ and is not affected by a change in the potential. 
Therefore also the intermediate result $p_{k+1/2}$ is not affected by a change in the potential.
The momentum update $p_k \rightarrow p_{k+1}$ is contained in the sequence of steps $\mathcal{BOB}$ 
(vertical line in Fig.~\ref{fig:stepsInPhaseSpace}), where the $\mathcal{B}$-step is affected by a change in the potential.
In order to obtain the same momentum update in both potentials, the following condition for the two random numbers needs to be fulfilled:
\begin{eqnarray}
    \left[\mathcal{B'}\mathcal{O}\mathcal{B}' - \widetilde{\mathcal{B}'}\widetilde{\mathcal{O}} \widetilde{\mathcal{B}'}\right]
    \left(\begin{array}{c} q_{k+1/2} \\ p_{k} \end{array}\right)
    &=& \left(\begin{array}{c} 0 \\ 0 \end{array}\right)
    \label{eq:ABOBA_cond}
\end{eqnarray}
The final $\mathcal{A}$-step only depends on $p_{k+1}$ and is the same in both potentials if $p_{k+1}$ remains the same.
Solving eq.~\ref{eq:ABOBA_cond} yields the same $\Delta \eta_k$ as defined in eq.~\ref{eq:ABOBA_Delta_eta}. 
Similarly, in the AOBOA method it suffices to solve 
\begin{eqnarray}
    \left[\mathcal{O'}\mathcal{B}\mathcal{O}' - \widetilde{\mathcal{O}'}\widetilde{\mathcal{B}} \widetilde{\mathcal{O}'}\right]
    \left(\begin{array}{c} q_{k+1/2} \\ p_{k} \end{array}\right)
    &=& \left(\begin{array}{c} 0 \\ 0 \end{array}\right)
    \label{eq:AOBOA_cond}
\end{eqnarray}
to obtain eq.~\ref{eq:AOBOA_Delta_eta} for $\Delta \eta_k^{\mathrm{cond}}$.
Consider the algorithms in the second column. 
In the BOAOB method, the update of the positions $q_k \rightarrow q_{k+1}$ is contained in a single $\mathcal{A}$-step, which depends on the intermediate momentum result $p_{k+1/2}$. 
If $p_{k+1/2}$ is altered, $q_{k+1}$ will change.
In order to obtain the same path at the simulation and at the target potential, $p_{k+1/2}$ has to be the same in both potentials. 
This is the case, if the updates $p_{k} \rightarrow p_{k+1/2}$ and $p_{k+1/2} \rightarrow p_{k+1}$
(vertical lines in Fig.~\ref{fig:stepsInPhaseSpace}) are the same in both potentials.
Thus, we have two separate conditions, one for each random number:
\begin{subequations}
\begin{eqnarray}
    \left[\mathcal{O}'\mathcal{B}' - \widetilde{\mathcal{O}'} \widetilde{\mathcal{B}'}\right]\left(\begin{array}{c} q_{k} \\ p_{k} \end{array}\right)
    &=& \left(\begin{array}{c} 0 \\ 0 \end{array}\right)  \\
    \label{eq:BOAOB_cond1} 
    \left[\mathcal{B}'\mathcal{O}' - \widetilde{\mathcal{B}'} \widetilde{\mathcal{O}'}\right]
    \left(\begin{array}{c} q_{k+1} \\ p_{k+1/2} \end{array}\right)
    &=& \left(\begin{array}{c} 0 \\ 0 \end{array}\right)   \label{eq:BOAOB_cond2} 
\end{eqnarray}
\end{subequations}
which lead to eqs.~\ref{eq:BOAOB_Delta_eta1} and \ref{eq:BOAOB_Delta_eta2} for $\eta_k^{(1)}$ and $\eta_k^{(2)}$.
(Note that the operators act from right to left, and thus their order is reversed compared to the name of the method.)
Similarly, the conditions for OBABO/BP are 
\begin{subequations}
\begin{eqnarray}
    \left[\mathcal{B}'\mathcal{O}' - \widetilde{\mathcal{B}'} \widetilde{\mathcal{O}'}\right]
    \left(\begin{array}{c} q_{k} \\ p_{k} \end{array}\right)
    &=& \left(\begin{array}{c} 0 \\ 0 \end{array}\right)
    \label{eq:OBABO_cond1} \\
    \left[\mathcal{O}'\mathcal{B}' - \widetilde{\mathcal{O}'} \widetilde{\mathcal{B}'}\right]
    \left(\begin{array}{c} q_{k+1} \\ p_{k+1/2} \end{array}\right)
    &=& \left(\begin{array}{c} 0 \\ 0 \end{array}\right) 
    \label{eq:OBABO_cond2}
\end{eqnarray}
\end{subequations}
which lead to eqs.~\ref{eq:OBABO_Delta_eta1} and \ref{eq:OBABO_Delta_eta2} for $\eta_k^{(1)}$ and $\eta_k^{(2)}$.
BAOAB and BAOA/GSD in the third column in Fig.~\ref{fig:stepsInPhaseSpace} do not allow for reweighting in phase space. 
Consider BAOA/GSD to understand why. 
The first two steps in the algorithm are deterministic and the $\mathcal{B}$-step is affected by a change in $V$.
Consequently, at $\widetilde{V}$ the intermediate state $(q_{k+1/2}, p_{k+1/2})$ will be different from the intermediate state at $V$.
One can now choose the value of $\eta_k$ such that the $\mathcal{O}$-step reaches $p_{k+1}$. 
In that case, the subsequent $\mathcal{A}$-step will not reach $q_{k+1}$.
Or one can scale to a momentum $\widetilde{p}_{k+1}$ that is sufficient to bridge the remaining gap between $q_{k+1/2}$ and $q_{k+1}$, but this $\widetilde{p}_{k+1}$ will not be equal to the original $p_{k+1}$.
Thus, one can reach either $p_{k+1}$ or $q_{k+1}$, but not both. 
The equations for the corresponding values of $\Delta \eta_k$ are reported in the supplementary material.
Note that this in principles allows for reweighting path expected values, in which the path observable $s[\mathbf{x}]$ only depends on the positions or only depends on the momenta.
However, the resulting reweighted estimators might have a large variance or even a bias.
An analogous reasoning applies to BAOAB. 
We provide the equations for $\Delta \eta_k$ for separately reweighting in position and momentum space in the supplementary material, and give the same warning.
For the OBABO method reweighting should in principle be possible, because we have two random numbers to adjust the updated position and the updated momentum. 
However, since pathological cases in which the image of the update function changes with the potential can be constructed (see supplementary material), we do not derive the equations for the random number differences here.

%
%

\section{Conclusion}
\label{sec:conclusion}
We have introduced a strategy to derive the relative path-probability ratio for Langevin integrators. 
To achieve this, we adapted the method to derive the random number difference $\Delta\eta_k$ as presented in Ref.~\onlinecite{Kieninger:2021}.
With the random numbers differences presented in Tab.~\ref{tab:integrators} it is now possible to use Girsanov reweighting with underdamped Langevin dynamics. 
This removes a major road block to study rare events by combining biased simulations with dynamical reweighting. 
Besides reweighting enhanced sampling simulations, the relative path probability ratio for underdamped Langevin integrators can also be use in other contexts such as searching and sampling in path space \cite{Fujisaki:2013, Lee:2017, Peter:2020} or force-field optimization \cite{Bolhuis:2002}. 
It might also help in understanding the relation between molecular models that rely on path sampling and models that are derived from a direct discretization of the Fokker-Planck equation \cite{Donati:2022}.
The analysis revealed that some algorithms violate absolute continuity, because the image of their update changes when the potential is change. 
As a consequence single-step transitions $(q_k, p_k) \rightarrow (q_{k+1}, p_{k+1})$ that are possible at the simulation potential are no longer possible at the target potential.
Our condition for the absolute continuity of the single-step transitions (eq.~\ref{eq:absoluteContinuity_transition}) is a very strict condition and guarantees the Girsanov reweighting in the full phase space is possible.
However, path observables usually either depend on the positions or on the momenta, rarely on both. 
Thus, for algorithms that violate eq.~\ref{eq:absoluteContinuity_transition} reweighting in position space or in path space might still be possible. 
In fact, we report random number difference for these approaches for BAOA/GSD in the supplementary material. 
Whether these equations yield accurate results has yet to be tested.
Another interesting question is how the absolute continuity of a Langevin integrator relates to the accuracy with which it reproduces transport properties \cite{Gronbech:2020, Finkelstein:2021}.
Finally, in some cases one might be able to relax the absolute continuity over a single integration time-step to an absolute continuity over a larger lag time of several integration time steps, and thereby link Girsanov reweighting to dynamical reweighting methods that assume a Markov state model \cite{Wu:2014, Mey:2014, Rosta:2015, Wu:2016, Stelzl:2017}.

%

%
%
    %
    %
    %

%
%
\section{Acknowledgements}

This research has been funded by Deutsche Forschungsgemeinschaft (DFG) 
through grant SFB 1114 "Scaling Cascades in Complex Systems" - project number 235221301, and
through grant GRK 2473 ”Bioactive Peptides” - project number 392923329.
S.G.~acknowledges funding by the Einstein Center of Catalysis/BIG-NSE.

%
%
\section{References}
\bibliography{literature_theory}

\begin{thebibliography}{86}%
\makeatletter
\providecommand \@ifxundefined [1]{%
 \@ifx{#1\undefined}
}%
\providecommand \@ifnum [1]{%
 \ifnum #1\expandafter \@firstoftwo
 \else \expandafter \@secondoftwo
 \fi
}%
\providecommand \@ifx [1]{%
 \ifx #1\expandafter \@firstoftwo
 \else \expandafter \@secondoftwo
 \fi
}%
\providecommand \natexlab [1]{#1}%
\providecommand \enquote  [1]{``#1''}%
\providecommand \bibnamefont  [1]{#1}%
\providecommand \bibfnamefont [1]{#1}%
\providecommand \citenamefont [1]{#1}%
\providecommand \href@noop [0]{\@secondoftwo}%
\providecommand \href [0]{\begingroup \@sanitize@url \@href}%
\providecommand \@href[1]{\@@startlink{#1}\@@href}%
\providecommand \@@href[1]{\endgroup#1\@@endlink}%
\providecommand \@sanitize@url [0]{\catcode `\\12\catcode `\$12\catcode
  `\&12\catcode `\#12\catcode `\^12\catcode `\_12\catcode `\%12\relax}%
\providecommand \@@startlink[1]{}%
\providecommand \@@endlink[0]{}%
\providecommand \url  [0]{\begingroup\@sanitize@url \@url }%
\providecommand \@url [1]{\endgroup\@href {#1}{\urlprefix }}%
\providecommand \urlprefix  [0]{URL }%
\providecommand \Eprint [0]{\href }%
\providecommand \doibase [0]{https://doi.org/}%
\providecommand \selectlanguage [0]{\@gobble}%
\providecommand \bibinfo  [0]{\@secondoftwo}%
\providecommand \bibfield  [0]{\@secondoftwo}%
\providecommand \translation [1]{[#1]}%
\providecommand \BibitemOpen [0]{}%
\providecommand \bibitemStop [0]{}%
\providecommand \bibitemNoStop [0]{.\EOS\space}%
\providecommand \EOS [0]{\spacefactor3000\relax}%
\providecommand \BibitemShut  [1]{\csname bibitem#1\endcsname}%
\let\auto@bib@innerbib\@empty
\bibitem [{\citenamefont {Karplus}\ and\ \citenamefont
  {McCammon}(2002)}]{Karplus:2002}%
  \BibitemOpen
  \bibfield  {author} {\bibinfo {author} {\bibfnamefont {M.}~\bibnamefont
  {Karplus}}\ and\ \bibinfo {author} {\bibfnamefont {J.~A.}\ \bibnamefont
  {McCammon}},\ }\bibfield  {title} {\enquote {\bibinfo {title} {Molecular
  dynamics simulations of biomolecules},}\ }\href@noop {} {\bibfield  {journal}
  {\bibinfo  {journal} {Nat. Struct. Biol.}\ }\textbf {\bibinfo {volume} {9}},\
  \bibinfo {pages} {646--652} (\bibinfo {year} {2002})}\BibitemShut {NoStop}%
\bibitem [{\citenamefont {Dror}\ \emph {et~al.}(2012)\citenamefont {Dror},
  \citenamefont {Dirks}, \citenamefont {Grossman}, \citenamefont {Xu},\ and\
  \citenamefont {Shaw}}]{Dror:2012}%
  \BibitemOpen
  \bibfield  {author} {\bibinfo {author} {\bibfnamefont {R.~O.}\ \bibnamefont
  {Dror}}, \bibinfo {author} {\bibfnamefont {R.~M.}\ \bibnamefont {Dirks}},
  \bibinfo {author} {\bibfnamefont {J.}~\bibnamefont {Grossman}}, \bibinfo
  {author} {\bibfnamefont {H.}~\bibnamefont {Xu}},\ and\ \bibinfo {author}
  {\bibfnamefont {D.~E.}\ \bibnamefont {Shaw}},\ }\bibfield  {title} {\enquote
  {\bibinfo {title} {Biomolecular simulation: a computational microscope for
  molecular biology},}\ }\href@noop {} {\bibfield  {journal} {\bibinfo
  {journal} {Annu Rev Biophys}\ }\textbf {\bibinfo {volume} {41}},\ \bibinfo
  {pages} {429--452} (\bibinfo {year} {2012})}\BibitemShut {NoStop}%
\bibitem [{\citenamefont {Lane}\ \emph {et~al.}(2013)\citenamefont {Lane},
  \citenamefont {Shukla}, \citenamefont {Beauchamp},\ and\ \citenamefont
  {Pande}}]{Lane:2013}%
  \BibitemOpen
  \bibfield  {author} {\bibinfo {author} {\bibfnamefont {T.~J.}\ \bibnamefont
  {Lane}}, \bibinfo {author} {\bibfnamefont {D.}~\bibnamefont {Shukla}},
  \bibinfo {author} {\bibfnamefont {K.~A.}\ \bibnamefont {Beauchamp}},\ and\
  \bibinfo {author} {\bibfnamefont {V.~S.}\ \bibnamefont {Pande}},\ }\bibfield
  {title} {\enquote {\bibinfo {title} {To milliseconds and beyond: challenges
  in the simulation of protein folding},}\ }\href@noop {} {\bibfield  {journal}
  {\bibinfo  {journal} {Curr. Opin. Struct. Biol.}\ }\textbf {\bibinfo {volume}
  {23}},\ \bibinfo {pages} {58--65} (\bibinfo {year} {2013})}\BibitemShut
  {NoStop}%
\bibitem [{\citenamefont {Huber}, \citenamefont {Torda},\ and\ \citenamefont
  {van Gunsteren}(1994)}]{Huber:1994}%
  \BibitemOpen
  \bibfield  {author} {\bibinfo {author} {\bibfnamefont {T.}~\bibnamefont
  {Huber}}, \bibinfo {author} {\bibfnamefont {A.}~\bibnamefont {Torda}},\ and\
  \bibinfo {author} {\bibfnamefont {W.}~\bibnamefont {van Gunsteren}},\
  }\bibfield  {title} {\enquote {\bibinfo {title} {Local elevation: A method
  for improving the searching properties of molecular dynamics simulation},}\
  }\href@noop {} {\bibfield  {journal} {\bibinfo  {journal} {J. Comput. Aided
  Mol. Des.}\ }\textbf {\bibinfo {volume} {8}},\ \bibinfo {pages} {695}
  (\bibinfo {year} {1994})}\BibitemShut {NoStop}%
\bibitem [{\citenamefont {Grubm{\"u}ller}(1995)}]{Grubmuller:1995}%
  \BibitemOpen
  \bibfield  {author} {\bibinfo {author} {\bibfnamefont {H.}~\bibnamefont
  {Grubm{\"u}ller}},\ }\bibfield  {title} {\enquote {\bibinfo {title}
  {Predicting slow structural transitions in macromolecular systems:
  Conformational flooding},}\ }\href@noop {} {\bibfield  {journal} {\bibinfo
  {journal} {Physical Review E}\ }\textbf {\bibinfo {volume} {52}},\ \bibinfo
  {pages} {2893} (\bibinfo {year} {1995})}\BibitemShut {NoStop}%
\bibitem [{\citenamefont {Darve}\ and\ \citenamefont
  {Pohorille}(2001)}]{Darve:2001}%
  \BibitemOpen
  \bibfield  {author} {\bibinfo {author} {\bibfnamefont {E.}~\bibnamefont
  {Darve}}\ and\ \bibinfo {author} {\bibfnamefont {A.}~\bibnamefont
  {Pohorille}},\ }\bibfield  {title} {\enquote {\bibinfo {title} {Calculating
  free energies using average force},}\ }\href@noop {} {\bibfield  {journal}
  {\bibinfo  {journal} {The Journal of chemical physics}\ }\textbf {\bibinfo
  {volume} {115}},\ \bibinfo {pages} {9169--9183} (\bibinfo {year}
  {2001})}\BibitemShut {NoStop}%
\bibitem [{\citenamefont {Laio}\ and\ \citenamefont
  {Parrinello}(2002)}]{Laio:2002}%
  \BibitemOpen
  \bibfield  {author} {\bibinfo {author} {\bibfnamefont {A.}~\bibnamefont
  {Laio}}\ and\ \bibinfo {author} {\bibfnamefont {M.}~\bibnamefont
  {Parrinello}},\ }\bibfield  {title} {\enquote {\bibinfo {title} {Escaping
  free-energy minima},}\ }\href@noop {} {\bibfield  {journal} {\bibinfo
  {journal} {Proc. Natl. Acad. Sci. U.S.A.}\ }\textbf {\bibinfo {volume}
  {99}},\ \bibinfo {pages} {12562--12566} (\bibinfo {year} {2002})}\BibitemShut
  {NoStop}%
\bibitem [{\citenamefont {Barducci}, \citenamefont {Bussi},\ and\ \citenamefont
  {Parrinello}(2008)}]{Barducci:2008}%
  \BibitemOpen
  \bibfield  {author} {\bibinfo {author} {\bibfnamefont {A.}~\bibnamefont
  {Barducci}}, \bibinfo {author} {\bibfnamefont {G.}~\bibnamefont {Bussi}},\
  and\ \bibinfo {author} {\bibfnamefont {M.}~\bibnamefont {Parrinello}},\
  }\bibfield  {title} {\enquote {\bibinfo {title} {Well-tempered metadynamics:
  a smoothly converging and tunable free-energy method},}\ }\href@noop {}
  {\bibfield  {journal} {\bibinfo  {journal} {Phys. Rev. Lett.}\ }\textbf
  {\bibinfo {volume} {100}},\ \bibinfo {pages} {020603} (\bibinfo {year}
  {2008})}\BibitemShut {NoStop}%
\bibitem [{\citenamefont {Torrie}\ and\ \citenamefont
  {Valleau}(1977)}]{Torrie:1977}%
  \BibitemOpen
  \bibfield  {author} {\bibinfo {author} {\bibfnamefont {G.~M.}\ \bibnamefont
  {Torrie}}\ and\ \bibinfo {author} {\bibfnamefont {J.~P.}\ \bibnamefont
  {Valleau}},\ }\bibfield  {title} {\enquote {\bibinfo {title} {Nonphysical
  sampling distributions in monte carlo free-energy estimation: Umbrella
  sampling},}\ }\href@noop {} {\bibfield  {journal} {\bibinfo  {journal} {J.
  Comput. Phys.}\ }\textbf {\bibinfo {volume} {23}},\ \bibinfo {pages}
  {187--199} (\bibinfo {year} {1977})}\BibitemShut {NoStop}%
\bibitem [{\citenamefont {K{\"a}stner}\ and\ \citenamefont
  {Thiel}(2005)}]{Kaestner:2005}%
  \BibitemOpen
  \bibfield  {author} {\bibinfo {author} {\bibfnamefont {J.}~\bibnamefont
  {K{\"a}stner}}\ and\ \bibinfo {author} {\bibfnamefont {W.}~\bibnamefont
  {Thiel}},\ }\bibfield  {title} {\enquote {\bibinfo {title} {Bridging the gap
  between thermodynamic integration and umbrella sampling provides a novel
  analysis method: “umbrella integration"},}\ }\href@noop {} {\bibfield
  {journal} {\bibinfo  {journal} {J. Chem. Phys.}\ }\textbf {\bibinfo {volume}
  {123}},\ \bibinfo {pages} {144104} (\bibinfo {year} {2005})}\BibitemShut
  {NoStop}%
\bibitem [{\citenamefont {Swendsen}, \citenamefont {Wang},\ and\ \citenamefont
  {Ferrenberg}(1992)}]{Swendsen:1992}%
  \BibitemOpen
  \bibfield  {author} {\bibinfo {author} {\bibfnamefont {R.~H.}\ \bibnamefont
  {Swendsen}}, \bibinfo {author} {\bibfnamefont {J.-S.}\ \bibnamefont {Wang}},\
  and\ \bibinfo {author} {\bibfnamefont {A.~M.}\ \bibnamefont {Ferrenberg}},\
  }\bibfield  {title} {\enquote {\bibinfo {title} {New monte carlo methods for
  improved efficiency of computer simulations in statistical mechanics},}\
  }\href@noop {} {\bibfield  {journal} {\bibinfo  {journal} {The Monte Carlo
  method in condensed matter physics}\ ,\ \bibinfo {pages} {75--91}} (\bibinfo
  {year} {1992})}\BibitemShut {NoStop}%
\bibitem [{\citenamefont {Kumar}\ \emph {et~al.}(1992)\citenamefont {Kumar},
  \citenamefont {Rosenberg}, \citenamefont {Bouzida}, \citenamefont
  {Swendsen},\ and\ \citenamefont {Kollman}}]{Kumar:1992}%
  \BibitemOpen
  \bibfield  {author} {\bibinfo {author} {\bibfnamefont {S.}~\bibnamefont
  {Kumar}}, \bibinfo {author} {\bibfnamefont {J.~M.}\ \bibnamefont
  {Rosenberg}}, \bibinfo {author} {\bibfnamefont {D.}~\bibnamefont {Bouzida}},
  \bibinfo {author} {\bibfnamefont {R.~H.}\ \bibnamefont {Swendsen}},\ and\
  \bibinfo {author} {\bibfnamefont {P.~A.}\ \bibnamefont {Kollman}},\
  }\bibfield  {title} {\enquote {\bibinfo {title} {The weighted histogram
  analysis method for free-energy calculations on biomolecules. i. the
  method},}\ }\href@noop {} {\bibfield  {journal} {\bibinfo  {journal} {Journal
  of computational chemistry}\ }\textbf {\bibinfo {volume} {13}},\ \bibinfo
  {pages} {1011--1021} (\bibinfo {year} {1992})}\BibitemShut {NoStop}%
\bibitem [{\citenamefont {Kieninger}, \citenamefont {Donati},\ and\
  \citenamefont {Keller}(2020)}]{Kieninger:2020}%
  \BibitemOpen
  \bibfield  {author} {\bibinfo {author} {\bibfnamefont {S.}~\bibnamefont
  {Kieninger}}, \bibinfo {author} {\bibfnamefont {L.}~\bibnamefont {Donati}},\
  and\ \bibinfo {author} {\bibfnamefont {B.~G.}\ \bibnamefont {Keller}},\
  }\bibfield  {title} {\enquote {\bibinfo {title} {Dynamical reweighting
  methods for markov models},}\ }\href@noop {} {\bibfield  {journal} {\bibinfo
  {journal} {Curr. Opin. Struct. Biol.}\ }\textbf {\bibinfo {volume} {61}},\
  \bibinfo {pages} {124--131} (\bibinfo {year} {2020})}\BibitemShut {NoStop}%
\bibitem [{\citenamefont {De~Oliveira}, \citenamefont {Hamelberg},\ and\
  \citenamefont {McCammon}(2007)}]{deOliveira:2007}%
  \BibitemOpen
  \bibfield  {author} {\bibinfo {author} {\bibfnamefont {C.~A.~F.}\
  \bibnamefont {De~Oliveira}}, \bibinfo {author} {\bibfnamefont
  {D.}~\bibnamefont {Hamelberg}},\ and\ \bibinfo {author} {\bibfnamefont
  {J.~A.}\ \bibnamefont {McCammon}},\ }\bibfield  {title} {\enquote {\bibinfo
  {title} {Estimating kinetic rates from accelerated molecular dynamics
  simulations: Alanine dipeptide in explicit solvent as a case study},}\
  }\href@noop {} {\bibfield  {journal} {\bibinfo  {journal} {J. Chem. Phys.}\
  }\textbf {\bibinfo {volume} {127}},\ \bibinfo {pages} {11B605} (\bibinfo
  {year} {2007})}\BibitemShut {NoStop}%
\bibitem [{\citenamefont {Doshi}\ and\ \citenamefont
  {Hamelberg}(2011)}]{Doshi:2011}%
  \BibitemOpen
  \bibfield  {author} {\bibinfo {author} {\bibfnamefont {U.}~\bibnamefont
  {Doshi}}\ and\ \bibinfo {author} {\bibfnamefont {D.}~\bibnamefont
  {Hamelberg}},\ }\bibfield  {title} {\enquote {\bibinfo {title} {Extracting
  realistic kinetics of rare activated processes from accelerated molecular
  dynamics using kramers’ theory},}\ }\href@noop {} {\bibfield  {journal}
  {\bibinfo  {journal} {Journal of Chemical Theory and Computation}\ }\textbf
  {\bibinfo {volume} {7}},\ \bibinfo {pages} {575--581} (\bibinfo {year}
  {2011})}\BibitemShut {NoStop}%
\bibitem [{\citenamefont {Tiwary}\ and\ \citenamefont
  {Parrinello}(2013)}]{Tiwary:2013}%
  \BibitemOpen
  \bibfield  {author} {\bibinfo {author} {\bibfnamefont {P.}~\bibnamefont
  {Tiwary}}\ and\ \bibinfo {author} {\bibfnamefont {M.}~\bibnamefont
  {Parrinello}},\ }\bibfield  {title} {\enquote {\bibinfo {title} {From
  metadynamics to dynamics},}\ }\href@noop {} {\bibfield  {journal} {\bibinfo
  {journal} {Phys. Rev. Lett.}\ }\textbf {\bibinfo {volume} {111}},\ \bibinfo
  {pages} {230602} (\bibinfo {year} {2013})}\BibitemShut {NoStop}%
\bibitem [{\citenamefont {Frank}\ and\ \citenamefont
  {Andricioaei}(2016)}]{Frank:2016}%
  \BibitemOpen
  \bibfield  {author} {\bibinfo {author} {\bibfnamefont {A.~T.}\ \bibnamefont
  {Frank}}\ and\ \bibinfo {author} {\bibfnamefont {I.}~\bibnamefont
  {Andricioaei}},\ }\bibfield  {title} {\enquote {\bibinfo {title} {Reaction
  coordinate-free approach to recovering kinetics from potential-scaled
  simulations: application of kramers’ rate theory},}\ }\href@noop {}
  {\bibfield  {journal} {\bibinfo  {journal} {The Journal of Physical Chemistry
  B}\ }\textbf {\bibinfo {volume} {120}},\ \bibinfo {pages} {8600--8605}
  (\bibinfo {year} {2016})}\BibitemShut {NoStop}%
\bibitem [{\citenamefont {Palacio-Rodriguez}\ \emph {et~al.}(2022)\citenamefont
  {Palacio-Rodriguez}, \citenamefont {Vroylandt}, \citenamefont {Stelzl},
  \citenamefont {Pietrucci}, \citenamefont {Hummer},\ and\ \citenamefont
  {Cossio}}]{Palacio:2022b}%
  \BibitemOpen
  \bibfield  {author} {\bibinfo {author} {\bibfnamefont {K.}~\bibnamefont
  {Palacio-Rodriguez}}, \bibinfo {author} {\bibfnamefont {H.}~\bibnamefont
  {Vroylandt}}, \bibinfo {author} {\bibfnamefont {L.~S.}\ \bibnamefont
  {Stelzl}}, \bibinfo {author} {\bibfnamefont {F.}~\bibnamefont {Pietrucci}},
  \bibinfo {author} {\bibfnamefont {G.}~\bibnamefont {Hummer}},\ and\ \bibinfo
  {author} {\bibfnamefont {P.}~\bibnamefont {Cossio}},\ }\bibfield  {title}
  {\enquote {\bibinfo {title} {Transition rates and efficiency of collective
  variables from time-dependent biased simulations},}\ }\href@noop {}
  {\bibfield  {journal} {\bibinfo  {journal} {The Journal of Physical Chemistry
  Letters}\ }\textbf {\bibinfo {volume} {13}},\ \bibinfo {pages} {7490--7496}
  (\bibinfo {year} {2022})}\BibitemShut {NoStop}%
\bibitem [{\citenamefont {Bicout}\ and\ \citenamefont
  {Szabo}(1998)}]{Bicout:1998}%
  \BibitemOpen
  \bibfield  {author} {\bibinfo {author} {\bibfnamefont {D.}~\bibnamefont
  {Bicout}}\ and\ \bibinfo {author} {\bibfnamefont {A.}~\bibnamefont {Szabo}},\
  }\bibfield  {title} {\enquote {\bibinfo {title} {Electron transfer reaction
  dynamics in non-debye solvents},}\ }\href@noop {} {\bibfield  {journal}
  {\bibinfo  {journal} {The Journal of chemical physics}\ }\textbf {\bibinfo
  {volume} {109}},\ \bibinfo {pages} {2325--2338} (\bibinfo {year}
  {1998})}\BibitemShut {NoStop}%
\bibitem [{\citenamefont {Wu}\ \emph {et~al.}(2014)\citenamefont {Wu},
  \citenamefont {Mey}, \citenamefont {Rosta},\ and\ \citenamefont
  {No{\'e}}}]{Wu:2014}%
  \BibitemOpen
  \bibfield  {author} {\bibinfo {author} {\bibfnamefont {H.}~\bibnamefont
  {Wu}}, \bibinfo {author} {\bibfnamefont {A.~S.}\ \bibnamefont {Mey}},
  \bibinfo {author} {\bibfnamefont {E.}~\bibnamefont {Rosta}},\ and\ \bibinfo
  {author} {\bibfnamefont {F.}~\bibnamefont {No{\'e}}},\ }\bibfield  {title}
  {\enquote {\bibinfo {title} {Statistically optimal analysis of
  state-discretized trajectory data from multiple thermodynamic states},}\
  }\href@noop {} {\bibfield  {journal} {\bibinfo  {journal} {The Journal of
  Chemical Physics}\ }\textbf {\bibinfo {volume} {141}},\ \bibinfo {pages}
  {12B629\_1} (\bibinfo {year} {2014})}\BibitemShut {NoStop}%
\bibitem [{\citenamefont {Mey}, \citenamefont {Wu},\ and\ \citenamefont
  {No{\'e}}(2014)}]{Mey:2014}%
  \BibitemOpen
  \bibfield  {author} {\bibinfo {author} {\bibfnamefont {A.~S.}\ \bibnamefont
  {Mey}}, \bibinfo {author} {\bibfnamefont {H.}~\bibnamefont {Wu}},\ and\
  \bibinfo {author} {\bibfnamefont {F.}~\bibnamefont {No{\'e}}},\ }\bibfield
  {title} {\enquote {\bibinfo {title} {xtram: Estimating equilibrium
  expectations from time-correlated simulation data at multiple thermodynamic
  states},}\ }\href@noop {} {\bibfield  {journal} {\bibinfo  {journal}
  {Physical Review X}\ }\textbf {\bibinfo {volume} {4}},\ \bibinfo {pages}
  {041018} (\bibinfo {year} {2014})}\BibitemShut {NoStop}%
\bibitem [{\citenamefont {Rosta}\ and\ \citenamefont
  {Hummer}(2015)}]{Rosta:2015}%
  \BibitemOpen
  \bibfield  {author} {\bibinfo {author} {\bibfnamefont {E.}~\bibnamefont
  {Rosta}}\ and\ \bibinfo {author} {\bibfnamefont {G.}~\bibnamefont {Hummer}},\
  }\bibfield  {title} {\enquote {\bibinfo {title} {Free energies from dynamic
  weighted histogram analysis using unbiased markov state model},}\ }\href@noop
  {} {\bibfield  {journal} {\bibinfo  {journal} {J. Chem. Theory Comput.}\
  }\textbf {\bibinfo {volume} {11}},\ \bibinfo {pages} {276--285} (\bibinfo
  {year} {2015})}\BibitemShut {NoStop}%
\bibitem [{\citenamefont {Wu}\ \emph {et~al.}(2016)\citenamefont {Wu},
  \citenamefont {Paul}, \citenamefont {Wehmeyer},\ and\ \citenamefont
  {No\'{e}}}]{Wu:2016}%
  \BibitemOpen
  \bibfield  {author} {\bibinfo {author} {\bibfnamefont {H.}~\bibnamefont
  {Wu}}, \bibinfo {author} {\bibfnamefont {F.}~\bibnamefont {Paul}}, \bibinfo
  {author} {\bibfnamefont {C.}~\bibnamefont {Wehmeyer}},\ and\ \bibinfo
  {author} {\bibfnamefont {F.}~\bibnamefont {No\'{e}}},\ }\bibfield  {title}
  {\enquote {\bibinfo {title} {Multiensemble markov models of molecular
  thermodynamics and kinetics},}\ }\href@noop {} {\bibfield  {journal}
  {\bibinfo  {journal} {Proc. Natl. Acad. Sci. U.S.A}\ }\textbf {\bibinfo
  {volume} {113}},\ \bibinfo {pages} {E3221} (\bibinfo {year}
  {2016})}\BibitemShut {NoStop}%
\bibitem [{\citenamefont {Stelzl}\ \emph {et~al.}(2017)\citenamefont {Stelzl},
  \citenamefont {Kells}, \citenamefont {Rosta},\ and\ \citenamefont
  {Hummer}}]{Stelzl:2017}%
  \BibitemOpen
  \bibfield  {author} {\bibinfo {author} {\bibfnamefont {L.~S.}\ \bibnamefont
  {Stelzl}}, \bibinfo {author} {\bibfnamefont {A.}~\bibnamefont {Kells}},
  \bibinfo {author} {\bibfnamefont {E.}~\bibnamefont {Rosta}},\ and\ \bibinfo
  {author} {\bibfnamefont {G.}~\bibnamefont {Hummer}},\ }\bibfield  {title}
  {\enquote {\bibinfo {title} {Dynamic histogram analysis to determine free
  energies and rates from biased simulations},}\ }\href@noop {} {\bibfield
  {journal} {\bibinfo  {journal} {Journal of chemical theory and computation}\
  }\textbf {\bibinfo {volume} {13}},\ \bibinfo {pages} {6328--6342} (\bibinfo
  {year} {2017})}\BibitemShut {NoStop}%
\bibitem [{\citenamefont {Onsager}\ and\ \citenamefont
  {Machlup}(1953)}]{Onsager:1953}%
  \BibitemOpen
  \bibfield  {author} {\bibinfo {author} {\bibfnamefont {L.}~\bibnamefont
  {Onsager}}\ and\ \bibinfo {author} {\bibfnamefont {S.}~\bibnamefont
  {Machlup}},\ }\bibfield  {title} {\enquote {\bibinfo {title} {Fluctuations
  and irreversible processes},}\ }\href@noop {} {\bibfield  {journal} {\bibinfo
   {journal} {Phys. Rev.}\ }\textbf {\bibinfo {volume} {91}},\ \bibinfo {pages}
  {1505} (\bibinfo {year} {1953})}\BibitemShut {NoStop}%
\bibitem [{\citenamefont {Girsanov}(1960)}]{Girsanov1960}%
  \BibitemOpen
  \bibfield  {author} {\bibinfo {author} {\bibfnamefont {I.~V.}\ \bibnamefont
  {Girsanov}},\ }\bibfield  {title} {\enquote {\bibinfo {title} {On
  transforming a certain class of stochastic processes by absolutely continuous
  substitution of measures},}\ }\href@noop {} {\bibfield  {journal} {\bibinfo
  {journal} {Theory Probab. Its Appl.}\ }\textbf {\bibinfo {volume} {5}},\
  \bibinfo {pages} {285} (\bibinfo {year} {1960})}\BibitemShut {NoStop}%
\bibitem [{\citenamefont {{\O}ksendal}(2003)}]{Oeksendal:2003}%
  \BibitemOpen
  \bibfield  {author} {\bibinfo {author} {\bibfnamefont {B.}~\bibnamefont
  {{\O}ksendal}},\ }\href@noop {} {\emph {\bibinfo {title} {Stochastic
  Differential Equations: An Introduction with Applications}}},\ \bibinfo
  {edition} {6th}\ ed.\ (\bibinfo  {publisher} {Springer Verlag, Berlin},\
  \bibinfo {year} {2003})\BibitemShut {NoStop}%
\bibitem [{\citenamefont {Kloeden}\ and\ \citenamefont
  {Platen}(1992)}]{Kloeden:1992}%
  \BibitemOpen
  \bibfield  {author} {\bibinfo {author} {\bibfnamefont {P.~E.}\ \bibnamefont
  {Kloeden}}\ and\ \bibinfo {author} {\bibfnamefont {E.}~\bibnamefont
  {Platen}},\ }\href@noop {} {\emph {\bibinfo {title} {Numerical Solution of
  Stochastic Differential Equations}}},\ \bibinfo {edition} {1st}\ ed.\
  (\bibinfo  {publisher} {Springer, Berlin},\ \bibinfo {year}
  {1992})\BibitemShut {NoStop}%
\bibitem [{\citenamefont {Mazonka}, \citenamefont {Jarzy{\'n}ski},\ and\
  \citenamefont {B{\l}ocki}(1998)}]{Mazonka:1998}%
  \BibitemOpen
  \bibfield  {author} {\bibinfo {author} {\bibfnamefont {O.}~\bibnamefont
  {Mazonka}}, \bibinfo {author} {\bibfnamefont {C.}~\bibnamefont
  {Jarzy{\'n}ski}},\ and\ \bibinfo {author} {\bibfnamefont {J.}~\bibnamefont
  {B{\l}ocki}},\ }\bibfield  {title} {\enquote {\bibinfo {title} {Computing
  probabilities of very rare events for langevin processes: a new method based
  on importance sampling},}\ }\href@noop {} {\bibfield  {journal} {\bibinfo
  {journal} {Nuclear Physics A}\ }\textbf {\bibinfo {volume} {641}},\ \bibinfo
  {pages} {335--354} (\bibinfo {year} {1998})}\BibitemShut {NoStop}%
\bibitem [{\citenamefont {Woolf}(1998)}]{Woolf:1998}%
  \BibitemOpen
  \bibfield  {author} {\bibinfo {author} {\bibfnamefont {T.~B.}\ \bibnamefont
  {Woolf}},\ }\bibfield  {title} {\enquote {\bibinfo {title} {Path corrected
  functionals of stochastic trajectories: towards relative free energy and
  reaction coordinate calculations},}\ }\href@noop {} {\bibfield  {journal}
  {\bibinfo  {journal} {Chem. Phys. Lett.}\ }\textbf {\bibinfo {volume}
  {289}},\ \bibinfo {pages} {433--441} (\bibinfo {year} {1998})}\BibitemShut
  {NoStop}%
\bibitem [{\citenamefont {Zuckerman}\ and\ \citenamefont
  {Woolf}(1999)}]{Zuckerman:1999}%
  \BibitemOpen
  \bibfield  {author} {\bibinfo {author} {\bibfnamefont {D.~M.}\ \bibnamefont
  {Zuckerman}}\ and\ \bibinfo {author} {\bibfnamefont {T.~B.}\ \bibnamefont
  {Woolf}},\ }\bibfield  {title} {\enquote {\bibinfo {title} {Dynamic reaction
  paths and rates through importance-sampled stochastic dynamics},}\
  }\href@noop {} {\bibfield  {journal} {\bibinfo  {journal} {J. Chem. Phys.}\
  }\textbf {\bibinfo {volume} {111}},\ \bibinfo {pages} {9475--9484} (\bibinfo
  {year} {1999})}\BibitemShut {NoStop}%
\bibitem [{\citenamefont {Zuckerman}\ and\ \citenamefont
  {Woolf}(2000)}]{Zuckerman:2000}%
  \BibitemOpen
  \bibfield  {author} {\bibinfo {author} {\bibfnamefont {D.~M.}\ \bibnamefont
  {Zuckerman}}\ and\ \bibinfo {author} {\bibfnamefont {T.~B.}\ \bibnamefont
  {Woolf}},\ }\bibfield  {title} {\enquote {\bibinfo {title} {Efficient dynamic
  importance sampling of rare events in one dimension},}\ }\href@noop {}
  {\bibfield  {journal} {\bibinfo  {journal} {Phys. Rev. E}\ }\textbf {\bibinfo
  {volume} {63}},\ \bibinfo {pages} {016702} (\bibinfo {year}
  {2000})}\BibitemShut {NoStop}%
\bibitem [{\citenamefont {Adib}(2008)}]{Adib:2008}%
  \BibitemOpen
  \bibfield  {author} {\bibinfo {author} {\bibfnamefont {A.~B.}\ \bibnamefont
  {Adib}},\ }\bibfield  {title} {\enquote {\bibinfo {title} {Stochastic actions
  for diffusive dynamics: Reweighting, sampling, and minimization},}\
  }\href@noop {} {\bibfield  {journal} {\bibinfo  {journal} {J. Phys. Chem. B}\
  }\textbf {\bibinfo {volume} {112}},\ \bibinfo {pages} {5910--5916} (\bibinfo
  {year} {2008})}\BibitemShut {NoStop}%
\bibitem [{\citenamefont {Jang}\ and\ \citenamefont {Woolf}(2006)}]{Jang:2006}%
  \BibitemOpen
  \bibfield  {author} {\bibinfo {author} {\bibfnamefont {H.}~\bibnamefont
  {Jang}}\ and\ \bibinfo {author} {\bibfnamefont {T.~B.}\ \bibnamefont
  {Woolf}},\ }\bibfield  {title} {\enquote {\bibinfo {title} {Multiple pathways
  in conformational transitions of the alanine dipeptide: an application of
  dynamic importance sampling},}\ }\href@noop {} {\bibfield  {journal}
  {\bibinfo  {journal} {Journal of computational chemistry}\ }\textbf {\bibinfo
  {volume} {27}},\ \bibinfo {pages} {1136--1141} (\bibinfo {year}
  {2006})}\BibitemShut {NoStop}%
\bibitem [{\citenamefont {Sch{\"u}tte}, \citenamefont {Nielsen},\ and\
  \citenamefont {Weber}(2015)}]{Schuette:2015}%
  \BibitemOpen
  \bibfield  {author} {\bibinfo {author} {\bibfnamefont {C.}~\bibnamefont
  {Sch{\"u}tte}}, \bibinfo {author} {\bibfnamefont {A.}~\bibnamefont
  {Nielsen}},\ and\ \bibinfo {author} {\bibfnamefont {M.}~\bibnamefont
  {Weber}},\ }\bibfield  {title} {\enquote {\bibinfo {title} {Markov state
  models and molecular alchemy},}\ }\href@noop {} {\bibfield  {journal}
  {\bibinfo  {journal} {Mol. Phys.}\ }\textbf {\bibinfo {volume} {113}},\
  \bibinfo {pages} {69--78} (\bibinfo {year} {2015})}\BibitemShut {NoStop}%
\bibitem [{\citenamefont {Bolhuis}, \citenamefont {Brotzakis},\ and\
  \citenamefont {Keller}(2022)}]{Bolhuis:2022}%
  \BibitemOpen
  \bibfield  {author} {\bibinfo {author} {\bibfnamefont {P.}~\bibnamefont
  {Bolhuis}}, \bibinfo {author} {\bibfnamefont {Z.}~\bibnamefont {Brotzakis}},\
  and\ \bibinfo {author} {\bibfnamefont {B.}~\bibnamefont {Keller}},\
  }\bibfield  {title} {\enquote {\bibinfo {title} {Force field optimization by
  imposing kinetic constraints with path reweighting},}\ }\href@noop {}
  {\bibfield  {journal} {\bibinfo  {journal} {arXiv preprint arXiv:2207.04558}\
  } (\bibinfo {year} {2022})}\BibitemShut {NoStop}%
\bibitem [{\citenamefont {Huber}\ and\ \citenamefont
  {McCammon}(2019)}]{Huber:2019}%
  \BibitemOpen
  \bibfield  {author} {\bibinfo {author} {\bibfnamefont {G.~A.}\ \bibnamefont
  {Huber}}\ and\ \bibinfo {author} {\bibfnamefont {J.~A.}\ \bibnamefont
  {McCammon}},\ }\bibfield  {title} {\enquote {\bibinfo {title} {Brownian
  dynamics simulations of biological molecules},}\ }\href@noop {} {\bibfield
  {journal} {\bibinfo  {journal} {Trends in chemistry}\ }\textbf {\bibinfo
  {volume} {1}},\ \bibinfo {pages} {727--738} (\bibinfo {year}
  {2019})}\BibitemShut {NoStop}%
\bibitem [{\citenamefont {Cholko}\ \emph {et~al.}(2022)\citenamefont {Cholko},
  \citenamefont {Kaushik}, \citenamefont {Wu}, \citenamefont {Montes},\ and\
  \citenamefont {Chang}}]{Cholko:2022}%
  \BibitemOpen
  \bibfield  {author} {\bibinfo {author} {\bibfnamefont {T.}~\bibnamefont
  {Cholko}}, \bibinfo {author} {\bibfnamefont {S.}~\bibnamefont {Kaushik}},
  \bibinfo {author} {\bibfnamefont {K.~Y.}\ \bibnamefont {Wu}}, \bibinfo
  {author} {\bibfnamefont {R.}~\bibnamefont {Montes}},\ and\ \bibinfo {author}
  {\bibfnamefont {C.-e.~A.}\ \bibnamefont {Chang}},\ }\bibfield  {title}
  {\enquote {\bibinfo {title} {Geombd3: Brownian dynamics simulation software
  for biological and engineered systems},}\ }\href@noop {} {\bibfield
  {journal} {\bibinfo  {journal} {Journal of chemical information and
  modeling}\ }\textbf {\bibinfo {volume} {62}},\ \bibinfo {pages} {2257--2263}
  (\bibinfo {year} {2022})}\BibitemShut {NoStop}%
\bibitem [{\citenamefont {H{\"u}nenberger}(2005)}]{Huenenberger:2005}%
  \BibitemOpen
  \bibfield  {author} {\bibinfo {author} {\bibfnamefont {P.~H.}\ \bibnamefont
  {H{\"u}nenberger}},\ }\enquote {\bibinfo {title} {Thermostat algorithms for
  molecular dynamics simulations},}\ in\ \href@noop {} {\emph {\bibinfo
  {booktitle} {Advanced Computer Simulation: Approaches for Soft Matter
  Sciences I}}}\ (\bibinfo  {publisher} {Springer Berlin Heidelberg},\ \bibinfo
  {year} {2005})\ pp.\ \bibinfo {pages} {105--149}\BibitemShut {NoStop}%
\bibitem [{\citenamefont {Kwon}\ and\ \citenamefont {Lee}(2022)}]{Kwon:2022}%
  \BibitemOpen
  \bibfield  {author} {\bibinfo {author} {\bibfnamefont {C.}~\bibnamefont
  {Kwon}}\ and\ \bibinfo {author} {\bibfnamefont {H.~K.}\ \bibnamefont {Lee}},\
  }\bibfield  {title} {\enquote {\bibinfo {title} {Thermodynamic uncertainty
  relation for underdamped dynamics driven by time-dependent protocols},}\
  }\href@noop {} {\bibfield  {journal} {\bibinfo  {journal} {New Journal of
  Physics}\ }\textbf {\bibinfo {volume} {24}},\ \bibinfo {pages} {013029}
  (\bibinfo {year} {2022})}\BibitemShut {NoStop}%
\bibitem [{\citenamefont {Athenes}(2004)}]{Athenes:2004}%
  \BibitemOpen
  \bibfield  {author} {\bibinfo {author} {\bibfnamefont {M.}~\bibnamefont
  {Athenes}},\ }\bibfield  {title} {\enquote {\bibinfo {title} {A path-sampling
  scheme for computing thermodynamic properties of a many-body system in a
  generalized ensemble},}\ }\href@noop {} {\bibfield  {journal} {\bibinfo
  {journal} {The European Physical Journal B-Condensed Matter and Complex
  Systems}\ }\textbf {\bibinfo {volume} {38}},\ \bibinfo {pages} {651--663}
  (\bibinfo {year} {2004})}\BibitemShut {NoStop}%
\bibitem [{\citenamefont {Xing}\ and\ \citenamefont
  {Andricioaei}(2006)}]{Xing:2006}%
  \BibitemOpen
  \bibfield  {author} {\bibinfo {author} {\bibfnamefont {C.}~\bibnamefont
  {Xing}}\ and\ \bibinfo {author} {\bibfnamefont {I.}~\bibnamefont
  {Andricioaei}},\ }\bibfield  {title} {\enquote {\bibinfo {title} {On the
  calculation of time correlation functions by potential scaling},}\
  }\href@noop {} {\bibfield  {journal} {\bibinfo  {journal} {J. Chem. Phys.}\
  }\textbf {\bibinfo {volume} {124}},\ \bibinfo {pages} {034110} (\bibinfo
  {year} {2006})}\BibitemShut {NoStop}%
\bibitem [{\citenamefont {Donati}, \citenamefont {Hartmann},\ and\
  \citenamefont {Keller}(2017)}]{Donati:2017}%
  \BibitemOpen
  \bibfield  {author} {\bibinfo {author} {\bibfnamefont {L.}~\bibnamefont
  {Donati}}, \bibinfo {author} {\bibfnamefont {C.}~\bibnamefont {Hartmann}},\
  and\ \bibinfo {author} {\bibfnamefont {B.~G.}\ \bibnamefont {Keller}},\
  }\bibfield  {title} {\enquote {\bibinfo {title} {Girsanov reweighting for
  path ensembles and markov state models},}\ }\href@noop {} {\bibfield
  {journal} {\bibinfo  {journal} {J. Chem. Phys.}\ }\textbf {\bibinfo {volume}
  {146}},\ \bibinfo {pages} {244112} (\bibinfo {year} {2017})}\BibitemShut
  {NoStop}%
\bibitem [{\citenamefont {Donati}\ and\ \citenamefont
  {Keller}(2018)}]{Donati:2018}%
  \BibitemOpen
  \bibfield  {author} {\bibinfo {author} {\bibfnamefont {L.}~\bibnamefont
  {Donati}}\ and\ \bibinfo {author} {\bibfnamefont {B.~G.}\ \bibnamefont
  {Keller}},\ }\bibfield  {title} {\enquote {\bibinfo {title} {Girsanov
  reweighting for metadynamics simulations},}\ }\href@noop {} {\bibfield
  {journal} {\bibinfo  {journal} {J. Chem. Phys.}\ }\textbf {\bibinfo {volume}
  {149}},\ \bibinfo {pages} {072335} (\bibinfo {year} {2018})}\BibitemShut
  {NoStop}%
\bibitem [{\citenamefont {Kieninger}\ and\ \citenamefont
  {Keller}(2021)}]{Kieninger:2021}%
  \BibitemOpen
  \bibfield  {author} {\bibinfo {author} {\bibfnamefont {S.}~\bibnamefont
  {Kieninger}}\ and\ \bibinfo {author} {\bibfnamefont {B.~G.}\ \bibnamefont
  {Keller}},\ }\bibfield  {title} {\enquote {\bibinfo {title} {Path probability
  ratios for langevin dynamics—exact and approximate},}\ }\href@noop {}
  {\bibfield  {journal} {\bibinfo  {journal} {J. Chem. Phys.}\ }\textbf
  {\bibinfo {volume} {154}},\ \bibinfo {pages} {094102} (\bibinfo {year}
  {2021})}\BibitemShut {NoStop}%
\bibitem [{\citenamefont {Bussi}\ and\ \citenamefont
  {Parrinello}(2007)}]{Bussi:2007}%
  \BibitemOpen
  \bibfield  {author} {\bibinfo {author} {\bibfnamefont {G.}~\bibnamefont
  {Bussi}}\ and\ \bibinfo {author} {\bibfnamefont {M.}~\bibnamefont
  {Parrinello}},\ }\bibfield  {title} {\enquote {\bibinfo {title} {{Accurate
  sampling using Langevin dynamics.}}}\ }\href@noop {} {\bibfield  {journal}
  {\bibinfo  {journal} {Phys. Rev. E Stat. Nonlin. Soft Matter Phys.}\ }\textbf
  {\bibinfo {volume} {75}},\ \bibinfo {pages} {056707} (\bibinfo {year}
  {2007})}\BibitemShut {NoStop}%
\bibitem [{\citenamefont {Leimkuhler}\ and\ \citenamefont
  {Matthews}(2012)}]{Leimkuhler:2012}%
  \BibitemOpen
  \bibfield  {author} {\bibinfo {author} {\bibfnamefont {B.}~\bibnamefont
  {Leimkuhler}}\ and\ \bibinfo {author} {\bibfnamefont {C.}~\bibnamefont
  {Matthews}},\ }\bibfield  {title} {\enquote {\bibinfo {title} {{Rational
  Construction of Stochastic Numerical Methods for Molecular Sampling}},}\
  }\href@noop {} {\bibfield  {journal} {\bibinfo  {journal} {Appl Math Res
  Express}\ }\textbf {\bibinfo {volume} {48}},\ \bibinfo {pages} {278}
  (\bibinfo {year} {2012})}\BibitemShut {NoStop}%
\bibitem [{\citenamefont {Leimkuhler}\ and\ \citenamefont
  {Matthews}(2013)}]{Leimkuhler:2013}%
  \BibitemOpen
  \bibfield  {author} {\bibinfo {author} {\bibfnamefont {B.}~\bibnamefont
  {Leimkuhler}}\ and\ \bibinfo {author} {\bibfnamefont {C.}~\bibnamefont
  {Matthews}},\ }\bibfield  {title} {\enquote {\bibinfo {title} {{Robust and
  efficient configurational molecular sampling via Langevin dynamics}},}\
  }\href@noop {} {\bibfield  {journal} {\bibinfo  {journal} {J. Chem. Phys.}\
  }\textbf {\bibinfo {volume} {138}},\ \bibinfo {pages} {174102} (\bibinfo
  {year} {2013})}\BibitemShut {NoStop}%
\bibitem [{\citenamefont {Sivak}, \citenamefont {Chodera},\ and\ \citenamefont
  {Crooks}(2013)}]{Sivak:2013}%
  \BibitemOpen
  \bibfield  {author} {\bibinfo {author} {\bibfnamefont {D.~A.}\ \bibnamefont
  {Sivak}}, \bibinfo {author} {\bibfnamefont {J.~D.}\ \bibnamefont {Chodera}},\
  and\ \bibinfo {author} {\bibfnamefont {G.~E.}\ \bibnamefont {Crooks}},\
  }\bibfield  {title} {\enquote {\bibinfo {title} {{Using Nonequilibrium
  Fluctuation Theorems to Understand and Correct Errors in Equilibrium and
  Nonequilibrium Simulations of Discrete Langevin Dynamics}},}\ }\href@noop {}
  {\bibfield  {journal} {\bibinfo  {journal} {Phys. Rev. X}\ }\textbf {\bibinfo
  {volume} {3}},\ \bibinfo {pages} {011007} (\bibinfo {year}
  {2013})}\BibitemShut {NoStop}%
\bibitem [{\citenamefont {Kieninger}\ and\ \citenamefont
  {Keller}(2022)}]{Kieninger:2022}%
  \BibitemOpen
  \bibfield  {author} {\bibinfo {author} {\bibfnamefont {S.}~\bibnamefont
  {Kieninger}}\ and\ \bibinfo {author} {\bibfnamefont {B.~G.}\ \bibnamefont
  {Keller}},\ }\bibfield  {title} {\enquote {\bibinfo {title} {Gromacs
  stochastic dynamics and baoab are equivalent configurational sampling
  algorithms},}\ }\href@noop {} {\bibfield  {journal} {\bibinfo  {journal}
  {Journal of Chemical Theory and Computation}\ }\textbf {\bibinfo {volume}
  {18}},\ \bibinfo {pages} {5792--5798} (\bibinfo {year} {2022})}\BibitemShut
  {NoStop}%
\bibitem [{\citenamefont {Bou-Rabee}\ and\ \citenamefont
  {Owhadi}(2010)}]{BouRabee:2010}%
  \BibitemOpen
  \bibfield  {author} {\bibinfo {author} {\bibfnamefont {N.}~\bibnamefont
  {Bou-Rabee}}\ and\ \bibinfo {author} {\bibfnamefont {H.}~\bibnamefont
  {Owhadi}},\ }\bibfield  {title} {\enquote {\bibinfo {title} {Long-run
  accuracy of variational integrators in the stochastic context},}\ }\href@noop
  {} {\bibfield  {journal} {\bibinfo  {journal} {SIAM J Numer Anal}\ }\textbf
  {\bibinfo {volume} {48}},\ \bibinfo {pages} {278--297} (\bibinfo {year}
  {2010})}\BibitemShut {NoStop}%
\bibitem [{\citenamefont {Goga}\ \emph {et~al.}(2012)\citenamefont {Goga},
  \citenamefont {Rzepiela}, \citenamefont {De~Vries}, \citenamefont {Marrink},\
  and\ \citenamefont {Berendsen}}]{Goga:2012}%
  \BibitemOpen
  \bibfield  {author} {\bibinfo {author} {\bibfnamefont {N.}~\bibnamefont
  {Goga}}, \bibinfo {author} {\bibfnamefont {A.}~\bibnamefont {Rzepiela}},
  \bibinfo {author} {\bibfnamefont {A.}~\bibnamefont {De~Vries}}, \bibinfo
  {author} {\bibfnamefont {S.}~\bibnamefont {Marrink}},\ and\ \bibinfo {author}
  {\bibfnamefont {H.}~\bibnamefont {Berendsen}},\ }\bibfield  {title} {\enquote
  {\bibinfo {title} {Efficient algorithms for langevin and dpd dynamics},}\
  }\href@noop {} {\bibfield  {journal} {\bibinfo  {journal} {J. Chem. Theory
  Comput.}\ }\textbf {\bibinfo {volume} {8}},\ \bibinfo {pages} {3637--3649}
  (\bibinfo {year} {2012})}\BibitemShut {NoStop}%
\bibitem [{\citenamefont {Eastman}\ \emph {et~al.}(2017)\citenamefont
  {Eastman}, \citenamefont {Swails}, \citenamefont {Chodera}, \citenamefont
  {McGibbon}, \citenamefont {Zhao}, \citenamefont {Beauchamp}, \citenamefont
  {Wang}, \citenamefont {Simmonett}, \citenamefont {Harrigan}, \citenamefont
  {Stern} \emph {et~al.}}]{Eastman:2017}%
  \BibitemOpen
  \bibfield  {author} {\bibinfo {author} {\bibfnamefont {P.}~\bibnamefont
  {Eastman}}, \bibinfo {author} {\bibfnamefont {J.}~\bibnamefont {Swails}},
  \bibinfo {author} {\bibfnamefont {J.~D.}\ \bibnamefont {Chodera}}, \bibinfo
  {author} {\bibfnamefont {R.~T.}\ \bibnamefont {McGibbon}}, \bibinfo {author}
  {\bibfnamefont {Y.}~\bibnamefont {Zhao}}, \bibinfo {author} {\bibfnamefont
  {K.~A.}\ \bibnamefont {Beauchamp}}, \bibinfo {author} {\bibfnamefont {L.-P.}\
  \bibnamefont {Wang}}, \bibinfo {author} {\bibfnamefont {A.~C.}\ \bibnamefont
  {Simmonett}}, \bibinfo {author} {\bibfnamefont {M.~P.}\ \bibnamefont
  {Harrigan}}, \bibinfo {author} {\bibfnamefont {C.~D.}\ \bibnamefont {Stern}},
  \emph {et~al.},\ }\bibfield  {title} {\enquote {\bibinfo {title} {Openmm 7:
  Rapid development of high performance algorithms for molecular dynamics},}\
  }\href@noop {} {\bibfield  {journal} {\bibinfo  {journal} {PLoS Comput.
  Biol.}\ }\textbf {\bibinfo {volume} {13}},\ \bibinfo {pages} {e1005659}
  (\bibinfo {year} {2017})}\BibitemShut {NoStop}%
\bibitem [{\citenamefont {Case}\ \emph {et~al.}(2005)\citenamefont {Case},
  \citenamefont {Cheatham~III}, \citenamefont {Darden}, \citenamefont {Gohlke},
  \citenamefont {Luo}, \citenamefont {Merz~Jr}, \citenamefont {Onufriev},
  \citenamefont {Simmerling}, \citenamefont {Wang},\ and\ \citenamefont
  {Woods}}]{Case:2005}%
  \BibitemOpen
  \bibfield  {author} {\bibinfo {author} {\bibfnamefont {D.~A.}\ \bibnamefont
  {Case}}, \bibinfo {author} {\bibfnamefont {T.~E.}\ \bibnamefont
  {Cheatham~III}}, \bibinfo {author} {\bibfnamefont {T.}~\bibnamefont
  {Darden}}, \bibinfo {author} {\bibfnamefont {H.}~\bibnamefont {Gohlke}},
  \bibinfo {author} {\bibfnamefont {R.}~\bibnamefont {Luo}}, \bibinfo {author}
  {\bibfnamefont {K.~M.}\ \bibnamefont {Merz~Jr}}, \bibinfo {author}
  {\bibfnamefont {A.}~\bibnamefont {Onufriev}}, \bibinfo {author}
  {\bibfnamefont {C.}~\bibnamefont {Simmerling}}, \bibinfo {author}
  {\bibfnamefont {B.}~\bibnamefont {Wang}},\ and\ \bibinfo {author}
  {\bibfnamefont {R.~J.}\ \bibnamefont {Woods}},\ }\bibfield  {title} {\enquote
  {\bibinfo {title} {The amber biomolecular simulation programs},}\ }\href@noop
  {} {\bibfield  {journal} {\bibinfo  {journal} {J. Comput. Chem}\ }\textbf
  {\bibinfo {volume} {26}},\ \bibinfo {pages} {1668--1688} (\bibinfo {year}
  {2005})}\BibitemShut {NoStop}%
\bibitem [{\citenamefont {Van Der~Spoel}\ \emph {et~al.}(2005)\citenamefont
  {Van Der~Spoel}, \citenamefont {Lindahl}, \citenamefont {Hess}, \citenamefont
  {Groenhof}, \citenamefont {Mark},\ and\ \citenamefont
  {Berendsen}}]{VanDerSpoel:2005}%
  \BibitemOpen
  \bibfield  {author} {\bibinfo {author} {\bibfnamefont {D.}~\bibnamefont {Van
  Der~Spoel}}, \bibinfo {author} {\bibfnamefont {E.}~\bibnamefont {Lindahl}},
  \bibinfo {author} {\bibfnamefont {B.}~\bibnamefont {Hess}}, \bibinfo {author}
  {\bibfnamefont {G.}~\bibnamefont {Groenhof}}, \bibinfo {author}
  {\bibfnamefont {A.~E.}\ \bibnamefont {Mark}},\ and\ \bibinfo {author}
  {\bibfnamefont {H.~J.}\ \bibnamefont {Berendsen}},\ }\bibfield  {title}
  {\enquote {\bibinfo {title} {Gromacs: fast, flexible, and free},}\
  }\href@noop {} {\bibfield  {journal} {\bibinfo  {journal} {J. Comput. Chem}\
  }\textbf {\bibinfo {volume} {26}},\ \bibinfo {pages} {1701--1718} (\bibinfo
  {year} {2005})}\BibitemShut {NoStop}%
\bibitem [{\citenamefont {Leimkuhler}\ and\ \citenamefont
  {Matthews}(2015)}]{Leimkuhler:2015}%
  \BibitemOpen
  \bibfield  {author} {\bibinfo {author} {\bibfnamefont {B.}~\bibnamefont
  {Leimkuhler}}\ and\ \bibinfo {author} {\bibfnamefont {C.}~\bibnamefont
  {Matthews}},\ }\href@noop {} {\emph {\bibinfo {title} {Molecular Dynamics,
  With Deterministic and Stochastic Numerical Methods}}},\ \bibinfo {edition}
  {1st}\ ed.\ (\bibinfo  {publisher} {Springer Cham Heidelberg New York
  Dordrecht London},\ \bibinfo {year} {2015})\BibitemShut {NoStop}%
\bibitem [{\citenamefont {Sivak}, \citenamefont {Chodera},\ and\ \citenamefont
  {Crooks}(2014)}]{Sivak:2014}%
  \BibitemOpen
  \bibfield  {author} {\bibinfo {author} {\bibfnamefont {D.~A.}\ \bibnamefont
  {Sivak}}, \bibinfo {author} {\bibfnamefont {J.~D.}\ \bibnamefont {Chodera}},\
  and\ \bibinfo {author} {\bibfnamefont {G.~E.}\ \bibnamefont {Crooks}},\
  }\bibfield  {title} {\enquote {\bibinfo {title} {{Time step rescaling
  recovers continuous-time dynamical properties for discrete-time Langevin
  integration of nonequilibrium systems.}}}\ }\href@noop {} {\bibfield
  {journal} {\bibinfo  {journal} {J. Phys. Chem. B}\ }\textbf {\bibinfo
  {volume} {118}},\ \bibinfo {pages} {6466--6474} (\bibinfo {year}
  {2014})}\BibitemShut {NoStop}%
\bibitem [{\citenamefont {Leimkuhler}, \citenamefont {Matthews},\ and\
  \citenamefont {Stoltz}(2016)}]{Leimkuhler:2016b}%
  \BibitemOpen
  \bibfield  {author} {\bibinfo {author} {\bibfnamefont {B.}~\bibnamefont
  {Leimkuhler}}, \bibinfo {author} {\bibfnamefont {C.}~\bibnamefont
  {Matthews}},\ and\ \bibinfo {author} {\bibfnamefont {G.}~\bibnamefont
  {Stoltz}},\ }\bibfield  {title} {\enquote {\bibinfo {title} {The computation
  of averages from equilibrium and nonequilibrium langevin molecular
  dynamics},}\ }\href@noop {} {\bibfield  {journal} {\bibinfo  {journal} {IMA
  Journal of Numerical Analysis}\ }\textbf {\bibinfo {volume} {36}},\ \bibinfo
  {pages} {13--79} (\bibinfo {year} {2016})}\BibitemShut {NoStop}%
\bibitem [{\citenamefont {Fass}\ \emph {et~al.}(2018)\citenamefont {Fass},
  \citenamefont {Sivak}, \citenamefont {Crooks}, \citenamefont {Beauchamp},
  \citenamefont {Leimkuhler},\ and\ \citenamefont {Chodera}}]{Fass:2018}%
  \BibitemOpen
  \bibfield  {author} {\bibinfo {author} {\bibfnamefont {J.}~\bibnamefont
  {Fass}}, \bibinfo {author} {\bibfnamefont {D.~A.}\ \bibnamefont {Sivak}},
  \bibinfo {author} {\bibfnamefont {G.~E.}\ \bibnamefont {Crooks}}, \bibinfo
  {author} {\bibfnamefont {K.~A.}\ \bibnamefont {Beauchamp}}, \bibinfo {author}
  {\bibfnamefont {B.}~\bibnamefont {Leimkuhler}},\ and\ \bibinfo {author}
  {\bibfnamefont {J.~D.}\ \bibnamefont {Chodera}},\ }\bibfield  {title}
  {\enquote {\bibinfo {title} {{Quantifying Configuration-Sampling Error in
  Langevin Simulations of Complex Molecular Systems}},}\ }\href@noop {}
  {\bibfield  {journal} {\bibinfo  {journal} {Entropy}\ }\textbf {\bibinfo
  {volume} {20}},\ \bibinfo {pages} {318} (\bibinfo {year} {2018})}\BibitemShut
  {NoStop}%
\bibitem [{\citenamefont {Br{\"u}nger}, \citenamefont {Brooks~III},\ and\
  \citenamefont {Karplus}(1984)}]{Brunger:1984}%
  \BibitemOpen
  \bibfield  {author} {\bibinfo {author} {\bibfnamefont {A.}~\bibnamefont
  {Br{\"u}nger}}, \bibinfo {author} {\bibfnamefont {C.~L.}\ \bibnamefont
  {Brooks~III}},\ and\ \bibinfo {author} {\bibfnamefont {M.}~\bibnamefont
  {Karplus}},\ }\bibfield  {title} {\enquote {\bibinfo {title} {Stochastic
  boundary conditions for molecular dynamics simulations of st2 water},}\
  }\href@noop {} {\bibfield  {journal} {\bibinfo  {journal} {Chem. Phys.
  Lett.}\ }\textbf {\bibinfo {volume} {105}},\ \bibinfo {pages} {495--500}
  (\bibinfo {year} {1984})}\BibitemShut {NoStop}%
\bibitem [{\citenamefont {Van~Gunsteren}\ and\ \citenamefont
  {Berendsen}(1988)}]{vanGunsteren:1988}%
  \BibitemOpen
  \bibfield  {author} {\bibinfo {author} {\bibfnamefont {W.~F.}\ \bibnamefont
  {Van~Gunsteren}}\ and\ \bibinfo {author} {\bibfnamefont {H.~J.}\ \bibnamefont
  {Berendsen}},\ }\bibfield  {title} {\enquote {\bibinfo {title} {A leap-frog
  algorithm for stochastic dynamics},}\ }\href@noop {} {\bibfield  {journal}
  {\bibinfo  {journal} {Molecular Simulation}\ }\textbf {\bibinfo {volume}
  {1}},\ \bibinfo {pages} {173--185} (\bibinfo {year} {1988})}\BibitemShut
  {NoStop}%
\bibitem [{\citenamefont {Pastor}, \citenamefont {Brooks},\ and\ \citenamefont
  {Szabo}(1988)}]{Pastor:1988}%
  \BibitemOpen
  \bibfield  {author} {\bibinfo {author} {\bibfnamefont {R.~W.}\ \bibnamefont
  {Pastor}}, \bibinfo {author} {\bibfnamefont {B.~R.}\ \bibnamefont {Brooks}},\
  and\ \bibinfo {author} {\bibfnamefont {A.}~\bibnamefont {Szabo}},\ }\bibfield
   {title} {\enquote {\bibinfo {title} {An analysis of the accuracy of langevin
  and molecular dynamics algorithms},}\ }\href@noop {} {\bibfield  {journal}
  {\bibinfo  {journal} {Molecular Physics}\ }\textbf {\bibinfo {volume} {65}},\
  \bibinfo {pages} {1409--1419} (\bibinfo {year} {1988})}\BibitemShut {NoStop}%
\bibitem [{\citenamefont {Hershkovitz}(1998)}]{Hershkovitz:1998}%
  \BibitemOpen
  \bibfield  {author} {\bibinfo {author} {\bibfnamefont {E.}~\bibnamefont
  {Hershkovitz}},\ }\bibfield  {title} {\enquote {\bibinfo {title} {A
  fourth-order numerical integrator for stochastic langevin equations},}\
  }\href@noop {} {\bibfield  {journal} {\bibinfo  {journal} {J. Chem. Phys.}\
  }\textbf {\bibinfo {volume} {108}},\ \bibinfo {pages} {9253} (\bibinfo {year}
  {1998})}\BibitemShut {NoStop}%
\bibitem [{\citenamefont {Paterlini}\ and\ \citenamefont
  {Ferguson}(1998)}]{Paterlini:1998}%
  \BibitemOpen
  \bibfield  {author} {\bibinfo {author} {\bibfnamefont {M.~G.}\ \bibnamefont
  {Paterlini}}\ and\ \bibinfo {author} {\bibfnamefont {D.~M.}\ \bibnamefont
  {Ferguson}},\ }\bibfield  {title} {\enquote {\bibinfo {title} {Constant
  temperature simulations using the langevin equation with velocity verlet
  integration},}\ }\href@noop {} {\bibfield  {journal} {\bibinfo  {journal}
  {Chem. Phys.}\ }\textbf {\bibinfo {volume} {236}},\ \bibinfo {pages} {243}
  (\bibinfo {year} {1998})}\BibitemShut {NoStop}%
\bibitem [{\citenamefont {Skeel}\ and\ \citenamefont
  {Izaguirre}(2002)}]{Skeel:2002}%
  \BibitemOpen
  \bibfield  {author} {\bibinfo {author} {\bibfnamefont {R.~D.}\ \bibnamefont
  {Skeel}}\ and\ \bibinfo {author} {\bibfnamefont {J.~A.}\ \bibnamefont
  {Izaguirre}},\ }\bibfield  {title} {\enquote {\bibinfo {title} {An impulse
  integrator for langevin dynamics},}\ }\href@noop {} {\bibfield  {journal}
  {\bibinfo  {journal} {Mol. Phys.}\ }\textbf {\bibinfo {volume} {100}},\
  \bibinfo {pages} {3885} (\bibinfo {year} {2002})}\BibitemShut {NoStop}%
\bibitem [{\citenamefont {Ricci}\ and\ \citenamefont
  {Ciccotti}(2003)}]{Ricci:2003}%
  \BibitemOpen
  \bibfield  {author} {\bibinfo {author} {\bibfnamefont {A.}~\bibnamefont
  {Ricci}}\ and\ \bibinfo {author} {\bibfnamefont {G.}~\bibnamefont
  {Ciccotti}},\ }\bibfield  {title} {\enquote {\bibinfo {title} {Algorithms for
  brownian dynamics},}\ }\href@noop {} {\bibfield  {journal} {\bibinfo
  {journal} {Molecular Physics}\ }\textbf {\bibinfo {volume} {101}},\ \bibinfo
  {pages} {1927--1931} (\bibinfo {year} {2003})}\BibitemShut {NoStop}%
\bibitem [{\citenamefont {Vanden-Eijnden}\ and\ \citenamefont
  {Ciccotti}(2006)}]{VandenEijnden:2006}%
  \BibitemOpen
  \bibfield  {author} {\bibinfo {author} {\bibfnamefont {E.}~\bibnamefont
  {Vanden-Eijnden}}\ and\ \bibinfo {author} {\bibfnamefont {G.}~\bibnamefont
  {Ciccotti}},\ }\bibfield  {title} {\enquote {\bibinfo {title} {Second-order
  integrators for langevin equations with holonomic constraints},}\ }\href@noop
  {} {\bibfield  {journal} {\bibinfo  {journal} {Chem. Phys. Lett.}\ }\textbf
  {\bibinfo {volume} {429}},\ \bibinfo {pages} {310} (\bibinfo {year}
  {2006})}\BibitemShut {NoStop}%
\bibitem [{\citenamefont {Melchionna}(2007)}]{Melchionna:2007}%
  \BibitemOpen
  \bibfield  {author} {\bibinfo {author} {\bibfnamefont {S.}~\bibnamefont
  {Melchionna}},\ }\bibfield  {title} {\enquote {\bibinfo {title} {Design of
  quasisymplectic propagators for langevin dynamics},}\ }\href@noop {}
  {\bibfield  {journal} {\bibinfo  {journal} {J. Chem. Phys.}\ }\textbf
  {\bibinfo {volume} {127}},\ \bibinfo {pages} {044108} (\bibinfo {year}
  {2007})}\BibitemShut {NoStop}%
\bibitem [{\citenamefont {Izaguirre}, \citenamefont {Sweet},\ and\
  \citenamefont {Pande}(2010)}]{Izaguirre:2010}%
  \BibitemOpen
  \bibfield  {author} {\bibinfo {author} {\bibfnamefont {J.~A.}\ \bibnamefont
  {Izaguirre}}, \bibinfo {author} {\bibfnamefont {C.~R.}\ \bibnamefont
  {Sweet}},\ and\ \bibinfo {author} {\bibfnamefont {V.~S.}\ \bibnamefont
  {Pande}},\ }\bibfield  {title} {\enquote {\bibinfo {title} {{Multiscale
  dynamics of macromolecules using normal mode Langevin.}}}\ }\href@noop {}
  {\bibfield  {journal} {\bibinfo  {journal} {Pacific Symp. Biocomput.}\ ,\
  \bibinfo {pages} {240--251}} (\bibinfo {year} {2010})}\BibitemShut {NoStop}%
\bibitem [{\citenamefont {Jia}\ and\ \citenamefont {Li}(2011)}]{Hongen:2011}%
  \BibitemOpen
  \bibfield  {author} {\bibinfo {author} {\bibfnamefont {H.}~\bibnamefont
  {Jia}}\ and\ \bibinfo {author} {\bibfnamefont {K.}~\bibnamefont {Li}},\
  }\bibfield  {title} {\enquote {\bibinfo {title} {A third accurate operator
  splitting method},}\ }\href@noop {} {\bibfield  {journal} {\bibinfo
  {journal} {Math Comput Model.}\ }\textbf {\bibinfo {volume} {53}},\ \bibinfo
  {pages} {387--396} (\bibinfo {year} {2011})}\BibitemShut {NoStop}%
\bibitem [{\citenamefont {Gr{\o}nbech-Jensen}\ and\ \citenamefont
  {Farago}(2013)}]{GronbechJensen:2013}%
  \BibitemOpen
  \bibfield  {author} {\bibinfo {author} {\bibfnamefont {N.}~\bibnamefont
  {Gr{\o}nbech-Jensen}}\ and\ \bibinfo {author} {\bibfnamefont
  {O.}~\bibnamefont {Farago}},\ }\bibfield  {title} {\enquote {\bibinfo {title}
  {A simple and effective verlet-type algorithm for simulating langevin
  dynamics},}\ }\href@noop {} {\bibfield  {journal} {\bibinfo  {journal} {Mol.
  Phys.}\ }\textbf {\bibinfo {volume} {111}},\ \bibinfo {pages} {983} (\bibinfo
  {year} {2013})}\BibitemShut {NoStop}%
\bibitem [{\citenamefont {Peters}, \citenamefont {Goga},\ and\ \citenamefont
  {Berendsen}(2014)}]{Peters:2014}%
  \BibitemOpen
  \bibfield  {author} {\bibinfo {author} {\bibfnamefont {E.~J.~F.}\
  \bibnamefont {Peters}}, \bibinfo {author} {\bibfnamefont {N.}~\bibnamefont
  {Goga}},\ and\ \bibinfo {author} {\bibfnamefont {H.~J.}\ \bibnamefont
  {Berendsen}},\ }\bibfield  {title} {\enquote {\bibinfo {title} {Stochastic
  dynamics with correct sampling for constrained systems},}\ }\href@noop {}
  {\bibfield  {journal} {\bibinfo  {journal} {Journal of chemical theory and
  computation}\ }\textbf {\bibinfo {volume} {10}},\ \bibinfo {pages}
  {4208--4220} (\bibinfo {year} {2014})}\BibitemShut {NoStop}%
\bibitem [{\citenamefont {Zhang}\ \emph {et~al.}(2019)\citenamefont {Zhang},
  \citenamefont {Liu}, \citenamefont {Yan}, \citenamefont {Tuckerman},\ and\
  \citenamefont {Liu}}]{Zhang:2019}%
  \BibitemOpen
  \bibfield  {author} {\bibinfo {author} {\bibfnamefont {Z.}~\bibnamefont
  {Zhang}}, \bibinfo {author} {\bibfnamefont {X.}~\bibnamefont {Liu}}, \bibinfo
  {author} {\bibfnamefont {K.}~\bibnamefont {Yan}}, \bibinfo {author}
  {\bibfnamefont {M.~E.}\ \bibnamefont {Tuckerman}},\ and\ \bibinfo {author}
  {\bibfnamefont {J.}~\bibnamefont {Liu}},\ }\bibfield  {title} {\enquote
  {\bibinfo {title} {Unified efficient thermostat scheme for the canonical
  ensemble with holonomic or isokinetic constraints via molecular dynamics},}\
  }\href@noop {} {\bibfield  {journal} {\bibinfo  {journal} {J. Phys. Chem. A}\
  }\textbf {\bibinfo {volume} {123}},\ \bibinfo {pages} {6056} (\bibinfo {year}
  {2019})}\BibitemShut {NoStop}%
\bibitem [{\citenamefont {Gr{\o}nbech-Jensen}(2020)}]{Gronbech:2020}%
  \BibitemOpen
  \bibfield  {author} {\bibinfo {author} {\bibfnamefont {N.}~\bibnamefont
  {Gr{\o}nbech-Jensen}},\ }\bibfield  {title} {\enquote {\bibinfo {title}
  {Complete set of stochastic verlet-type thermostats for correct langevin
  simulations},}\ }\href@noop {} {\bibfield  {journal} {\bibinfo  {journal}
  {Molecular Physics}\ }\textbf {\bibinfo {volume} {118}},\ \bibinfo {pages}
  {e1662506} (\bibinfo {year} {2020})}\BibitemShut {NoStop}%
\bibitem [{\citenamefont {Finkelstein}\ \emph {et~al.}(2021)\citenamefont
  {Finkelstein}, \citenamefont {Cheng}, \citenamefont {Fiorin}, \citenamefont
  {Seibold},\ and\ \citenamefont {Gr\o{}nbech-Jensen}}]{Finkelstein:2021}%
  \BibitemOpen
  \bibfield  {author} {\bibinfo {author} {\bibfnamefont {J.}~\bibnamefont
  {Finkelstein}}, \bibinfo {author} {\bibfnamefont {C.}~\bibnamefont {Cheng}},
  \bibinfo {author} {\bibfnamefont {G.}~\bibnamefont {Fiorin}}, \bibinfo
  {author} {\bibfnamefont {B.}~\bibnamefont {Seibold}},\ and\ \bibinfo {author}
  {\bibfnamefont {N.}~\bibnamefont {Gr\o{}nbech-Jensen}},\ }\bibfield  {title}
  {\enquote {\bibinfo {title} {Bringing discrete-time langevin splitting
  methods into agreement with thermodynamics},}\ }\href@noop {} {\bibfield
  {journal} {\bibinfo  {journal} {J. Chem. Phys.}\ }\textbf {\bibinfo {volume}
  {155}},\ \bibinfo {pages} {184104} (\bibinfo {year} {2021})}\BibitemShut
  {NoStop}%
\bibitem [{\citenamefont {Beutler}\ \emph {et~al.}(1994)\citenamefont
  {Beutler}, \citenamefont {Mark}, \citenamefont {van Schaik}, \citenamefont
  {Gerber},\ and\ \citenamefont {Van~Gunsteren}}]{Beutler:1994}%
  \BibitemOpen
  \bibfield  {author} {\bibinfo {author} {\bibfnamefont {T.~C.}\ \bibnamefont
  {Beutler}}, \bibinfo {author} {\bibfnamefont {A.~E.}\ \bibnamefont {Mark}},
  \bibinfo {author} {\bibfnamefont {R.~C.}\ \bibnamefont {van Schaik}},
  \bibinfo {author} {\bibfnamefont {P.~R.}\ \bibnamefont {Gerber}},\ and\
  \bibinfo {author} {\bibfnamefont {W.~F.}\ \bibnamefont {Van~Gunsteren}},\
  }\bibfield  {title} {\enquote {\bibinfo {title} {Avoiding singularities and
  numerical instabilities in free energy calculations based on molecular
  simulations},}\ }\href@noop {} {\bibfield  {journal} {\bibinfo  {journal}
  {Chemical physics letters}\ }\textbf {\bibinfo {volume} {222}},\ \bibinfo
  {pages} {529--539} (\bibinfo {year} {1994})}\BibitemShut {NoStop}%
\bibitem [{\citenamefont {Zacharias}, \citenamefont {Straatsma},\ and\
  \citenamefont {McCammon}(1994)}]{Zacharias:1994}%
  \BibitemOpen
  \bibfield  {author} {\bibinfo {author} {\bibfnamefont {M.}~\bibnamefont
  {Zacharias}}, \bibinfo {author} {\bibfnamefont {T.}~\bibnamefont
  {Straatsma}},\ and\ \bibinfo {author} {\bibfnamefont {J.}~\bibnamefont
  {McCammon}},\ }\bibfield  {title} {\enquote {\bibinfo {title}
  {Separation-shifted scaling, a new scaling method for lennard-jones
  interactions in thermodynamic integration},}\ }\href@noop {} {\bibfield
  {journal} {\bibinfo  {journal} {The Journal of chemical physics}\ }\textbf
  {\bibinfo {volume} {100}},\ \bibinfo {pages} {9025--9031} (\bibinfo {year}
  {1994})}\BibitemShut {NoStop}%
\bibitem [{\citenamefont {Donati}, \citenamefont {Weber},\ and\ \citenamefont
  {Keller}(2022)}]{Donati:2022}%
  \BibitemOpen
  \bibfield  {author} {\bibinfo {author} {\bibfnamefont {L.}~\bibnamefont
  {Donati}}, \bibinfo {author} {\bibfnamefont {M.}~\bibnamefont {Weber}},\ and\
  \bibinfo {author} {\bibfnamefont {B.~G.}\ \bibnamefont {Keller}},\ }\bibfield
   {title} {\enquote {\bibinfo {title} {A review of girsanov reweighting and of
  square root approximation for building molecular markov state models},}\
  }\href@noop {} {\bibfield  {journal} {\bibinfo  {journal} {Journal of
  Mathematical Physics}\ }\textbf {\bibinfo {volume} {63}},\ \bibinfo {pages}
  {123306} (\bibinfo {year} {2022})}\BibitemShut {NoStop}%
\bibitem [{\citenamefont {Trotter}(1959)}]{Trotter:1959}%
  \BibitemOpen
  \bibfield  {author} {\bibinfo {author} {\bibfnamefont {H.~F.}\ \bibnamefont
  {Trotter}},\ }\bibfield  {title} {\enquote {\bibinfo {title} {On the product
  of semi-groups of operators},}\ }\href@noop {} {\bibfield  {journal}
  {\bibinfo  {journal} {Proc Am Math Soc}\ }\textbf {\bibinfo {volume} {10}},\
  \bibinfo {pages} {545--551} (\bibinfo {year} {1959})}\BibitemShut {NoStop}%
\bibitem [{\citenamefont {Tuckerman}, \citenamefont {Berne},\ and\
  \citenamefont {Martyna}(1992)}]{Tuckerman:1992}%
  \BibitemOpen
  \bibfield  {author} {\bibinfo {author} {\bibfnamefont {M.}~\bibnamefont
  {Tuckerman}}, \bibinfo {author} {\bibfnamefont {B.~J.}\ \bibnamefont
  {Berne}},\ and\ \bibinfo {author} {\bibfnamefont {G.~J.}\ \bibnamefont
  {Martyna}},\ }\bibfield  {title} {\enquote {\bibinfo {title} {{Reversible
  multiple time scale molecular dynamics}},}\ }\href@noop {} {\bibfield
  {journal} {\bibinfo  {journal} {J. Chem. Phys.}\ }\textbf {\bibinfo {volume}
  {97}},\ \bibinfo {pages} {1990--2001} (\bibinfo {year} {1992})}\BibitemShut
  {NoStop}%
\bibitem [{\citenamefont {Leimkuhler}\ and\ \citenamefont
  {Matthews}(2016)}]{Leimkuhler:2016}%
  \BibitemOpen
  \bibfield  {author} {\bibinfo {author} {\bibfnamefont {B.}~\bibnamefont
  {Leimkuhler}}\ and\ \bibinfo {author} {\bibfnamefont {C.}~\bibnamefont
  {Matthews}},\ }\bibfield  {title} {\enquote {\bibinfo {title} {Efficient
  molecular dynamics using geodesic integration and solvent--solute
  splitting},}\ }\href@noop {} {\bibfield  {journal} {\bibinfo  {journal}
  {Proc. R. Soc. A: Math. Phys. Eng. Sci.}\ }\textbf {\bibinfo {volume}
  {472}},\ \bibinfo {pages} {20160138} (\bibinfo {year} {2016})}\BibitemShut
  {NoStop}%
\bibitem [{\citenamefont {Arens}\ \emph {et~al.}(2018)\citenamefont {Arens},
  \citenamefont {Hettlich}, \citenamefont {Karpfinger}, \citenamefont
  {Kockelkorn}, \citenamefont {Lichtenegger},\ and\ \citenamefont
  {Stachel}}]{Arens2018}%
  \BibitemOpen
  \bibfield  {author} {\bibinfo {author} {\bibfnamefont {T.}~\bibnamefont
  {Arens}}, \bibinfo {author} {\bibfnamefont {F.}~\bibnamefont {Hettlich}},
  \bibinfo {author} {\bibfnamefont {C.}~\bibnamefont {Karpfinger}}, \bibinfo
  {author} {\bibfnamefont {U.}~\bibnamefont {Kockelkorn}}, \bibinfo {author}
  {\bibfnamefont {K.}~\bibnamefont {Lichtenegger}},\ and\ \bibinfo {author}
  {\bibfnamefont {H.}~\bibnamefont {Stachel}},\ }\bibfield  {title} {\enquote
  {\bibinfo {title} {Kurven und fl{\"a}chen--von kr{\"u}mmung, torsion und
  l{\"a}ngenmessung},}\ }in\ \href@noop {} {\emph {\bibinfo {booktitle}
  {Mathematik}}}\ (\bibinfo  {publisher} {Springer},\ \bibinfo {year} {2018})\
  pp.\ \bibinfo {pages} {955--991}\BibitemShut {NoStop}%
\bibitem [{\citenamefont {Fujisaki}\ \emph {et~al.}(2013)\citenamefont
  {Fujisaki}, \citenamefont {Shiga}, \citenamefont {Moritsugu},\ and\
  \citenamefont {Kidera}}]{Fujisaki:2013}%
  \BibitemOpen
  \bibfield  {author} {\bibinfo {author} {\bibfnamefont {H.}~\bibnamefont
  {Fujisaki}}, \bibinfo {author} {\bibfnamefont {M.}~\bibnamefont {Shiga}},
  \bibinfo {author} {\bibfnamefont {K.}~\bibnamefont {Moritsugu}},\ and\
  \bibinfo {author} {\bibfnamefont {A.}~\bibnamefont {Kidera}},\ }\bibfield
  {title} {\enquote {\bibinfo {title} {Multiscale enhanced path sampling based
  on the onsager-machlup action: Application to a model polymer},}\ }\href@noop
  {} {\bibfield  {journal} {\bibinfo  {journal} {J. Chem. Phys.}\ }\textbf
  {\bibinfo {volume} {139}},\ \bibinfo {pages} {08B607\_1} (\bibinfo {year}
  {2013})}\BibitemShut {NoStop}%
\bibitem [{\citenamefont {Lee}\ \emph {et~al.}(2017)\citenamefont {Lee},
  \citenamefont {Lee}, \citenamefont {Joung}, \citenamefont {Lee},\ and\
  \citenamefont {Brooks}}]{Lee:2017}%
  \BibitemOpen
  \bibfield  {author} {\bibinfo {author} {\bibfnamefont {J.}~\bibnamefont
  {Lee}}, \bibinfo {author} {\bibfnamefont {I.-H.}\ \bibnamefont {Lee}},
  \bibinfo {author} {\bibfnamefont {I.}~\bibnamefont {Joung}}, \bibinfo
  {author} {\bibfnamefont {J.}~\bibnamefont {Lee}},\ and\ \bibinfo {author}
  {\bibfnamefont {B.~R.}\ \bibnamefont {Brooks}},\ }\bibfield  {title}
  {\enquote {\bibinfo {title} {Finding multiple reaction pathways via global
  optimization of action},}\ }\href@noop {} {\bibfield  {journal} {\bibinfo
  {journal} {Nature Communications}\ }\textbf {\bibinfo {volume} {8}},\
  \bibinfo {pages} {15443} (\bibinfo {year} {2017})}\BibitemShut {NoStop}%
\bibitem [{\citenamefont {Peter}, \citenamefont {Shea},\ and\ \citenamefont
  {Schug}(2020)}]{Peter:2020}%
  \BibitemOpen
  \bibfield  {author} {\bibinfo {author} {\bibfnamefont {E.~K.}\ \bibnamefont
  {Peter}}, \bibinfo {author} {\bibfnamefont {J.-E.}\ \bibnamefont {Shea}},\
  and\ \bibinfo {author} {\bibfnamefont {A.}~\bibnamefont {Schug}},\ }\bibfield
   {title} {\enquote {\bibinfo {title} {Core-md, a path correlated molecular
  dynamics simulation method},}\ }\href@noop {} {\bibfield  {journal} {\bibinfo
   {journal} {The Journal of Chemical Physics}\ }\textbf {\bibinfo {volume}
  {153}},\ \bibinfo {pages} {084114} (\bibinfo {year} {2020})}\BibitemShut
  {NoStop}%
\bibitem [{\citenamefont {Bolhuis}\ \emph {et~al.}(2002)\citenamefont
  {Bolhuis}, \citenamefont {Chandler}, \citenamefont {Dellago},\ and\
  \citenamefont {Geissler}}]{Bolhuis:2002}%
  \BibitemOpen
  \bibfield  {author} {\bibinfo {author} {\bibfnamefont {P.~G.}\ \bibnamefont
  {Bolhuis}}, \bibinfo {author} {\bibfnamefont {D.}~\bibnamefont {Chandler}},
  \bibinfo {author} {\bibfnamefont {C.}~\bibnamefont {Dellago}},\ and\ \bibinfo
  {author} {\bibfnamefont {P.~L.}\ \bibnamefont {Geissler}},\ }\bibfield
  {title} {\enquote {\bibinfo {title} {Transition path sampling: Throwing
  ropes},}\ }\href@noop {} {\bibfield  {journal} {\bibinfo  {journal} {Annu.
  Rev. Phys. Chem}\ }\textbf {\bibinfo {volume} {53}},\ \bibinfo {pages}
  {291--318} (\bibinfo {year} {2002})}\BibitemShut {NoStop}%
\end{thebibliography}%

%
%
\appendix

%
%
\section{Langevin integration methods}
\label{app:Langevin_integration_methods}
%
For the update operators we use the following abbreviations:
\begin{subequations}
\begin{eqnarray}
    a       &=& \Delta t \frac{1}{m} \label{eq:a}\\
    b(q_k)  &=& -\Delta t\nabla V(q_k) \label{eq:b}\\
    d       &=& e^{-\xi\Delta t}  \label{eq:d}\\
    f       &=& \sqrt{k_BTm(1-e^{-2\xi\Delta t})} \, . \label{eq:f}
\end{eqnarray}
\end{subequations}
and for half time steps
\begin{subequations}
\begin{eqnarray}
    a'       &=& \frac{\Delta t}{2} \frac{1}{m} \label{eq:a_prime}\\
    b'(q_k)  &=& -\frac{\Delta t}{2}\nabla V(q_k) \label{eq:b_prime}\\
    d'       &=& e^{-\xi\frac{\Delta t}{2}}  \label{eq:d_prime}\\
    f'       &=& \sqrt{k_BTm(1-e^{-2\xi\frac{\Delta t}{2}})} \, . \label{eq:f_prime}
\end{eqnarray}
\end{subequations}
We will further use that
\begin{eqnarray}
   \widetilde{\eta_k} - \eta_k &=& \Delta \eta_k 
\label{eq:DeltaEta_k}    
\end{eqnarray}
and that 
\begin{eqnarray}
    \widetilde{b}(q_k) - b(q_k) &=& -\Delta t\nabla \widetilde{V}(q_k) + \Delta t\nabla V(q_k) \cr
    &=& -\Delta t(\nabla V(q_k) + \nabla U(q_k)) + \Delta t\nabla V(q_k) \cr
    &=&-\Delta t \nabla U(q_k)\, 
\label{eq:DeltaGradU}       
\end{eqnarray}
and analogously for half-time steps.

\subsection{ABOBA}
\subsubsection{Algorithm}
\begin{subequations}
\begin{eqnarray}
    q_{k+1/2}   &=& q_k + \frac{\Delta t}{2 m} p_k    \label{eq:ABOBA01}\\
    p_{k+1/3}   &=& p_k - \frac{\Delta t}{2} \nabla V(q_{k+1/2})    \label{eq:ABOBA02}\\
    p_{k+2/3}   &=& e^{-\xi\Delta t} p_{k+1/3} + \sqrt{k_BTm(1-e^{-2\xi\Delta t})}\,\eta_{k} \label{eq:ABOBA03}\\
    p_{k+1}     &=& p_{k+2/3} - \frac{\Delta t}{2} \,\nabla V(q_{k+1/2})  \label{eq:ABOBA04}\\
    q_{k+1}     &=& q_{k+1/2} + \frac{\Delta t}{2 m} \,p_{k+1} \label{eq:ABOBA05}
\end{eqnarray}
\end{subequations}
The algorithm has been reported in Refs.~\onlinecite{Leimkuhler:2012} and in \onlinecite{Sivak:2014}.
Compared to Ref.~\onlinecite{Leimkuhler:2013}, we changed the notation as follows:
$n \rightarrow k$,
$R_n \rightarrow \eta_k$,
$\delta t \rightarrow \Delta t$, 
$M \rightarrow m$, 
$\gamma \rightarrow \xi$, 
$F \rightarrow -\nabla V$.

\subsubsection{Update operator}
\begin{eqnarray}
        \mathcal{A'B'OB'A'} \left(\begin{array}{l} q_{k}\\p_{k}\end{array}\right)
    &=& \mathcal{A'B'OB'} \left(\begin{array}{l} q_{k} + a'p_k\\p_{k}\end{array}\right) \cr
    &=& \mathcal{A'B'O} \left(\begin{array}{l} q_{k} + a'p_k\\p_{k}  + b'(q_{k} + a'p_k)\end{array}\right) \cr
    &=& \mathcal{A'B'} \left(\begin{array}{l} q_{k} + a'p_k\\dp_{k} + db'(q_{k} + a'p_k) + f\eta_k \end{array}\right) \cr
    &=& \mathcal{A'} \left(\begin{array}{l} q_{k} + a'p_k\\dp_{k} + db'(q_{k} + a'p_k) + f\eta_k + b'(q_{k} + a'p_k)\end{array}\right) \cr
    &=& \left(\begin{array}{l} q_{k} + a'p_k + a'[dp_{k} + db'(q_{k} + a'p_k) + f\eta_k + b'(q_{k} + a'p_k)]\\dp_{k} + db'(q_{k} + a'p_k) + f\eta_k + b'(q_{k} + a'p_k)\end{array}\right)
\end{eqnarray}

\subsubsection{Update function}
\begin{eqnarray}
     \left(\begin{array}{c} q_{k+1}\\p_{k+1}\end{array}\right) 
    =   \mathcal{U}_{\mathrm{ABOBA}}(\eta_k; x_k ,V)
    &=& \left(\begin{array}{l} \bar{q}_{k+1}\\\bar{p}_{k+1} \end{array}\right) +
        \left(\begin{array}{l} a'f\ \\ f\end{array}\right) \eta_k
\end{eqnarray}
with 
$\bar{q}_{k+1} = q_{k} + a'(1 + d)p_{k} + a'(d+1)b'(q_{k} + a'p_k)$
and
$\bar{p}_{k+1} = dp_{k} + (d+1)b'(q_{k} + a'p_k)$.
Thus, 
\begin{eqnarray}
    \mathcal{U}_{\mathrm{ABOBA}}:\;\; \mathbb{R} &\rightarrow& L_{1d} \subset \Gamma \cr
    \mathcal{U}_{\mathrm{ABOBA}}: \; \eta_k &\mapsto& x_{k+1} \, ,
\end{eqnarray}
where $L_{1d}$ denotes a line in $\Gamma$.

\subsubsection{Image at $V$ and $\widetilde{V}$}
\begin{eqnarray}
    \left(\begin{array}{c} 0\\0\end{array}\right) 
    &=& \mathcal{U}_{\mathrm{ABOBA}}(\eta_k; x_k ,V) - \mathcal{U}_{\mathrm{ABOBA}}(\widetilde{\eta_k}; x_k ,\widetilde{V}) \cr
    &=& \left(\begin{array}{l} q_{k} + a'p_k + a'(dp_{k} + db'(q_{k} + a'p_k) + f\eta_k + b'(q_{k} + a'p_k))\\dp_{k} + db'(q_{k} + a'p_k) + f\eta_k + b'(q_{k} + a'p_k)\end{array}\right) - \cr
    && \left(\begin{array}{l} q_{k} + a'p_k + a'(dp_{k} + d\widetilde{b}'(q_{k} + a'p_k) + f\widetilde{\eta}_k + \widetilde{b}'(q_{k} + a'p_k))\\dp_{k} + d\widetilde{b}'(q_{k} + a'p_k) + f\widetilde{\eta}_k + \widetilde{b}'(q_{k} + a'p_k)\end{array}\right) \cr
    &=& \left(\begin{array}{l} a'[d\cdot b'(q_{k} + a'p_k) + f\eta_k + b'(q_{k} + a'p_k)]\\db'(q_{k} + a'p_k) + f\eta_k + b'(q_{k} + a'p_k)\end{array}\right) - \cr
    && \left(\begin{array}{l} a'[d\cdot\widetilde{b}'(q_{k} + a'p_k) + f\widetilde{\eta}_k + \widetilde{b}'(q_{k} + a'p_k)]\\d\widetilde{b}'(q_{k} + a'p_k) + f\widetilde{\eta}_k + \widetilde{b}'(q_{k} + a'p_k)\end{array}\right) \cr
    &=& \left(\begin{array}{l} a'\\1\end{array}\right) [d\cdot b'(q_{k} + a'p_k) + f\eta_k + b'(q_{k} + a'p_k)]- \cr
    && \left(\begin{array}{l} a'\\1\end{array}\right) [d\cdot\widetilde{b}'(q_{k} + a'p_k) + f\widetilde{\eta}_k + \widetilde{b}'(q_{k} + a'p_k)] \cr
    &=& \left(\begin{array}{l} a'\\1\end{array}\right) [d\cdot b'(q_{k} + a'p_k) + f\eta_k + b'(q_{k} + a'p_k)- d\cdot\widetilde{b}'(q_{k} + a'p_k) - f\widetilde{\eta}_k - \widetilde{b}'(q_{k} + a'p_k)] \cr
    &=&\left(\begin{array}{l} a'\\1\end{array}\right) [ (d+1)\cdot b'(q_{k} + a'p_k) - (d+1)\cdot \widetilde{b}'(q_{k} + a'p_k) + f(\eta_k-\widetilde{\eta}_k) ] \cr    
    &=&\left(\begin{array}{l} a'\\1\end{array}\right) [ (d+1)\cdot \frac{\Delta t}{2} \nabla U(q_{k+1/2}) - f\Delta \eta_k ]   
\end{eqnarray}
is fulfilled if
\begin{eqnarray}
     \Delta \eta_k  &=& \frac{(d+1)}{f} \frac{\Delta t}{2} \nabla U(q_{k+1/2})\, ,
\end{eqnarray}
where we substituted $q_{k+1/2} = q_{k}+ a' p_k$ (eq.~\ref{eq:ABOBA01}) and used eqs.~\ref{eq:fDeltaEta_k} and eq.~\ref{eq:DeltaGradU}.
Thus, $\mathcal{U}_{\mathrm{ABOBA}}(\eta_k; x_k ,V)$ and $\mathcal{U}_{\mathrm{ABOBA}}(\widetilde{\eta_k}; x_k ,\widetilde{V})$ have the same image
$$  
    L_{1d} = \widetilde{L}_{1d} \, .
$$

\subsection{BAOAB}
\subsubsection{Algorithm}
\begin{subequations}
\begin{eqnarray}
    p_{k+1/3} &=& p_k - \frac{\Delta t}{2} \nabla V(q_k)        \label{eq:BAOAB01} \\
    q_{k+1/2} &=& q_k + \frac{\Delta t}{2 m} \, p_{k+1/3}   \label{eq:BAOAB02} \\
    p_{k+2/3} &=& e^{-\xi\Delta t}\, p_{k+1/3} +  \sqrt{k_BTm(1-e^{-2\xi\Delta t})}\, \eta_{k} \label{eq:BAOAB03} \\
    q_{k+1} &=& q_{k+1/2} +  \frac{\Delta t}{2m} \, p_{k+2/3} \label{eq:BAOAB04} \\
    p_{k+1} &=& p_{k+2/3}- \frac{\Delta t}{2} \nabla V(q_{k+1}) \label{eq:BAOAB05}
\end{eqnarray}
\end{subequations}
The algorithm has been reported in Refs.~\onlinecite{Leimkuhler:2012} and \onlinecite{Sivak:2014}.
Compared to Ref.~\onlinecite{Leimkuhler:2013}, we changed the notation as follows: 
$n \rightarrow k$,
$R_n \rightarrow \eta_k$,
$\delta t \rightarrow \Delta t$, 
$M \rightarrow m$, 
$\gamma \rightarrow \xi$, 
$F \rightarrow -\nabla V$.

\subsubsection{Update operator}
\begin{eqnarray}
        \mathcal{B'A'OA'B'} \left(\begin{array}{l} q_k\\p_k\end{array}\right)
    &=& \mathcal{B'A'OA'} \left(\begin{array}{l} q_k\\p_k + b'(q_k)\end{array}\right) \cr
    &=& \mathcal{B'A'O} \left(\begin{array}{l} q_k + a'p_k + a'b'(q_k)\\p_k + b'(q_k)\end{array}\right) \cr
    &=& \mathcal{B'A'} \left(\begin{array}{l} q_k + a'p_k + a'b'(q_k)\\dp_k + db'(q_k) + f\eta_k\end{array}\right) \cr
    &=& \mathcal{B'} \left(\begin{array}{l} q_k + a'p_k + a'b'(q_k) + a'dp_k + a'db'(q_k) + a'f\eta_k\\dp_k + db'(q_k) + f\eta_k\end{array}\right) \cr
    &=& \left(\begin{array}{l} q_k + a'p_k + a'b'(q_k) + a'dp_k + a'db'(q_k) + a'f\eta_k\\dp_k + db'(q_k) + f\eta_k + b'(q_k + a'p_k + a'b'(q_k) + a'dp_k + a'db'(q_k) + a'f\eta_k)\end{array}\right) 
\end{eqnarray}    

\subsubsection{Update function}
\begin{eqnarray}
     \left(\begin{array}{c} q_{k+1}\\p_{k+1}\end{array}\right) 
    =   \mathcal{U}_{\mathrm{BAOAB}}(\eta_k; x_k ,V)
    &=& \left(\begin{array}{l} \bar{q}_{k+1} \\ \bar{p}_{k+1}\end{array}  \right) +    
        \left(\begin{array}{l} 0\\b'(\bar{q}_{k+1} +  a'f\eta_k) \end{array}  \right)  + 
        \left(\begin{array}{l} a'f\\f\end{array}  \right) \eta_k 
\end{eqnarray}
with 
$\bar{q}_{k+1} = q_k + a'p_k + a'b'(q_k) + a'dp_k + a'db'(q_k) = q_k + a'(1+d)p_k + a'(1 + d)b'(q_k)$
and
$\bar{p}_{k+1} = dp_k + db'(q_k)$.
Thus, 
\begin{eqnarray}
    \mathcal{U}_{\mathrm{BAOAB}}:\;\; \mathbb{R} &\rightarrow& C_{1d} \subset \Gamma \cr
    \mathcal{U}_{\mathrm{BAOAB}}: \; \eta_k &\mapsto& x_{k+1} \, ,
\end{eqnarray}
where $C_{1d}$ denotes a curve in $\Gamma$.

\subsubsection{Image at $V$ and $\widetilde{V}$}
\begin{eqnarray}
    \left(\begin{array}{c} 0\\0\end{array}\right) 
    &=& \mathcal{U}_{\mathrm{BAOAB}}(\eta_k; x_k ,V) - \mathcal{U}_{\mathrm{BAOAB}}(\widetilde{\eta_k}; x_k ,\widetilde{V}) \cr
    &=& \left(\begin{array}{l} q_k + a'(1+d)p_k + a'(1 + d)b'(q_k) + a'f\eta_k\\dp_k + d\cdot b'(q_k) + f\eta_k + b'(q_k + a'(1+d)p_k + a'(1 + d)b'(q_k) + a'f\eta_k)\end{array}\right) - \cr
    && \left(\begin{array}{l} q_k + a'(1+d)p_k + a'(1 + d)\widetilde{b}'(q_k) + a'f\widetilde{\eta}_k\\dp_k + d\cdot\widetilde{b}'(q_k) + f\widetilde{\eta}_k + \widetilde{b}'(q_k + a'(1+d)p_k + a'(1 + d)\widetilde{b}'(q_k) + a'f\widetilde{\eta}_k)\end{array}\right)  \cr
    &=& \left(\begin{array}{l} a'(d+1)\cdot b'(q_k) + a'f\eta_k\\
        d\cdot b'(q_k) + b'(q_k + a'(d+1)p_k + a'(d+1)\cdot b'(q_k) + a'f\eta_k) + f\eta_k \end{array}\right) - \cr
    && \left(\begin{array}{l} a'(d+1)\cdot\widetilde{b}'(q_k) + a'f\widetilde{\eta}_k\\
        d\cdot\widetilde{b}'(q_k) + \widetilde{b}'(q_k + a'(d+1)p_k + a'(d+1)\cdot\widetilde{b}'(q_k)  + a'f\widetilde{\eta}_k)+ f\widetilde{\eta}_k\end{array}\right)  \, .
\end{eqnarray}    
$\eta_k$ appears linearly in the update of the positions. 
It appears as the argument of the function $b(q)$ in the update of the momenta which is typically a non-linear function (eq.~\ref{eq:b}). 
Therefore, one cannot find an analytical expression for $\Delta \eta_k$ that solves both lines of the equation.
One can however find an expression for $\Delta \eta_k^{\mathrm{pos}}$, for which the two update functions yield the same positions:
\begin{eqnarray}
    0               &=& a'(d+1)\cdot b'(q_k) + a'f\eta_k - a'(d+1)\cdot\widetilde{b}'(q_k) - a'f\widetilde{\eta}_k \cr
                    &=& a' (d+1)\cdot \frac{\Delta t}{2}\nabla U(q_k) - a' f \Delta \eta_k^{\mathrm{pos}} \cr
                    &\Updownarrow& \cr
    \Delta \eta_k^{\mathrm{pos}}   &=& \frac{d+1}{f} \frac{\Delta t}{2}\nabla U(q_k)  \, .
\end{eqnarray}
Then 
\begin{eqnarray}
    q_{k+1} &=& q_k + a'(d+1)p_k + a'(d+1)\cdot b'(q_k) + a'f\eta_k \cr
            &=& q_k + a'(d+1)p_k + a'(d+1)\cdot \widetilde{b}'(q_k) + a'f\widetilde{\eta}_k \, ,
\end{eqnarray}
and we can make the following substitution in the equation above
\begin{eqnarray}
    \left(\begin{array}{c} 0\\0\end{array}\right) 
    &=& \left(\begin{array}{l} 0  \\
        d\cdot b'(q_k) + b'(q_{k+1})  + f\eta_k - d\cdot \widetilde{b}'(q_k) - \widetilde{b}'(q_{k+1})  - f \widetilde{\eta}_k 
        \end{array}\right)\cr
    &=& \left(\begin{array}{l} 0  \\
        d \frac{\Delta t}{2}\nabla U(q_k) + \frac{\Delta t}{2}\nabla U(q_{k+1})  - f \Delta \eta_k^{\mathrm{pos}} 
        \end{array}\right) \, .
\end{eqnarray}  
One can now see that the equation for the momenta is not fulfilled, because
\begin{eqnarray}
    0 &\ne& d \frac{\Delta t}{2}\nabla U(q_k) + \frac{\Delta t}{2}\nabla U(q_{k+1})  - f \Delta \eta_k^{\mathrm{pos}}  \cr
    0 &\ne& d \frac{\Delta t}{2}\nabla U(q_k) + \frac{\Delta t}{2}\nabla U(q_{k+1})  - (d+1)\frac{\Delta t}{2}\nabla U(q_k)   \cr \cr
    \nabla U(q_{k+1}) &\ne&  \nabla U(q_k)   \, .
\end{eqnarray}
Thus, $\mathcal{U}_{\mathrm{BAOAB}}(\eta_k; x_k ,V)$ and $\mathcal{U}_{\mathrm{BAOAB}}(\widetilde{\eta_k}; x_k ,\widetilde{V})$ parameterise different curves in state space 
$$  
    C_{1d} \ne \widetilde{C}_{1d} \, .
$$

\subsection{BAOA/ Gromacs stochastic dynamics}
BAOA is equivalent to Gromacs stochastic dynamics (GSD)\cite{Goga:2012, Kieninger:2022}.
\subsubsection{Algorithm}
\begin{subequations}
\begin{eqnarray}
    p_{k+ \frac{1}{2}}           &=& p_{k}  -\Delta t \nabla V(q_k) \label{eq:BAOA01} \\
    q_{k+\frac{1}{2}} &=& q_k + \frac{\Delta t}{2m} p_{k+ \frac{1}{2}} \label{eq:BAOA02} \\
    p_{k + 1} &=& e^{-\xi \Delta t} p_{k+\frac{1}{2}}  + \sqrt{k_B Tm\left(1-e^{-2\xi \Delta t }\right)} \eta_k   \label{eq:BAOA03} \\
     q_{k+1} &=& q_{k+\frac{1}{2}} + \frac{\Delta t}{2m} p_{k+1}. \label{eq:BAOA04}
\end{eqnarray}
\end{subequations}

\subsubsection{Update operator}
\begin{eqnarray}
        \mathcal{A'OA'B} \left(\begin{array}{l} q_k\\p_k\end{array}\right)
    &=& \mathcal{A'OA'} \left(\begin{array}{l} q_k\\p_k + b(q_k)\end{array}\right) \cr
    &=& \mathcal{A'O} \left(\begin{array}{l} q_k + a'p_k + a'b(q_k) \\p_k + b(q_k)\end{array}\right) \cr
    &=& \mathcal{A'} \left(\begin{array}{l} q_k + a'p_k + a'b(q_k) \\dp_k + db(q_k) + f\eta_k\end{array}\right) \cr
    &=& \left(\begin{array}{l} q_k + a'p_k + a'b(q_k) + a'dp_k + a'db(q_k) + a'f\eta_k\\dp_k + db(q_k) + f\eta_k\end{array}\right) 
\end{eqnarray}

\subsubsection{Update function}
\begin{eqnarray}
     \left(\begin{array}{c} q_{k+1}\\p_{k+1}\end{array}\right) 
    =   \mathcal{U}_{\mathrm{BAOA}}(\eta_k; x_k ,V)
    &=& \left(\begin{array}{l} \bar{q}_{k+1}  \\ \bar{p}_{k+1}\end{array}\right) + 
        \left(\begin{array}{l} a'f\\ f\end{array}\right)\eta_k 
\end{eqnarray}
with 
$\bar{q}_{k+1} = q_k + a'p_k + a'b(q_k) + a'dp_k + a'db(q_k) = q_k + a'(1+d)p_k + a'(1+d)b(q_k)$ 
and 
$\bar{p}_{k+1} = dp_k + db(q_k)$.
Thus, 
\begin{eqnarray}
    \mathcal{U}_{\mathrm{BAOA}}:\;\; \mathbb{R} &\rightarrow& L_{1d} \subset \Gamma \cr
    \mathcal{U}_{\mathrm{BAOA}}: \; \eta_k &\mapsto& x_{k+1} \, ,
\end{eqnarray}
where $L_{1d}$ denotes a line in $\Gamma$.

\subsubsection{Image at $V$ and $\widetilde{V}$}
\begin{eqnarray}
    \left(\begin{array}{c} 0\\0\end{array}\right) 
    &=& \mathcal{U}_{\mathrm{BAOA}}(\eta_k; x_k ,V) - \mathcal{U}_{\mathrm{BAOA}}(\widetilde{\eta_k}; x_k ,\widetilde{V}) \cr
    &=& \left(\begin{array}{l} q_k + a'(1+d)p_k + a'(1+d)b(q_k) + a'f\eta_k\\dp_k + d\cdot b(q_k) + f\eta_k\end{array}\right) - \cr
    &&  \left(\begin{array}{l} q_k + a'(1+d)p_k + a'(1+d)\widetilde{b}(q_k) + a'f\widetilde{\eta}_k\\dp_k + d\cdot \widetilde{b}(q_k) + f\widetilde{\eta}_k\end{array}\right) \cr
    &=& \left(\begin{array}{l} a'(1+d)\cdot b(q_k)  + a'f\eta_k\\d\cdot b(q_k) + f\eta_k\end{array}\right) - 
        \left(\begin{array}{l} a'(1+d)\cdot \widetilde{b}(q_k) + a'f\widetilde{\eta}_k\\d\cdot \widetilde{b}(q_k) + f\widetilde{\eta}_k\end{array}\right) \cr
    &=& \left(\begin{array}{l} a'(1+d)\cdot \Delta t \nabla U(q_k)  - a'f\Delta \eta_k\\d\cdot \Delta t \nabla U(q_k) - f\Delta \eta_k\end{array}\right)
\label{eq:BAOA_equality}      
\end{eqnarray}    
The two update functions yield the same positions if
\begin{eqnarray}
    0   &=&  a'(1+d)\cdot \Delta t \nabla U(q_k)  - a'f\Delta \eta_k^\mathrm{pos} \cr
        &\Updownarrow& \cr
    \Delta \eta_k^{\mathrm{pos}} &=& \frac{1+d}{f}\cdot \Delta t \nabla U(q_k) \, .
\end{eqnarray}
Substituting $\eta_k^{\mathrm{pos}}$ for $\eta_k$ in eq.~\ref{eq:BAOA_equality} yields
\begin{eqnarray}
    \left(\begin{array}{c} 0\\0\end{array}\right) 
    &=&  \left(\begin{array}{c} 0\\d\cdot \Delta t \nabla U(q_k) - f\Delta \eta_k^{\mathrm{pos}}\end{array}\right) \cr
    &=&  \left(\begin{array}{c} 0\\d\cdot \Delta t \nabla U(q_k) - (1+d) \Delta t \nabla U(q_k)\end{array}\right) 
\end{eqnarray}
which shows that the equation for the momenta is not fulfilled with $\Delta \eta_k^{\mathrm{pos}}$, because
\begin{eqnarray}
    d\cdot \Delta t \nabla U(q_k) &\ne & (1+d) \Delta t \nabla U(q_k)\, .
\end{eqnarray}
Thus, $\mathcal{U}_{\mathrm{BAOA}}(\eta_k; x_k ,V)$ and $\mathcal{U}_{\mathrm{BAOA}}(\widetilde{\eta_k}; x_k ,\widetilde{V})$ parameterise different curves in state space
$$  
    L_{1d} \ne \widetilde{L}_{1d} \, .
$$

Because $\eta_k$ appears linearly in the update of the positions as well as in the update of the momenta, one can solve eq.~\ref{eq:BAOA_equality} for the momentum update
\begin{eqnarray}
    0   &=&  d\cdot \Delta t \nabla U(q_k) - f\Delta \eta_k^\mathrm{mom} \cr
        &\Updownarrow& \cr
    \Delta \eta_k^{\mathrm{mom}} &=& \frac{d}{f}\cdot \Delta t \nabla U(q_k) \, .
\end{eqnarray}
Note that $\Delta \eta_k^{\mathrm{mom}} \ne \Delta  \eta_k^{\mathrm{pos}}$.

%
%
\subsection{AOBOA}
\subsubsection{Algorithm}
\begin{subequations}
\begin{eqnarray}
    q_{k+1/2}   &=& q_k + \frac{\Delta t}{2 m} p_k    \label{eq:AOBOA01}\\
    p_{k+1/3}   &=& e^{-\frac{\xi \Delta t}{2}} p_{k} + \sqrt{k_BTm(1-e^{-\xi\Delta t})}\,\eta_{k}^{(1)}    \label{eq:AOBOA02}\\
    p_{k+2/3}   &=& p_{k+1/3}  - \Delta t \nabla V(q_{k+1/2})
    \label{eq:AOBOA03}\\
    p_{k+1}     &=& e^{-\frac{\xi \Delta t}{2}} p_{k+2/3} + \sqrt{k_BTm(1-e^{-\xi\Delta t})}\,\eta_{k}^{(2)} \label{eq:AOBOA04}\\
    q_{k+1}     &=& q_{k+1/2} + \frac{\Delta t}{2 m} \,p_{k+1} \, . \label{eq:AOBOA05}
\end{eqnarray}
\end{subequations}
Here, two random numbers $\eta_k^{(1)} \sim \mathcal{N}(0,1)$ and $\eta_k^{(2)} \sim \mathcal{N}(0,1)$ need to be drawn per full update cycle.
%

\subsubsection{Update operator}
\begin{eqnarray}
        \mathcal{A'O'BO'A'} \left(\begin{array}{l} q_{k}\\p_{k}\end{array}\right)
    &=& \mathcal{A'O'BO'} \left(\begin{array}{l} q_{k} + a' p_k\\p_{k}\end{array}\right)   \cr
    &=& \mathcal{A'O'B} \left(\begin{array}{l} q_{k} + a' p_k\\d'p_{k} + f'\eta_k^{(1)}\end{array}\right)   \cr
    &=& \mathcal{A'O'} \left(\begin{array}{l} q_{k} + a' p_k\\d'p_{k} + f'\eta_k^{(1)} + b(q_{k} + a' p_k)\end{array}\right)   \cr
    &=& \mathcal{A'} \left(\begin{array}{l} q_{k} + a' p_k\\d'd'p_{k} + d'f'\eta_k^{(1)} + d'b(q_{k} + a' p_k) +f'\eta_k^{(2)}\end{array}\right)   \cr
    &=& \left(\begin{array}{l} q_{k} + a' p_k + a'\Big(d'd'p_{k} + d'f'\eta_k^{(1)} + d'b(q_{k} + a' p_k) +f'\eta_k^{(2)}\Big)\\d'd'p_{k} + d'f'\eta_k^{(1)} + d'b(q_{k} + a' p_k) +f'\eta_k^{(2)}\end{array}\right) 
\end{eqnarray}        

\subsubsection{Update function}
\begin{eqnarray}
     \left(\begin{array}{c} q_{k+1}\\p_{k+1}\end{array}\right) 
    =   \mathcal{U}_{\mathrm{AOBOA}}(\eta_k^{\mathrm{comb}}; x_k ,V)
    &=& \left(\begin{array}{l} \bar{q}_{k+1}\\\bar{p}_{k+1} \end{array}\right) +
        \left(\begin{array}{l} a'\\ 1\end{array}\right) f'\eta^{\mathrm{comb}}_k
\end{eqnarray}
with
$\bar{q}_{k+1} = q_{k} + (a' + a'd'd')p_k  + a'd'b(q_{k} + a' p_k)$ and
$\bar{p}_{k+1} = d'd'p_{k} + d'b(q_{k} + a' p_k)$
and a combined random number
\begin{eqnarray}
    \eta^{\mathrm{comb}}_k &=& d'\eta_k^{(1)} + \eta_k^{(2)} \sim \mathcal{N}(0,d'^2+1) \, .
\end{eqnarray}
Thus, 
\begin{eqnarray}
    \mathcal{U}_{\mathrm{AOBOA}}:\qquad \mathbb{R} &\rightarrow& L_{1d} \subset \Gamma \cr
    \mathcal{U}_{\mathrm{AOBOA}}: \; \eta_k^\mathrm{comb} &\mapsto& x_{k+1} ,
\end{eqnarray}
where $L_{1d}$ denotes a line in $\Gamma$.

\subsubsection{Image at $V$ and $\widetilde{V}$}
\begin{eqnarray}
    \left(\begin{array}{c} 0\\0\end{array}\right) 
    &=& \mathcal{U}_{\mathrm{AOBOA}}(\eta_k^{\mathrm{comb}}; x_k ,V) - \mathcal{U}_{\mathrm{AOBOA}}(\widetilde{\eta}_k^{\mathrm{comb}}; x_k ,\widetilde{V}) \cr
    &=& \left(\begin{array}{l} q_{k} + (a' + a'd'd')p_k  + a'd'b(q_{k} + a' p_k) \\d'd'p_{k} + d'b(q_{k} + a' p_k) \end{array}\right) +
        \left(\begin{array}{l} a'\\ 1\end{array}\right) f'\eta^{\mathrm{comb}}_k - \cr
    &&  \left(\begin{array}{l} q_{k} + (a' + a'd'd')p_k  + a'd'\cdot \widetilde{b}(q_{k} + a' p_k) \\d'd'p_{k} + d'\cdot \widetilde{b}(q_{k} + a' p_k) \end{array}\right) -
        \left(\begin{array}{l} a' \\ 1 \end{array}\right) f'\widetilde{\eta}^{\mathrm{comb}}_k  \cr
    &=& \left(\begin{array}{l} a'd'b(q_{k} + a' p_k) \\d'b(q_{k} + a' p_k) \end{array}\right) -
        \left(\begin{array}{l} a'd'\cdot \widetilde{b}(q_{k} + a' p_k) \\d'\cdot \widetilde{b}(q_{k} + a' p_k) \end{array}\right) -
        \left(\begin{array}{l} a' \\ 1\end{array}\right)f' \Delta \eta^{\mathrm{comb}}_k \cr
    &=& \left(\begin{array}{l} a' \\ 1\end{array}\right) d'\left[b(q_{k}+ a' p_k) -  \widetilde{b}(q_{k} + a' p_k)\right] -
        \left(\begin{array}{l} a' \\ 1\end{array}\right)f' \Delta \eta^{\mathrm{comb}}_k \cr
    &=& \left(\begin{array}{l} a' \\ 1\end{array}\right) d'\Delta t \nabla U(q_{k}+ a' p_k) -
        \left(\begin{array}{l} a' \\ 1\end{array}\right) f'\Delta \eta^{\mathrm{comb}}_k      
\end{eqnarray}    
is fulfilled if
\begin{eqnarray}
    \Delta \eta^{\mathrm{comb}}_k &=& \frac{d'}{f'}\Delta t \nabla U(q_{k+1/2})
\end{eqnarray}
where we substituted $q_{k+1/2} = q_{k}+ a' p_k$ (eq.~\ref{eq:AOBOA01}).
Thus, $\mathcal{U}_{\mathrm{AOBOA}}(\eta_k; x_k ,V)$ and $\mathcal{U}_{\mathrm{AOBOA}}(\widetilde{\eta_k}; x_k ,\widetilde{V})$ have the same image
$$  
    L_{1d} = \widetilde{L}_{1d} \, .
$$

%
%
\subsection{BOAOB}
\subsubsection{Algorithm}
\begin{subequations}
\begin{eqnarray}
    p_{k+1/4}   &=& p_k - \frac{\Delta t}{2} \nabla V(q_k) \label{eq:BOAOB01}\\
    p_{k+2/4}   &=& e^{-\frac{\xi \Delta t}{2}}\,p_{k+1/4} 
                    + \sqrt{k_BTm \left(1- e^{-\xi\Delta t} \right)} \, \eta_k^{(1)} \label{eq:BOAOB02}\\
    q_{k+1}     &=& q_k + \frac{\Delta t}{m} p_{k+2/4} \label{eq:BOAOB03}\\
    p_{k+3/4}   &=& e^{-\frac{\xi \Delta t}{2}}\,p_{k+2/4}   
                    + \sqrt{k_BTm \left(1- e^{-\xi\Delta t} \right)} \, \eta_k^{(2)}  \label{eq:BOAOB04}\\
    p_{k+1}     &=& p_{k+3/4} - \frac{\Delta t}{2} \nabla V(q_{k+1})
    \, .\label{eq:BOAOB05}
\end{eqnarray}
\end{subequations}
Here, two random numbers $\eta_k^{(1)} \sim \mathcal{N}(0,1)$ and $\eta_k^{(2)} \sim \mathcal{N}(0,1)$ need to be drawn per full update cycle.
%

\subsubsection{Update operator}
\begin{eqnarray}
        \mathcal{B'O'AO'B'} \left(\begin{array}{l} q_{k}\\p_{k}\end{array}\right)
    &=& \mathcal{B'O'AO'} \left(\begin{array}{l} q_{k}\\p_{k} + b'(q_k)\end{array}\right) \cr
    &=& \mathcal{B'O'A} \left(\begin{array}{l} q_{k}\\d'p_{k} + d'b'(q_k) +f'\eta_k^{(1)}\end{array}\right) \cr
    &=& \mathcal{B'O'} \left(\begin{array}{l} q_{k} + ad'p_{k} + ad'b'(q_k) +af'\eta_k^{(1)}\\d'p_{k} + d'b'(q_k) +f'\eta_k^{(1)}\end{array}\right) \cr
    &=& \mathcal{B'} \left(\begin{array}{l} q_{k} + ad'p_{k} + ad'b'(q_k) +af'\eta_k^{(1)}\\d'd'p_{k} + d'd'b'(q_k) +d'f'\eta_k^{(1)} + f'\eta_k^{(2)}\end{array}\right) \cr
    &=& \left(\begin{array}{l} q_{k} + ad'p_{k} + ad'b'(q_k) +af'\eta_k^{(1)}\\d'd'p_{k} + d'd'b'(q_k) +d'f'\eta_k^{(1)} + f'\eta_k^{(2)} + b'\Big(q_{k} + ad'p_{k} + ad'b'(q_k) +af'\eta_k^{(1)}\Big)\end{array}\right) 
\end{eqnarray}

\subsubsection{Update function}
\begin{eqnarray}
     \left(\begin{array}{c} q_{k+1}\\p_{k+1}\end{array}\right) 
     &=&   \mathcal{U}_{\mathrm{BOAOB}}(\eta_k^{(1)}, \eta_k^{(2)}; x_k ,V) \cr
    &=&\left(\begin{array}{l} \bar{q}_{k+1}\\\bar{p}_{k+1} \end{array}\right) +
        \left(\begin{array}{l} 0 \\ b'\Big( \bar{q}_{k+1} + af'\eta^{(1)}_k \Big) \end{array}\right) + \left(\begin{array}{l} af'\\  d'f' \end{array}\right) \eta^{(1)}_k + \left(\begin{array}{l} 0\\  f' \end{array}\right) \eta^{(2)}_k
\label{eq:BOAOB_udpateFunction}        
\end{eqnarray}
with
$\bar{q}_{k+1} = q_{k} + ad'p_{k} + ad'b'(q_k)$
and
$\bar{p}_{k+1} = d'd'p_{k} + d'd'b'(q_k)$.
Thus, 
\begin{eqnarray}
    \mathcal{U}_{\mathrm{BOAOB}}:\qquad \;\;\;\;\; \mathbb{R}^2 &\rightarrow&  \Gamma \cr
    \mathcal{U}_{\mathrm{BOAOB}}: \; (\eta_k^{(1)},\eta_k^{(2)})  &\mapsto& x_{k+1} \, .
\end{eqnarray}

\subsubsection{Derivation of $\Delta \eta_k^{(1)}$ and $\Delta \eta_k^{(2)}$}
We derive $\Delta \eta_k^{(1)}$ and $\Delta \eta_k^{(2)}$ from the condition 
\begin{eqnarray}
    \left(\begin{array}{l} 0\\  0 \end{array}\right) &=& \mathcal{U}_{\mathrm{BOAOB}}(\eta_k^{(1)}, \eta_k^{(2)}; x_k ,V) -
    \mathcal{U}_{\mathrm{BOAOB}}(\widetilde{\eta}_k^{(1)}, \widetilde{\eta}_k^{(2)}; x_k ,\widetilde{V})
\end{eqnarray}
Inserting eq.~\ref{eq:BOAOB_udpateFunction} yields
\begin{eqnarray}
     \left(\begin{array}{l} 0\\  0 \end{array}\right)
    &=&\quad \left(\begin{array}{l} q_{k} + ad'p_{k} + ad'b'(q_k)\\d'd'p_{k} + d'd'b'(q_k) \end{array}\right) +
        \left(\begin{array}{l} 0 \\ b'(q_{k+1}) \end{array}\right) + \left(\begin{array}{l} af'\\  d'f' \end{array}\right) \eta^{(1)}_k + \left(\begin{array}{l} 0\\  f' \end{array}\right) \eta^{(2)}_k\cr
    && - \left(\begin{array}{l} q_{k} + ad'p_{k} + ad'\widetilde{b}'(q_k)\\d'd'p_{k} + d'd'\widetilde{b}'(q_k) \end{array}\right) -
        \left(\begin{array}{l} 0 \\ \widetilde{b}'(\widetilde{q}_{k+1}) \end{array}\right) - \left(\begin{array}{l} af'\\  d'f' \end{array}\right) \widetilde{\eta}^{(1)}_k - \left(\begin{array}{l} 0\\  f' \end{array}\right) \widetilde{\eta}^{(2)}_k \cr\cr 
    &=&\quad \left(\begin{array}{l} ad'\frac{\Delta t}{2}\nabla U(q_k)\\d'd'\frac{\Delta t}{2}\nabla U(q_k) \end{array}\right) +
        \left(\begin{array}{l} 0 \\ b'(q_{k+1}) - \widetilde{b}'(\widetilde{q}_{k+1}) \end{array}\right) - \left(\begin{array}{l} af'\\  d'f' \end{array}\right) \Delta \eta^{(1)}_k - \left(\begin{array}{l} 0\\  f' \end{array}\right) \Delta \eta^{(2)}_k \, , 
\end{eqnarray}
where we used eqs.~\ref{eq:DeltaEta_k} and \ref{eq:DeltaGradU}.
The second term evaluates the potential at the updated position which, a priori, might differ in $V$ and $\widetilde{V}$. 
Solving the line for the position in the above equation yields
\begin{eqnarray}
   \Delta \eta^{(1)}_k &=& \frac{d'}{ f'}\frac{\Delta t}{2}\nabla U(q_k) \, .
\end{eqnarray}
With this $q_{k+1} = \widetilde{q}_{k+1}$, and thus $b'(q_{k+1}) - \widetilde{b}'(\widetilde{q}_{k+1}) = \frac{\Delta t}{2} \nabla U(q_{k+1})$.
Then the line for the momentum yields
\begin{eqnarray}
    0 &=& d'd'\frac{\Delta t}{2}\nabla U(q_k) + \frac{\Delta t}{2} \nabla U(q_{k+1}) - d'f'\cdot \frac{d'}{ f'}\frac{\Delta t}{2}\nabla U(q_k) - f'\Delta \eta^{(2)}_k \cr
    &=& \frac{\Delta t}{2} \nabla U(q_{k+1})  - f'\Delta \eta^{(2)}_k \cr
     &\Updownarrow& \cr
    \Delta \eta^{(2)}_k &=& \frac{1}{f'} \frac{\Delta t}{2} \nabla U(q_{k+1}) 
\end{eqnarray}

%
%
\subsection{OBABO method / Bussi-Parrinello thermostat}
\subsubsection{Algorithm}
\begin{subequations}
\begin{eqnarray}
    p_{k+1/4}   &=& e^{-\frac{\xi \Delta t}{2}}\,p_k 
                    + \sqrt{k_BTm \left(1- e^{-\xi\Delta t} \right)} \, \eta_k^{(1)} \label{eq:OBABO01}\\
    p_{k+2/4}   &=& p_{k+1/4} - \frac{\Delta t}{2} \nabla V(q_k) \label{eq:OBABO02}\\
    q_{k+1}     &=& q_k + \frac{\Delta t}{m} p_{k+2/4} \label{eq:OBABO03}\\
    p_{k+3/4}   &=& p_{k+2/4} - \frac{\Delta t}{2} \nabla V(q_{k+1}) \label{eq:OBABO04}\\
    p_{k+1}     &=& e^{-\frac{\xi \Delta t}{2}}\,p_{k+3/4}   
                    + \sqrt{k_BTm \left(1- e^{-\xi\Delta t} \right)} \, \eta_k^{(2)} \, .\label{eq:OBABO05}
\end{eqnarray}
\end{subequations}
Here, two random numbers $\eta_k^{(1)} \sim \mathcal{N}(0,1)$ and $\eta_k^{(2)} \sim \mathcal{N}(0,1)$ need to be drawn per full update cycle.
The algorithm is equal to the Bussi-Parrinello thermostat \cite{Bussi:2007}.
Compared to Ref.~\onlinecite{Leimkuhler:2013}, we changed the notation as follows: 
$n \rightarrow k$,
$R_n \rightarrow \eta_k$,
$\delta t \rightarrow \Delta t$, 
$M \rightarrow m$, 
$\gamma \rightarrow \xi$, 
$F \rightarrow -\nabla V$.

\subsubsection{Update operator}
\begin{eqnarray}
        \mathcal{O'B'AB'O'} \left(\begin{array}{l} q_{k}\\p_{k}\end{array}\right)
    &=& \mathcal{O'B'AB'} \left(\begin{array}{l} q_{k}\\d'p_k + f'\eta^{(1)}_k\end{array}\right) \cr
    &=& \mathcal{O'B'A} \left(\begin{array}{l} q_{k}\\d'p_k + f'\eta^{(1)}_k + b'(q_k) \end{array}\right) \cr
    &=& \mathcal{O'B'} \left(\begin{array}{l} q_{k}+ad'p_k+af'\eta_k^{(1)}+ab'(q_k)\\d'p_k + f'\eta^{(1)}_k + b'(q_k) \end{array}\right) \cr
    &=& \mathcal{O'} \left(\begin{array}{l} q_{k}+ad'p_k+af'\eta_k^{(1)}+ab'(q_k)\\d'p_k + f'\eta^{(1)}_k + b'(q_k) + b' \Big( q_{k}+ad'p_k+af'\eta_k^{(1)}+ab'(q_k) \Big) \end{array}\right) \cr
    &=& \left(\begin{array}{l} q_{k}+ad'p_k+af'\eta_k^{(1)}+ab'(q_k)\\d'd'p_k + d'f'\eta^{(1)}_k + d'b'(q_k) + d'b' \Big( q_{k}+ad'p_k+af'\eta_k^{(1)}+ab'(q_k) \Big) +f'\eta_k^{(2)} \end{array}\right) 
\end{eqnarray}

\subsubsection{Update function}
\begin{eqnarray}
     \left(\begin{array}{c} q_{k+1}\\p_{k+1}\end{array}\right) 
     &=&   \mathcal{U}_{\mathrm{OBABO}}(\eta_k^{(1)},\eta_k^{(2)}; x_k ,V) \cr
    &=& \left(\begin{array}{l} \bar{q}_{k+1}\\\bar{p}_{k+1} \end{array}\right) +
        \left(\begin{array}{l} 0\\d'b'(\bar{q}_{k+1}+ af'\eta_k^{(1)})\end{array}\right) + 
        \left(\begin{array}{l} af'\\d'f'\end{array}\right)\eta_k^{(1)} + 
        \left(\begin{array}{l} 0\\ f' \end{array}\right)\eta_k^{(2)}
\label{eq:OBABO_udpateFunction}        
\end{eqnarray}
with 
$\bar{q}_{k+1} = q_{k} + ad'p_{k} + ab'(q_k)$
and
$\bar{p}_{k+1} = d'd'p_{k} + d'b'(q_k)$.
Thus, 
\begin{eqnarray}
    \mathcal{U}_{\mathrm{OBABO}}:\qquad\;\;\;\; \mathbb{R}^2 &\rightarrow&  \Gamma \cr
    \mathcal{U}_{\mathrm{OBABO}}: \; (\eta_k^{(1)},\eta_k^{(2)})  &\mapsto& x_{k+1} \, .
\end{eqnarray}

\subsubsection{Derivation of $\Delta \eta_k^{(1)}$ and $\Delta \eta_k^{(2)}$}
We derive $\Delta \eta_k^{(1)}$ and $\Delta \eta_k^{(2)}$ from the condition 
\begin{eqnarray}
    \left(\begin{array}{l} 0\\  0 \end{array}\right) &=& \mathcal{U}_{\mathrm{OBABO}}(\eta_k^{(1)}, \eta_k^{(2)}; x_k ,V) -
    \mathcal{U}_{\mathrm{OBABO}}(\widetilde{\eta}_k^{(1)}, \widetilde{\eta}_k^{(2)}; x_k ,\widetilde{V})
\end{eqnarray}
Inserting eq.~\ref{eq:OBABO_udpateFunction} yields
\begin{eqnarray}
     \left(\begin{array}{c} 0\\0\end{array}\right)
    &=& \quad\left(\begin{array}{l} q_{k} + ad'p_{k} + ab'(q_k)\\d'd'p_{k} + d'b'(q_k) \end{array}\right) +
        \left(\begin{array}{l} 0\\d'b'(q_{k+1})\end{array}\right) + 
        \left(\begin{array}{l} af'\\d'f'\end{array}\right)\eta_k^{(1)} + 
        \left(\begin{array}{l} 0\\ f' \end{array}\right)\eta_k^{(2)} \cr
    &&  -\left(\begin{array}{l} q_{k} + ad'p_{k} + a\widetilde{b}'(q_k)\\d'd'p_{k} + d'\widetilde{b}'(q_k) \end{array}\right) -
        \left(\begin{array}{l} 0\\d'\widetilde{b}'(\widetilde{q}_{k+1})\end{array}\right) - 
        \left(\begin{array}{l} af'\\d'f'\end{array}\right)\widetilde{\eta}_k^{(1)} - 
        \left(\begin{array}{l} 0\\ f' \end{array}\right)\widetilde{\eta}_k^{(2)} \cr      
    &=& \quad\left(\begin{array}{l} a\frac{\Delta t}{2}\nabla U(q_k)\\d'\frac{\Delta t}{2}\nabla U(q_k) \end{array}\right) +
        \left(\begin{array}{l} 0\\d'b'(q_{k+1}) - d'\widetilde{b}'(\widetilde{q}_{k+1} )\end{array}\right) -
        \left(\begin{array}{l} af'\\d'f'\end{array}\right)\Delta \eta_k^{(1)} - 
        \left(\begin{array}{l} 0\\ f' \end{array}\right)\Delta \eta_k^{(2)}
\end{eqnarray}
where we used eqs.~\ref{eq:DeltaGradU} and \ref{eq:DeltaEta_k}.
The second term evaluates the potential at the updated position which, a priori, might differ in $V$ and $\widetilde{V}$. 
Solving the line for the position in the above equation yields
\begin{eqnarray}
    \Delta \eta_k^{(1)} &=& \frac{1}{f'}\frac{\Delta t}{2}\nabla U(q_k)  \, .
\end{eqnarray}
With this $q_{k+1} = \widetilde{q}_{k+1}$, and thus $d'b'(q_{k+1}) - d'\widetilde{b}'(\widetilde{q}_{k+1}) = d'\frac{\Delta t}{2} \nabla U(q_{k+1})$.
Then the line for the momentum yields
\begin{eqnarray}
    0   &=& d'\frac{\Delta t}{2}\nabla U(q_k) + d'\frac{\Delta t}{2} \nabla U(q_{k+1}) 
          - d'f' \cdot \frac{1}{f'}\frac{\Delta t}{2}\nabla U(q_k) - f' \Delta \eta_k^{(2)} \cr
        &=& d'\frac{\Delta t}{2} \nabla U(q_{k+1}) - f' \Delta \eta_k^{(2)} \cr
        &\Updownarrow& \cr
    \Delta \eta_k^{(2)}    &=& \frac{d'}{f'}\frac{\Delta t}{2} \nabla U(q_{k+1})     
\end{eqnarray}

%
%
\subsection{OABAO}
\subsubsection{Algorithm}
\begin{subequations}
\begin{eqnarray}
    p_{k+1/3} &=& e^{-\frac{\xi \Delta t}{2}}\, p_k +  \sqrt{k_BTm(1-e^{-\xi\Delta t})}\, \eta_{k}^{(1)} \label{eq:OABAO01} \\
    q_{k+1/2} &=& q_k + \frac{\Delta t}{2 m} \, p_{k+1/3}   \label{eq:OABAO02} \\
    p_{k+2/3} &=&  p_{k+1/3} - \Delta t \nabla V(q_{k+1/2})   \label{eq:OABAO03} \\
    q_{k+1} &=& q_{k+1/2} +  \frac{\Delta t}{2m} \, p_{k+2/3} \label{eq:OABAO04} \\
    p_{k+1} &=& e^{-\frac{\xi \Delta t}{2}}\, p_{k+2/3} +  \sqrt{k_BTm(1-e^{-\xi\Delta t})}\, \eta_{k}^{(2)}  \label{eq:OABAO05}
\end{eqnarray}
\end{subequations}
Here, two random numbers $\eta_k^{(1)} \sim \mathcal{N}(0,1)$ and $\eta_k^{(2)} \sim \mathcal{N}(0,1)$ need to be drawn per full update cycle.
%

\subsubsection{Update operator}
\begin{eqnarray}
        \mathcal{O'A'BA'O'} \left(\begin{array}{l} q_k\\p_k\end{array}\right)
    &=& \mathcal{O'A'BA'} \left(\begin{array}{l} q_k\\d'p_k + f'\eta_k^{(1)}\end{array}\right) \cr
    &=& \mathcal{O'A'B} \left(\begin{array}{l} q_k + a'd'p_k + a'f'\eta_k^{(1)}\\d'p_k + f'\eta_k^{(1)}\end{array}\right) \cr
    &=& \mathcal{O'A'} \left(\begin{array}{l} q_k + a'd'p_k + a'f'\eta_k^{(1)}\\d'p_k + f'\eta_k^{(1)} + b(q_k + a'd'p_k + a'f'\eta_k^{(1)})\end{array}\right) \cr
    &=& \mathcal{O'} \left(\begin{array}{l} q_k + a'd'p_k + a'f'\eta_k^{(1)} + a'd'p_k + a'f'\eta_k^{(1)} + a'b(q_k + a'd'p_k + a'f'\eta_k^{(1)})\\d'p_k + f'\eta_k^{(1)} + b(q_k + a'd'p_k + a'f'\eta_k^{(1)})\end{array}\right) \cr
    &=& \left(\begin{array}{l} q_k + a'd'p_k + a'f'\eta_k^{(1)} + a'd'p_k + a'f'\eta_k^{(1)} + a'b(q_k + a'd'p_k + a'f'\eta_k^{(1)})\\d'd'p_k + d'f'\eta_k^{(1)} + d'b(q_k + a'd'p_k + a'f'\eta_k^{(1)}) + f'\eta_k^{(2)}\end{array}\right) 
\end{eqnarray}    

\subsubsection{Update function}
\begin{eqnarray}
     \left(\begin{array}{c} q_{k+1}\\p_{k+1}\end{array}\right) 
     &=&   \mathcal{U}_{\mathrm{OABAO}}(\eta_k^{(1)},\eta_k^{(2)}; x_k ,V) \cr
    &=& \left(\begin{array}{l} \bar{q}_{k+1} \\ \bar{p}_{k+1} \end{array}\right) + 
    \left(\begin{array}{l} a'b(q_k + a'd'p_k + a'f'\eta_k^{(1)})  \\  d'b(q_k + a'd'p_k + a'f'\eta_k^{(1)}) \end{array}\right) + 
    \left(\begin{array}{l} 2a'f'\\ d'f'\end{array}\right)\eta_k^{(1)} + 
    \left(\begin{array}{l} 0 \\  f'\end{array}\right) \eta_k^{(2)}
\end{eqnarray}
with 
$\bar{q}_{k+1} = q_k + 2a'd'p_k$ 
and
$\bar{p}_{k+1} = d'd'p_k $.
Thus for most potentials, 
\begin{eqnarray}
    \mathcal{U}_{\mathrm{OABAO}}:\qquad\;\; \mathbb{R}^2 &\rightarrow&  \Gamma \cr
    \mathcal{U}_{\mathrm{OABAO}}: \; (\eta_k^{(1)},\eta_k^{(2)})  &\mapsto& x_{k+1} \, .
\end{eqnarray}
If e.g. $a'b(q_k + a'd'p_k + a'f'\eta_k^{(1)}) = - 2a'f'\eta_k^{(1)}$, the contribution of $\eta_k^{(1)}$ to the position update cancels, and  the the image of $\mathcal{U}_{\mathrm{OABAO}}$ is a line parallel to the $p$-axis.

%
%

\end{document}